\documentclass[12pt]{article}

\usepackage[margin=1in]{geometry}

\usepackage{fullpage}
\usepackage[utf8]{inputenc}
\usepackage[english]{babel}
\usepackage[shortlabels]{enumitem}
\usepackage{mathtools,
  booktabs,
  braket,
  mathrsfs,
  latexsym,
  epsfig,
  xcolor,
  url,
  setspace,
  graphics,
  lipsum,
  float,
  lineno,
  placeins
}
\usepackage{subfigure}
\usepackage{float}
\usepackage{pdflscape}
\newcommand{\bland}{\begin{landscape}}
\newcommand{\eland}{\end{landscape}}

\usepackage[flushmargin]{footmisc}
\definecolor{CardinalRed}{cmyk}{0,1,0.65,0.34} 
\usepackage[colorlinks,linkcolor=CardinalRed,citecolor=CardinalRed,urlcolor = CardinalRed]{hyperref}
\usepackage{hyperref}

\makeatletter
\def\maxwidth{\ifdim\Gin@nat@width>\linewidth\linewidth\else\Gin@nat@width\fi}
\def\maxheight{\ifdim\Gin@nat@height>\textheight\textheight\else\Gin@nat@height\fi}
\makeatother

\setkeys{Gin}{width=\maxwidth,height=\maxheight,keepaspectratio}

\newcommand{\burl}[1]{\textcolor{blue}{\url{#1}}}

\usepackage[round]{natbib}

\usepackage[OT1]{fontenc}

\usepackage{todonotes}

\makeatletter
\providecommand\@dotsep{5}
\def\listtodoname{List of Todos}
\def\listoftodos{\@starttoc{tdo}\listtodoname}
\makeatother

\graphicspath{{./graphs/}}%

\usepackage[labelsep=period, font=sc, skip=0pt,justification=centering]{caption}
\usepackage[figuresright]{rotating}
\usepackage{titlesec}
\titlelabel{\thetitle.\ }

\usepackage{titling}
\usepackage{titlesec}
\titleformat{\section}
{\normalfont\fontsize{15}{15}\bfseries}{\thesection.}{0.5em}{}

\newenvironment{itemize*}%
  {\begin{itemize}%
    \setlength{\itemsep}{0pt}%
    \setlength{\parskip}{0pt}}%
  {\end{itemize}}
\newenvironment{enumerate*}%
  {\begin{enumerate}%
    \setlength{\itemsep}{0pt}%
    \setlength{\parskip}{0pt}}%
  {\end{enumerate}}

\usepackage{
  amscd,
  amsfonts,
  amsmath,
  amssymb,
  amsbsy,
  amsthm,
  bm, 
  dsfont,
  cancel, 
  latexsym,
  mathtools,
  graphicx,
  xcolor,
  xargs
}

\usepackage{longtable}

\newcommand{\beq}{\begin{equation}}
\newcommand{\eeq}{\end{equation}}

\newcommand*\Bigpar[1]{\left( #1 \right )}

\newcommand{\ba}{\begin{array}}
\newcommand{\ea}{\end{array}}
\newcommand{\be}{\begin{enumerate}}
\newcommand{\ee}{\end{enumerate}}
\newcommand{\bi}{\begin{itemize}}
\newcommand{\ei}{\end{itemize}}
\newcommand{\bs}{\begin{align}\begin{split}\nonumber}
\newcommand{\bsnumber}{\begin{align}\begin{split}}
\newcommand{\es}{\end{split}\end{align}}

\newcommandx{\deriv}[2][1=x,2=f]{\nabla \, #2 \Bigpar{ #1 } }

\newcommand\frakfamily{\usefont{U}{yfrak}{m}{n}}
\DeclareTextFontCommand{\textfrak}{\frakfamily}

\newcommand{\mc}[1]{\mathcal{#1}}
\def\mbf#1{\mathbf{#1}}

\newcommand{\E}{\mathbb{E}} 

\newcommand{\indic}[1]{\mbf{1}\left\{#1\right\}} 

\newcommand\indep{\protect\mathpalette{\protect\independenT}{\perp}}
\def\independenT#1#2{\mathrel{\rlap{$#1#2$}\mkern5mu{#1#2}}}

\newcommand{\hyp}[2]{
\ensuremath{H_0:#1 \ifhmode\quad\text{versus}\quad\fi\text{ vs. } H_1:#2}}

\newcommandx{\uniff}[1][1={a,b}]{\textrm{Unif}\left({#1}\right)}
\newcommandx{\unifd}[1][1={a,\ldots,b}]{\textrm{Unif}\left\{{#1}\right\}}
\newcommandx{\dunif}[3][1=x,2=a,3=b]{\frac{I(#2<#1<#3)}{#3-#2}}
\newcommandx{\dunifd}[3][1=x,2=a,3=b]{\frac{I(#2\le#1\le#3)}{#3-#2+1}}
\newcommandx{\punif}[3][1=x,2=a,3=b]{
\begin{cases} 0 & #1 < #2 \\ \frac{#1-#2}{#3-#2} & #2 < #1 < #3 \\ 1 & #1 > #3\\\end{cases}}
\newcommandx{\punifd}[3][1=x,2=a,3=b]{
\begin{cases} 0 & #1 < #2\\ \frac{\lfloor#1\rfloor-#2+1}{#3-#2} & #2 \le #1 \le #3 \\ 1 & #1 > #3\\ \end{cases}}

\newcommandx\bern[1][1=p]{\textrm{Bern}\left({#1}\right)}
\newcommandx\dbern[2][1=x,2=p]{#2^{#1} \left(1-#2\right)^{1-#1}}
\newcommandx\pbern[2][1=x,2=p]{\left(1-#2\right)^{1-#1}}

\newcommandx\bin[1][1={n,p}]{\textrm{Bin}\left(#1\right)}
\newcommandx\dbin[3][1=x,2=n,3=p]{\binom{#2}{#1}#3^#1\left(1-#3\right)^{#2-#1}}

\newcommandx\mult[1][1={n,p}]{\textrm{Mult}\left(#1\right)}
\newcommandx\dmult[3][1=x,2=n,3=p]{\frac{#2!}{#1_1!\ldots#1_k!}#3_1^{#1_1}\cdots#3_k^{#1_k}}

\newcommandx\hyper[1][1={N,m,n}]{\textrm{Hyp}\left({#1}\right)}
\newcommandx\dhyper[4][1=x,2=N,3=m,4=n]{\frac{\binom{#3}{#1}\binom{#2-#3}{#4-#1}}{\binom{#2}{#4}}}

\newcommandx\nbin[1][1={r,p}]{\textrm{NBin}\left({#1}\right)}
\newcommandx\dnbin[3][1=x,2=r,3=p]{\binom{#1+#2-1}{#2-1}#3^#2(1-#3)^#1}
\newcommandx\pnbin[3][1=x,2=r,3=p]{I_#3(#2,#1+1)}

\newcommandx\geo[1][1=p]{\textrm{Geo}\left(#1\right)}
\newcommandx\dgeo[2][1=x,2=p]{#2(1-#2)^{#1-1}}
\newcommandx\pgeo[2][1=x,2=p]{1-(1-#2)^#1}

\newcommandx\pois[1][1=\lambda]{\textrm{Po}\left({#1}\right)}
\newcommandx\dpois[2][1=x,2=\lambda]{\frac{#2^#1 e^{-#2}}{#1!}}
\newcommandx\ppois[2][1=x,2=\lambda]{e^{-#2}\sum_{i=0}^#1\frac{#2^i}{i!}}

\newcommandx\normall[1][1={\mu,\sigma^2}]{\mathcal{N}\left({#1}\right)}
\newcommandx\dnormall[3][1=x,2=\mu,3=\sigma]%
  {\frac{1}{#3\sqrt{2\pi}}\exp \Bigpar{-\frac{\left(#1-#2\right)^2}{2 #3^2}}}
\newcommandx\pnormall[1][1=x]{\Phi\left({#1}\right)}
\newcommandx\qnormall[1]{\Phi^{-1}\left({#1}\right)}

\newcommandx\mvn[1][1={\mu,\Sigma}]{\mathrm{MVN}\left({#1}\right)}

\newcommandx\ex[1][1=\lambda]{\textrm{Exp}\left(#1\right)}
\newcommandx\dex[2][1=x,2=\lambda]{#2e^{-#1 #2}}
\newcommandx\pex[2][1=x,2=\lambda]{1-e^{-#1 #2}}

\newcommandx\gam[1][1={\alpha,\lambda}]{\textrm{Gamma}\left({#1}\right)}
\newcommandx\dgamma[3][1=x,2=\alpha,3=\lambda]%
  {\frac{#3^{#2}}{\Gamma\left( #2 \right)} #1^{#2-1}e^{-#3#1}}

\newcommandx\invgamma[1][1={\alpha,\beta}]{\textrm{InvGamma}\left({#1}\right)}
\newcommandx\dinvgamma[3][1=x,2=\alpha,3=\beta]%
{\frac{#3^{#2}}{\Gamma\left(#2\right)}#1^{-#2-1}e^{-#3/#1}}
\newcommandx\pinvgamma[3][1=x,2=\alpha,3=\beta]%
{\frac{\Gamma\left(#2,\frac{#3}{#1}\right)}{\Gamma\left(#2\right)}}

\newcommandx\bet[1][1={\alpha,\beta}]{\textrm{Beta}\left(#1\right)}
\newcommandx\dbeta[3][1=x,2=\alpha,3=\beta]
{\frac{\Gamma\left(#2+#3\right)}{\Gamma\left(#2\right)\Gamma\left(#3\right)}#1^{#2-1}\left(1-#1\right)^{#3-1}}

\newcommandx\dir[1][1={\alpha}]{\textrm{Dir}\left(#1\right)}
\newcommandx\ddir[3][1=x,2=\alpha]{\frac{\Gamma\left(\sum_{i=1}^k #2_i\right)}{\prod_{i=1}^k\Gamma\left(#2_i\right)}\prod_{i=1}^k #1_i^{#2_i-1}}

\newcommandx\weibull[1][1={\alpha}]{\textrm{Dir}\left(#1\right)}
\newcommandx\dweibull[3][1=x,2=\lambda,3=k]{\frac{#3}{#2}
\left(\frac{#1}{#2}\right)^{#3-1} e^{-(#1/#2)^k}}

\newcommandx\chisq[1][1=k]{\chi_{#1}^2}

\newcommandx\zet[1][1=s]{\textrm{Zeta}\left(#1\right)}
\newcommandx\dzeta[2][1=x,2=s]{\frac{#1^{-#2}}{\zeta\left(#2\right)}}

\newtheoremstyle{mystyle}
  {12pt}
  {12pt}
  {}
  {}
  {\sffamily \bfseries }
  {.}
  {0.5em}
  {\thmname{#1}\thmnumber{ #2}\thmnote{ (#3)}}

\theoremstyle{mystyle}

\newenvironment{proof-sketch}{\noindent{\bf Sketch of Proof}
  \hspace*{1em}}{\qed\bigskip\\}
\newenvironment{proof-idea}{\noindent{\bf Proof Idea}
  \hspace*{1em}}{\qed\bigskip\\}
\newenvironment{proof-of-lemma}[1][{}]{\noindent{\bf Proof of Lemma {#1}}
  \hspace*{1em}}{\qed\bigskip\\}
\newenvironment{proof-of-proposition}[1][{}]{\noindent{\bf
    Proof of Proposition {#1}}
  \hspace*{1em}}{\qed\bigskip\\}
\newenvironment{proof-of-theorem}[1][{}]{\noindent{\bf Proof of Theorem {#1}}
  \hspace*{1em}}{\qed\bigskip\\}
\newenvironment{inner-proof}{\noindent{\bf Proof}\hspace{1em}}{
  $\bigtriangledown$\medskip\\}
\newenvironment{proof-attempt}{\noindent{\bf Proof Attempt}
  \hspace*{1em}}{\qed\bigskip\\}

\usepackage{array}
\newcolumntype{L}[1]{>{\raggedright\let\newline\\\arraybackslash\hspace{0pt}}m{#1}}
\newcolumntype{C}[1]{>{\centering\let\newline\\\arraybackslash\hspace{0pt}}m{#1}}
\newcolumntype{R}[1]{>{\raggedleft\let\newline\\\arraybackslash\hspace{0pt}}m{#1}}

\usepackage{multirow}

\def\E{\mathbb{E}} 

\def\Xit{X_{i,t}}

\def\Yit{Y_{i,t}} 
 
\def\Dit{D_{i,t}} 
\def\epit{\varepsilon_{i,t}}

\def\eventtime{E_{i,t}}

\usepackage{makecell} 
\graphicspath{{./figure/}{./figure/fect_entry/}}%

\begin{document}


\title{\Large\bf Causal Panel Analysis under  
Parallel Trends:\\Lessons from a Large Reanalysis Study%
\thanks{Albert Chiu, PhD student, Department of Political Science, Stanford University. Email: \url{altchiu@stanford.edu}. Xingchen Lan, PhD student, Wilf Family Department of Politics, New York University. Email: \url{xingchenlan@nyu.edu}. Ziyi Liu, PhD student, Haas School of Business, University of California, Berkeley. Email: \url{zyliu2023@berkeley.edu}. Yiqing Xu, Assistant Professor, Department of Political Science, Stanford University. Email: \url{yiqingxu@stanford.edu}. We thank Quintin Beazer, Kirill Borusyak, Kirk Bansak, Gary Cox, Avi Feller, Anthony Fowler, Francisco Garfias, Justin Grimmer, Jens Hainmueller, Erin Hartman, Guido Imbens, Julia Payson, Annamaria Prati, Jonathan Roth, Luwei Ying, Ye Wang, and participants at the 39th PolMeth, the 11th Asian PolMeth, Seminars at UCSD, UCLA, UC Berkeley, and University of Chicago for helpful comments and suggestions. We are also grateful to Neil Malhotra, Andrew Baker, Anton Strezhnev, and the Alethsia platform for providing pre-publication reviews of this paper. Further thanks go to two anonymous reviewers and the APSR editor, Andrew Eggers, for their helpful comments. We appreciate the authors of the studies we reviewed for their constructive feedback and for making their data publicly available.}
\\\bigskip}

\author{Albert Chiu \\(Stanford)\and Xingchen Lan\\(NYU)\and Ziyi Liu\\(Berkeley)\and Yiqing Xu\\(Stanford)}

\date{{\it American Political Science Review}\\{\small{Vol. 120, Iss. 1, Feb. 2026, pp. 245--266}}}

\maketitle

\vspace{-2em}
\begin{abstract}
\noindent Two-way fixed effects (TWFE) models are widely used in political science to establish causality, but recent methodological discussions highlight their limitations under heterogeneous treatment effects (HTE) and violations of the parallel trends (PT) assumption. This growing literature has introduced numerous new estimators and procedures, causing confusion among researchers about the reliability of existing results and best practices. To address these concerns, we replicated and reanalyzed 49 studies from leading journals that employ TWFE models for causal inference using observational panel data with binary treatments. Using six HTE-robust estimators, diagnostic tests, and sensitivity analyses, we find: (i) HTE-robust estimators yield qualitatively similar but highly variable results; (ii) while a few studies show clear signs of PT violations, many lack evidence to support this assumption; and (iii) many studies are underpowered when accounting for HTE and potential PT violations. We emphasize the importance of strong research designs and rigorous validation of key identifying assumptions.


\end{abstract}

\thispagestyle{empty}  
\clearpage
\newpage
\doublespace

\clearpage

\setcounter{page}{1}
\abovedisplayskip=5pt
\belowdisplayskip=5pt


\section{Introduction}

\noindent Over the past decade, political scientists have increasingly relied on panel data to draw causal conclusions \citep{Xu2023}. A favored method for such analyses is the two-way fixed effects (TWFE) model because of its ability to control for unobserved time-invariant confounders and common time trends. In our survey of 102 articles published from 2017 to 2023 in three top political science journals using observational panel data with \emph{binary} treatments, 64 studies (63\%) assume a TWFE model with the following functional form or a close variant:\footnote{The remaining 38 studies can be categorized into five groups: studies focusing on interaction effects (8~studies), studies using nonlinear links such as logit and Poisson (5~studies), studies employing instrumental variables or regression discontinuity designs (8~studies), and studies using other linear specifications, such as only one-way fixed effects or lagged dependent variables (17 studies).}
\vspace{-1em}\begin{equation}\label{eq:twfe}
  \Yit = \tau^{TWFE}\Dit + \Xit{'}\beta + \alpha_i + \xi_t + \epit,\quad\text{for all } i, t,
\end{equation}
where $\Yit$ and $\Dit$ are the outcome and treatment variables for unit $i$ at time $t$; $\Xit$ is a vector of time--varying covariates; $\alpha_{i}$ and $\xi_{t}$ are unit and time fixed effects; and $\epit$ is idiosyncratic errors.\footnote{In some studies classified as using TWFE models, ``unit'' fixed effects are specified at the group level $g$, where multiple units $i$ are nested (e.g., county fixed effects when $i$ indexes cities), or time fixed effects are at a higher level $p$ (e.g., year fixed effects when $t$ indexes days). For simplicity, we use the notation $\alpha_i$ and $\xi_t$ rather than the more general $\alpha_g$ and $\xi_p$.} Researchers typically interpret $\tau^{TWFE}$ as the treatment effect and estimate the model using ordinary least squares. The resulting estimator for $\tau^{TWFE}$ is commonly known as the TWFE estimator. Moreover, researchers frequently conflate this model with a difference-in-differences (DID) design, and use the two terms interchangeably.\footnote{We use the phrase ``DID design'' in reference to DID research design, which differs from the typical usage in the statistics literature that refers to treatment assignment mechanism \citep{xu2024factorial}.} 

Recent methodological discussions have raised concerns about the validity of TWFE models and the associated identifying assumptions, leaving many researchers in a quandary. First, existing findings based on the TWFE models may not hold given recent developments. Second, with the introduction of numerous new estimators and diagnostics, there is confusion about the current best practices. This paper seeks to bridge this gap by reviewing new estimation, inference, and diagnostics methods from the methodological literature and by reanalyzing published studies using both new estimators and the TWFE estimator. Based on the findings, we offer several practical recommendations for researchers.

These criticisms of the use of TWFE models mainly come from two directions. First, causal identification using TWFE models require the \textit{strict exogeneity} assumption, which critics argue is stronger than many researchers realize and is often unrealistic in real-world settings \citep[e.g.,][]{Imai2019-nw}. Strict exogeneity states that:\vspace{0.5em}
$$\hspace{-5em}\text{(Strict exogeneity)}\qquad \E[\varepsilon_{i,t} \mid \mathbf{D_{i}}, \mathbf{X_{i}}, \alpha_{i}, \xi_{t}] = \E[\varepsilon_{i,t} \mid D_{i,t}, X_{i,t}, \alpha_{i}, \xi_{t}] = 0,\quad \forall i, t,\vspace{0.5em}$$
in which $\mathbf{D_{i}} = \{D_{i,1}, D_{i, 2}, \cdots, D_{i,T}\}$ and $\mathbf{X_{i}} = \{X_{i,1}, X_{i, 2}, \cdots, X_{i,T}\}.$
It means that once current treatment status, covariates, and fixed effects are accounted for, treatment status in any other periods has no additional effect on $Y_{i,t}$  \citep[][p. 253]{wooldridge2010econometric}. Under Equation~(\ref{eq:twfe}), strict exogeneity implies a parallel trends (PT) assumption:\vspace{0.5em}
\begin{align*}
\hspace{-7em}\text{(Parallel trends)}\qquad & \E[Y_{i,t}(0) - Y_{i,s}(0)\mid D_{i,t} = 1, D_{i,s} = 0, X_{i,t} - X_{i,s} = x_{0}] \nonumber \\ 
=\ & \E[Y_{j,t}(0) - Y_{j,s}(0)\mid D_{j,t} = 0, D_{j,s} = 0, X_{j,t} - X_{j,s} = x_{0}],\vspace{0.5em}
\end{align*}
in which $Y_{i,t}(0) = Y_{i,t}(d_{i,t} = 0)$ represents the untreated potential outcome for unit $i$ at time~$t$. It states that the change in untreated potential outcomes between any two periods is mean independent of the change in observed treatment status during those periods, once changes in covariate values are controlled for. Threats to PT, such as the presence of time-varying confounders and feedback from past outcomes to current treatment assignment, also invalidate strict exogeneity. Therefore, throughout the rest of the article, we use the term ``PT violations'' to encompass violations of strict exogeneity.\footnote{We discuss the relationship between strict exogeneity and PT, as well as other assumptions, under Equation~(\ref{eq:twfe}) in Section A.1 of the Supplementary Materials (SM). Note that the PT assumption invoked by many HTE-robust estimators does not depend on Equation~(\ref{eq:twfe}).}

The second group of criticisms concerns the consequences of heterogeneous treatment effects (HTE), that is, $\tau^{TWFE}$ is not a constant \citep[e.g.,][]{Goodman-Bacon2021-xb, CDH2020, Strezhnev2018-ku, sun2021-event,callaway2021-did, athey2022design,borusyak2024revisiting}. Researchers have shown that, under HTE, TWFE estimates in general do not converge to a convex combination of the individual treatment effects for observations under the treatment condition, even when the PT assumption is valid. The so-called ``negative weighting'' problem, as described in \citet{CDH2020}, is an alarming theoretical result because it implies that a TWFE estimand can be negative (positive) even when all individual treatment effects are positive (negative). To address this issue, researchers have proposed many new estimators that are ``HTE-robust''---that is, estimators that converge to some convex combinations of individual treatment effects under their identifying assumptions. 

This article thus pursues two goals. First, we explain and compare six recent proposals to amend TWFE models, including the interaction weighted (IW) estimator \citep{sun2021-event}, stacked DID \citep{BLW2022}, CSDID \citep{callaway2021-did}, DID multiple \citep{CDH2020,de2024difference}, PanelMatch (\citealt{IKW2021}, hereafter IKW, \citeyear{IKW2021}), and the imputation method (\citealt{borusyak2024revisiting}, hereafter BJS, \citeyear{borusyak2024revisiting}; \citealt{liu2024practical}, hereafter LWX, \citeyear{liu2024practical}). These estimators produce causally interpretable estimates under HTE and PT (or its variants). Second, we replicate and reanalyze 49 studies published in the {\it American Political Science Review} (APSR), {\it American Journal of Political Science} (AJPS), and {\it The Journal of Politics} (JOP) from 2017 to 2023 which rely on a TWFE model to draw causal conclusions.\footnote{Replication materials for this article are available for download at the APSR Harvard Dataverse \citep{chiu2025replication}.} Our aim is to assess the consequences of using or not using HTE-robust estimators and shed light on the severity of PT violations in political science research.

Our reanalysis shows that, in most studies, the HTE-robust estimators yield qualitatively similar estimates to TWFE models. However, there is considerable variation in how closely these estimators align with TWFE. In three cases, at least one HTE-robust estimator produces an estimate with an opposite sign to the TWFE estimate; in one of these cases, the opposite-sign estimate is also statistically significant at the 5\% level. There is also a more widespread problem of power: HTE-robust estimators tend to have larger measures of uncertainty, which, combined with even small fluctuations in point estimates, can weaken statistical confidence. This is especially relevant for results that originally teeter on the brink of significance.  

The primary concern, however, is the validity of the PT assumption. While only a few studies show clear signs of PT violations, which likely lead to spurious findings, most studies lack the power to rule out that realistic PT violations could explain a nonzero estimated causal effect. In such cases, even mild PT violations (informed by pre-treatment estimates) prevent us from concluding that the original treatment effect is nonzero. This does not mean that these studies are wrong; rather, it indicates that the available data do not have sufficient power to reject the null hypothesis of no effects when the PT assumption is not perfectly met.

Overall, we find that about a small minority of the studies in our sample meet our criteria of being highly credible. In these studies, we can statistically distinguish the treatment effect from zero using an HTE-robust estimator, even when allowing for mild PT violations benchmarked against placebo estimates using pre-treatment data. We recognize this as a high standard, as most researchers do not account for the power needed for the sensitivity analysis we perform. To be clear, our intent is not to criticize the authors of the studies we have replicated, since many of the methods we used were not available at the time their studies were conducted. Our goal is to guide and improve future research.

In light of these findings, we urge researchers to prioritize a strong research design and sufficient power in causal panel studies. Credible observational studies should feature a well-defined treatment-outcome pair, shock-induced variation in treatment assignment, and sufficient power to ensure results are not undermined by small perturbations of key identifying assumptions. Research design has often been overlooked in causal panel analyses, likely because researchers have become accustomed to accepting the strong parametric and exogeneity assumptions behind TWFE models. Recent studies have emphasized the importance of (quasi-)randomness in treatment assignment for the robustness of DID findings \citep[e.g.,][]{Roth2021-hl}. 

This article makes several contributions. First, we propose a typology of various estimators for causal panel analysis. Our typology is based on the settings in which an estimator can be used and how controls are chosen. We also provide a comprehensive comparison of these estimators and show how several proposals are equivalent in some circumstances. We hope this discussion helps researchers deepen their understanding of these estimators. Second, we adapt the robust confidence set approach for sensitivity analysis proposed by \citet{rambachan2023more} to the setting of imputation estimators. We find it highly useful as it avoids the issue of conditional inference--where hypothesis testing conditional on passing a pretest (e.g., a pretrend test) can distort estimation and inference \citep{roth2022pretest}. Third, our reanalysis instills confidence in existing political science research that uses TWFE models correctly while also cautioning against potential risks, such as the failure of the PT assumption and insufficient power. Based on these findings, we provide recommendations to improve practices, including the choice of estimators and the use of proper diagnostics. Finally, we contribute to the ongoing conversation on replication and reproducibility in political science \citep[e.g.,][]{eggers2015validity, lall2016multiple, hainmueller2019much,lal2021much}.

Our work is closely related to \citet{BLW2022}, who evaluate the credibility of a handful of studies with staggered adoption treatments in finance and accounting. It differs in that: (i) we use a wider range of estimators and diagnostic tests on a larger and more diverse set of empirical applications, many of which involve treatment reversals; (ii) our review suggests that while the weighting issue under HTE is important, the main threats to causal inference with panel data are PT violations and insufficient power. Our work also relates to \citet{roth2023s}, \citet{Xu2023}, \citet{de2023credible}, \citet{Arkhangelsky2023-zy}, and \citet{baker2025difference}, who review and synthesize the recent methodological advancements in the DID literature. What sets this paper apart is our application of these innovations to data, allowing us to evaluate the practical relevance of the theoretical critiques.

This research has a few limitations. First, we do not examine methods based on sequential ignorability, an alternative identification framework that assumes no unobserved confounding but allows for dynamic treatment selection up to the current time period. Second, our analysis does not encompass studies that use continuous treatments, which is common in political science research. Finally, as the methodological literature continues to evolve rapidly, our recommendations should be regarded as reflecting current best practices.


\section{TWFE and Its Pitfalls} 

In this section, we review the pitfalls of TWFE models identified in the literature. In the classic two-group and two-period case, the TWFE estimator $\hat\tau^{TWFE}$ is equivalent to the DID estimator, which consistently estimates the average treatment effect on the treated (ATT) under no anticipation and PT even with HTE. These results do not hold more generally in more complex settings with differential treatment adoption times (known as staggered adoption) or treatment reversal, as we will discuss below.

Our survey of the top journals reveals that the TWFE model under Equation~(\ref{eq:twfe}) is the most commonly adopted approach for estimating causal effects using panel data in political science. Fixed effects models began their rise to prominence in political science in the early 2000s, and criticism promptly followed. In a debate with \citet{Green2001-dy}, \citet{Beck2001-as} and \citet{King2001-br} argue that linear fixed effects models often lead to misleading findings because they throw away valuable information, ignore rich temporal dynamics, and are incapable of capturing complex time-varying heterogeneities. Moreover, since both treatment and outcome variables are often serially correlated in panel data, researchers have advised against using standard error (SE) estimators suitable for cross-sectional data, such as Huber-White robust SEs \citep{Bertrand2004-rc}. Scholars also recommend using bootstrap procedures to more effectively control Type I error rates when the number of clusters (units) is small \citep{Cameron2008-ou}.

\subsection{Recent Criticisms} 

In the past few years, a surge of studies has renewed investigation into the properties of the TWFE estimator and the assumptions it requires to achieve causal identification. One group of work urges researchers to better understand TWFE models (and their assumptions) from a design-based perspective, with a focus on restrictions on treatment assignment mechanisms. For example, \citet{Imai2019-nw} point out that the strict exogeneity assumption not only implies the well-known no time-varying confounder requirement, but it also forbids a ``feedback'' effect from past outcomes to treatment assignment. \citet{Blackwell2018-br} clarify that such an assumption is closely related to baseline randomization in which the treatment vector is generated prior to, or independent of, the realization of the outcome. Another body of work cautions researchers that the PT assumption, even in a simple $2\times 2$ DID setting, is sensitive to functional form. For example, \citet{kahn2020promise} emphasize the implicit functional form restrictions imposed by PT, encouraging researchers to justify it from a (quasi-)experimental perspective and address pre-treatment covariate imbalance. \citet{Roth2021-hl} point out that strong assumptions are needed for PT to be scale-independent, ensuring that a monotonic transformation of the outcome does not invalidate it. In practice, we find that many political science studies do not provide strong evidence to justify strict exogeneity or PT. 

\begin{figure}[!ht]
\caption{Toy Examples: TWFE Assumptions Satisfied vs. Violated}\label{fg:toy}
\vspace{-1em}
\centering
\begin{minipage}{1\linewidth}{
\centering
\hspace{0em}
\subfigure[Staggered DID: without Treatment Reversal]{\includegraphics[width = 0.4\textwidth]{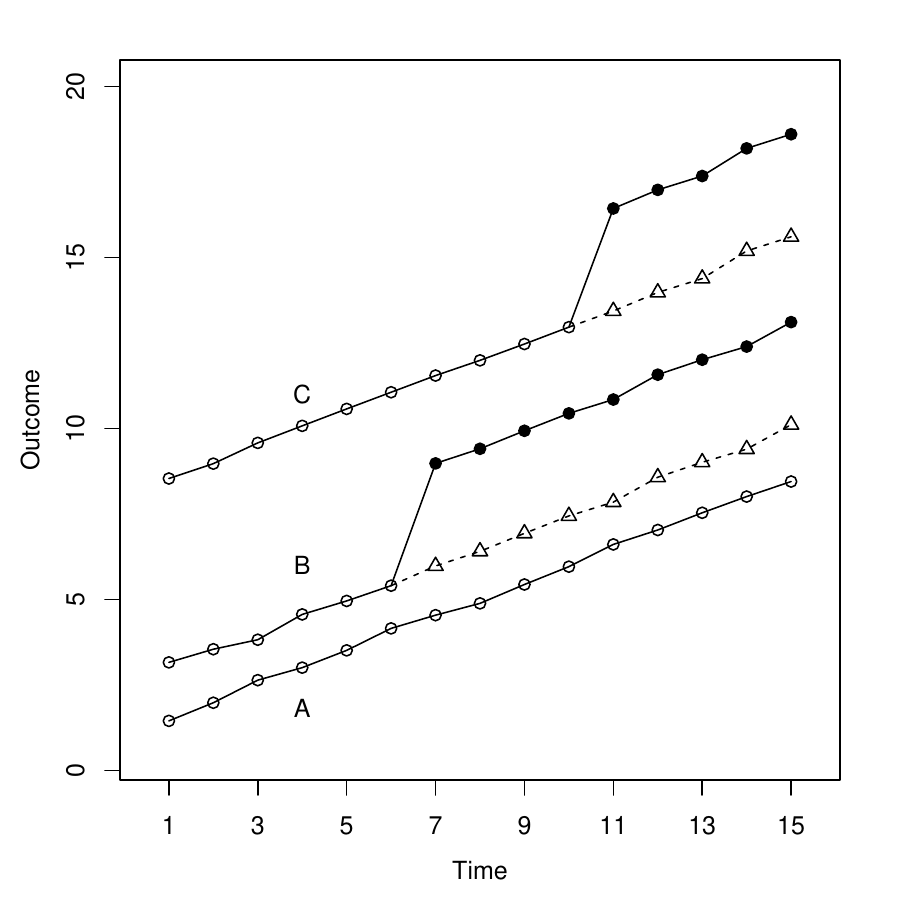}\hspace{1em}
\includegraphics[width = 0.4\textwidth]{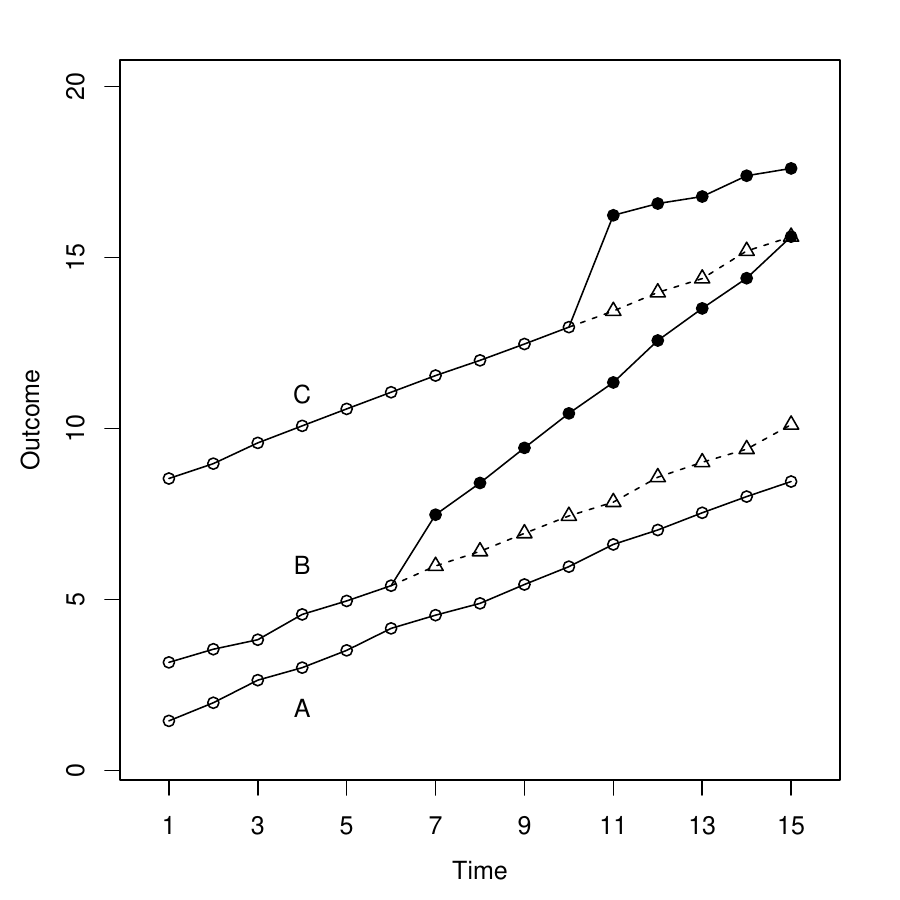}}
\subfigure[General Setting: with Treatment Reversal]{\includegraphics[width = 0.4\textwidth]{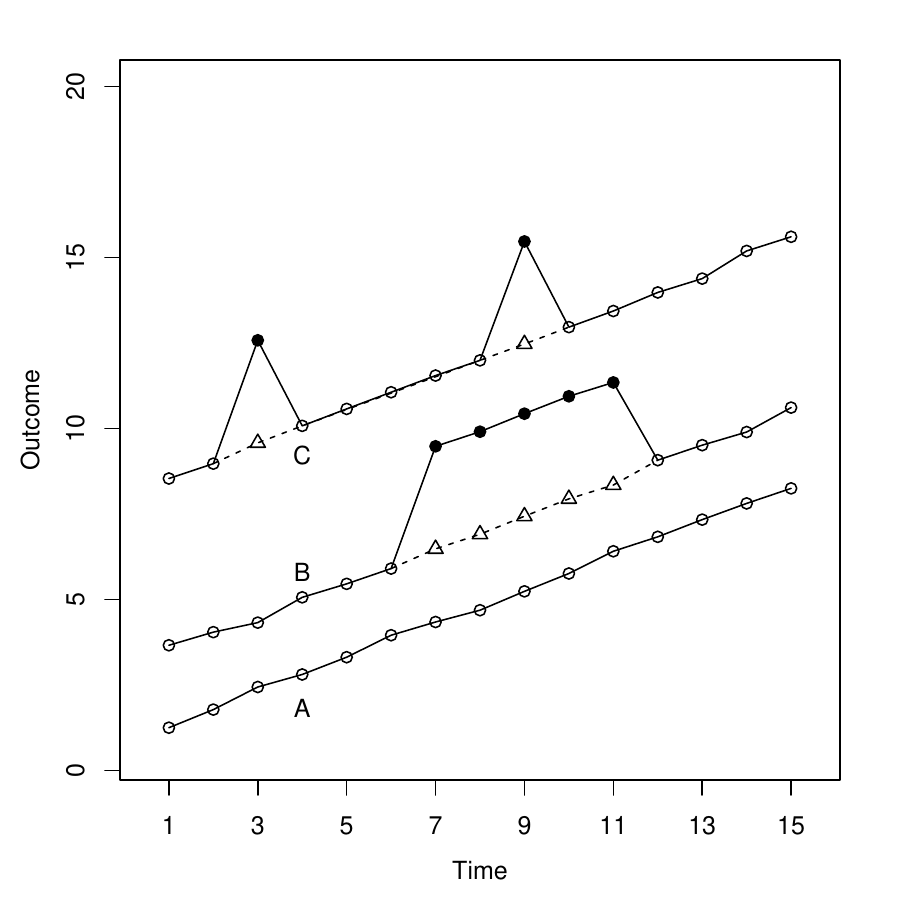}\hspace{1em}
\includegraphics[width = 0.4\textwidth]{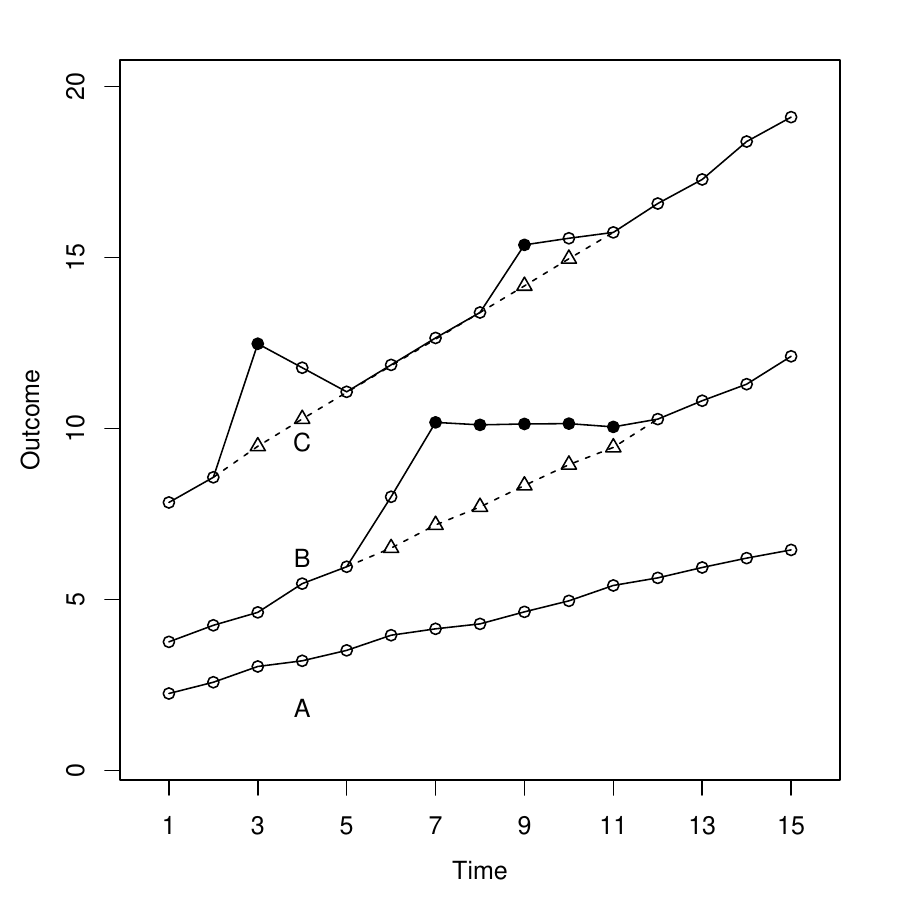}}\\
}
\footnotesize\textbf{Note:} The above panels show outcome trajectories of units in a staggered adoption setting (a) and in a general setting (b). Solid and hollow circles represent observed outcomes under the treatment and control conditions during the current period, respectively, while triangles represent counterfactual outcomes (in the absence of the treatment across all periods), $Y_{i,t}(\mathbf{d_i}=\mathbf{0})$. The data on the \emph{left} panels in both  (a) and (b) are generated by DGPs that satisfy TWFE assumptions while the data on the \emph{right} are not. The divergence between hollow circles and triangles in the right panel of (b), both of which are under the control condition, is caused by anticipation or carryover effects. 
\end{minipage}\vspace{-0.5em}
\end{figure}

A second body of research explores the implications of HTE with binary treatments within TWFE models \citep[e.g.,][]{Goodman-Bacon2021-xb, CDH2020, Strezhnev2018-ku, callaway2021-did, athey2022design,  borusyak2024revisiting}. Most of this literature assumes staggered adoption, but the insights from that setting are still relevant when there are treatment reversals. In Figure~\ref{fg:toy}, we present two simplified examples for the staggered adoption and general settings. The left panel of Figure~\ref{fg:toy}(a) represents outcome trajectories in line with standard TWFE assumptions, which not only include PT but also require that the treatment effect be contemporaneous and unvarying across units and over time. In contrast, the right panel portrays a scenario where PT holds, but the constant treatment effect assumption is not met. Various decompositions by the aforementioned researchers reveal that even under PT, when treatments begin at different times (such as in staggered adoption) and treatment effects evolve over time and vary across units, the TWFE estimand is generally not a convex combination of the individual treatment effects for observations subjected to the treatment. The basic intuition behind this theoretical result is that TWFE models use post-treatment data from units who adopt treatment earlier in the panel as controls for those who adopt the treatment later \citep[e.g.,][]{Goodman-Bacon2021-xb}. HTE-robust estimators capitalize on this insight by avoiding these ``invalid'' comparisons between two treated observations. 

A third limitation of the canonical TWFE specification is its presumption of no temporal and spatial interference. In most uses of TWFE models, researchers assume that there are no spatial spillovers and that treatment effects occur contemporaneously, hence no anticipation or carryover effects. No anticipation effects means that future treatments do not affect today's potential outcomes, while no carryover effects means that today's treatment does not affect future potential outcomes: For any $i, t$,
\begin{align*}
    \text{(No anticipation effects)}\qquad & Y_{i,t}(\mathbf{d}_{i}) = Y_{i,t}(d_{i,1}, d_{i,2}, \ldots, d_{i,t}) \\
    \text{(No carryover effects)}\qquad & Y_{i,t}(\mathbf{d}_{i}) = Y_{i,t}(d_{i,t}, d_{i,t+1}\ldots, d_{i,T})
\end{align*}
in which $\mathbf{d}_{i} = (d_{i,1}, d_{i,2}, \ldots, d_{i,T})$ is unit $i$'s vector of potential treatment conditions in all periods (from period $1$ to~$T$) and $Y_{i,t}(\mathbf{d}_{i})$ is the potential outcome in period $t$ given $\mathbf{d}_{i}$. The TWFE model specified in Equation~(\ref{eq:twfe}) satisfies these two assumptions because $Y_{i,t}(\mathbf{d}_{i}) = Y_{i,t}(d_{i,t})$. These assumptions are obviously very strong, but they are rarely questioned or tested in practice \citep{Imai2019-nw, athey2022design, Wang2021-xo}. Although some recent methods permit arbitrary carryover effects in staggered adoption settings, these effects are not distinguishable from contemporaneous effects.\footnote{See Section A.4 of the SM in \citet{liu2024practical} for more details.} This limitation becomes more complex when treatment reversal is possible, as demonstrated in Figure~\ref{fg:toy}(b). In Figure~\ref{fg:toy}(b), data in the left panel are consistent with TWFE assumptions, while the right panel illustrates deviations from PT, constant treatment effect, and the absence of anticipation or carryover effects. Real-world data often encounter the challenging scenarios depicted in the right panel rather than the idealized conditions in the left. Scholars have proposed methods to handle limited carryover effects in the general setting (IKW, \citeyear{IKW2021}; LWX, \citeyear{liu2024practical}). The challenge of addressing spatial spillover effects without strong structural assumptions still persists \citep{Aronow2020-jc, Wang2021-xo}, but its resolution is beyond the scope of this article.  

\subsection{Causal Estimands} 

To define the estimands clearly, consider the panel setting where multiple units $i\in\{1,\dots,N\}$ are observed at each time period $t\in\{1,\dots,T\}$. Each unit-time pair $(i,t)$ uniquely identifies an observation. Define $\eventtime$ as unit $i$'s ``event time'' at time $t$. For each $i$, let $\eventtime=\max\{t':t' \leq t, D_{i,t'}=1, D_{i,t'-1}=0\}$ if $\exists s \leq t: D_{i,s}=1$, and $\eventtime=\min\{t':D_{i,t'}=1, D_{i,t'-1}=0\}$ otherwise. That is to say, $\eventtime$ is the most recent time at which unit $i$ switched into treatment or, if $i$ has not yet been treated at any point up until time $t$, the first time $i$ switches into treatment. If $i$ is never treated, we let $\eventtime=\infty$. In the staggered setting, the event time for each unit is constant, $E_i=\eventtime$, and $\Dit=\indic{t \geq \eventtime}$, where $\indic{\cdot}$ is the indicator function. In such settings, we can partition units into distinct ``cohorts'' $g\in\{1,\dots,G\}$ according to the timing of treatment adoption $E_i$. Units transitioning to treatment at period $g$ ($i:\eventtime=g$) form cohort $g$, whereas units that never undergo treatment belong to the ``never-treated'' cohort ($i:\eventtime=\infty$). $Z_{i,t}$ ($Z_{i,g,t}$) represents the variable $Z$ for unit $i$ (part of cohort $g$) at time $t$. We use $\Yit(1)$ and $\Yit(0)$ to denote the potential outcomes under treatment and control, respectively, and $\Yit=\Dit \Yit(1) + (1-\Dit)\Yit(0)$ to denote the observed outcome.%
\footnote{The current notation will not cause confusion because we do not allow feedback or temporal spillover. In some of the studies we refer to, potential outcomes are defined in terms of treatment history, as opposed to current treatment status. We adopt similar notations for these frameworks. For instance, we use $\Yit(\Dit~=~1, 
\{D_{i,s}\}_{s<t}=0)$ to represent the potential outcome under the specified treatment history. } %

The finest estimand is the individual treatment effect, $\tau_{i,t}=\Yit(1)-\Yit(0)$, of which there exists one for each observation $(i,t)$.%
\footnote{This is without loss of generality when feedback and interference are excluded. In staggered DID designs, carryover effects are permissible. When potential outcomes are defined in terms of treatment history, $\tau_{i,t}$ is defined as $\Yit(1)-\Yit(\infty)$ where $\Yit(\infty)$ signifies the untreated potential outcome when unit $i$ never undergoes treatment.} %
Most political science research, however, typically focuses on estimating a single summary statistic. Commonly, this is the ATT, which represents individual treatment effects averaged over all observations exposed to the treatment condition. In between these extremes of granularity and coarseness are time-varying dynamic treatment effects, which are across-unit averages of individual treatment effects at each
time period relative to treatment adoption. In the staggered adoption setting, we can further subdivide by cohort. We use $\tau_{l}$ ($\tau_{g,l}$) to denote
the dynamic treatment effect $l$ periods after treatment adoption (for treatment cohort $g$), with $l=1$ representing the period immediately after treatment adoption.%
\footnote{Some of the authors we reference denote this first post-treatment period with $l=0$.} 
$\tau_{g,l}$ is also what some authors refer to as a cohort average treatment effect on the treated \citep{Strezhnev2018-ku, sun2021-event} or group-time average treatment effect \citep{callaway2021-did}.%

Each of the estimators we discuss can be used to estimate $\tau_{l}$. The outcome model analogous to TWFE for estimating dynamic effects is a lags-and-leads specification. For simplicity, we first describe the staggered setting. Let $K_{i,t}=(t-\eventtime+1)$ be the number of periods until (when $K_{i,t}\leq 0$) or since unit $i$'s event time at time $t$ (e.g., $K_{i,t} = 1$ if unit $i$ switches into treatment at time $t$). Consider a regression based on the following specification:
\vspace{-0.5em}\begin{align}\label{eq:event-studies} 
\Yit = \alpha_i + \xi_t + X^{\prime}_{i,t}\beta + \sum_{\substack{l=-a\\l\neq 0}}^{b} \tau_{l}^{TWFE}
\cdot \indic{K_{i,t}=l} + \tau_{b+}^{TWFE}\indic{K_{i,t}> b}\cdot D_{i,t} + \epit, \vspace{-1em}
\end{align} 
where $a$ and $b$ are the number of lag and lead terms (BJS, \citeyear{borusyak2024revisiting}). In the social science literature, the typical practice is to exclude $l= 0$, which corresponds to the time period immediately before the transition into the treatment phase, and use it as a reference period \citep{roth2022pretest}. Conventionally, $\hat\tau^{TWFE}_{l}$ is interpreted as an estimate of $\tau_{l}$ or as a meaningful weighted average of pertinent individual treatment effects. Meanwhile, $\hat\tau^{TWFE}_{b+}$ is viewed as an estimate for the long-term effect.


\section{HTE-Robust Estimators}\label{sc:methods}

In this section, we offer a brief overview and comparison of several recently introduced HTE-robust estimators. We use the term ``HTE'' to refer to individual treatment effects that are arbitrarily heterogeneous, that is, $\tau_{i,t} \neq \tau_{j,s}$ for some $i, j, s, t$. HTE-robust estimators are defined as those that produce causally interpretable estimates under their respective identifying assumptions. For a more comprehensive discussion on these estimators, please refer to the SM. 

\subsection{Summary of HTE-Robust Estimators} 

Table~\ref{tb:estimators} summarizes the estimators we discuss in this article. The primary difference resides in the mechanics of their estimation strategies: there are methods based on canonical DIDs and methods based on imputation. We refer
to the former as \emph{DID extensions} and the latter as \emph{imputation methods}. DID extensions use dynamic treatment effects, estimated from local, $2\times2$ DIDs between treated and control observations, as building blocks. Imputation methods use individual treatment effects, estimated as the difference between an imputed outcome under control and the observed outcome (under treatment), as building blocks. The imputation estimator we use in this article employs TWFE, fitted with all observations under the control condition, to impute treated counterfactuals. Different strategies also entail different assumptions. Each DID extension, for example, relies on a particular type of PT assumption, whereas imputation methods presuppose a TWFE model for untreated potential outcomes and require a zero mean for the error terms, which is implied by strict exogeneity.

Another noteworthy difference lies in the settings in which these estimators are applicable: Some estimators can only be used in settings with staggered treatment adoption, while others can accommodate scenarios with treatment reversals. In the latter setting, all estimators we discuss also require no anticipation and no or limited carryover effects. Furthermore, the estimators diverge in terms of (1) how they select untreated observations as controls for treated units, (2) how they incorporate pre-treatment or exogenous covariates, and (3) the choice of the reference period. We discuss these details further below and in Section A.1 of the SM.

\begin{table}[!ht]
    \centering
    \caption{Summary of HTE-Robust Estimators}\label{tb:estimators}
\resizebox{\textwidth}{!}{\small
\begin{tabular}{L{2.5cm} | C{2.5cm} C{2.5cm} C{2.5cm} | C{2.5cm} C{2.5cm} || C{2.6cm} C{2.5cm}}        \hline\hline
{\bf{Type}} & \multicolumn{5}{c||}{DID Extensions: uses $2\times 2$ DIDs as building blocks} & \multicolumn{2}{c}{Imputation Methods} \\ \hline
{\bf{Setting}} & \multicolumn{3}{C{7.5cm}|}{Staggered: treatment reversals not allowed} & \multicolumn{4}{c}{General: treatment reversals allowed} \\ \hline
{\bf{Research article}}    & \citet{sun2021-event} & \citet{callaway2021-did} & \citet{BLW2022} & \citet{CDH2020,de2024difference} & IKW \citeyearpar{IKW2021} & BJS \citeyearpar{borusyak2024revisiting} & LWX \citeyearpar{liu2024practical} \\ \hline
{\bf{Method known as}}  & \texttt{IW} & \texttt{CSDID} & \texttt{StackedDID} & \texttt{did\_mulitple} & \texttt{PanelMatch} &  \texttt{DID$_{impute}$} & \texttt{FEct} \\ \hline
{\bf{Key ID assumption}}  & Parallel trends & Parallel trends & Parallel trends & Parallel
trends &  Parallel trends & Strict exogeneity & Strict exogeneity \\
\hline
{\bf{Finest estimand}}  & $\tau_{g,l}$ & $\tau_{g,l}$  & $\tau_{l}^{vw}$ & $\tau_{g,l}$ & $\tau_{l}$ &  $\tau_{i,t}$ & $\tau_{i,t}$ \\ \hline
{\bf{Common aggregated estimand}}  & ATT & ATT  & ATT & \multicolumn{2}{c||}{ATT for switchers for $l$ periods} &  ATT & ATT \\ \hline
{\bf{Comparison group}} & Never-treated or last-treated & Never-treated or 
not-yet-treated & Never-treated & Matched stable group (not-yet-treated) & Matched stable group
(not-yet-treated) & Imputed counterfactuals (not-yet-treated) & Imputed counterfactuals (not-yet-treated) \\ \hline
{\bf{Reference period(s)}} & Period 0 & An arbitrary pre-treatment period & Period 0 & Period 0 & Period 0 & All pre-treatment periods & All pre-treatment periods \\ \hline
{\bf{Covariate adjustment}} & Possible extension & Outcome \& propensity score
modeling & Outcome modeling & Possible extension & Refined matched set and outcome modeling & Outcome modeling &
Outcome modeling \\ \hline
\end{tabular}}
\end{table}

\subsection{DID Extensions} 

DID extensions are all built from local, $2 \times
2$ DID estimates. The overarching strategy for these estimators is to estimate the dynamic treatment effects, $\tau_{l}$ (or $\tau_{g,l}$ for each cohort $g$ in the staggered setting), for each period since the most recent initiation of treatment, $l$, using one or more \emph{valid} $2 \times 2$ DIDs. By ``valid,'' we mean that the DID includes (1) a pre-period and a post-period and (2) a treated group and a comparison group. The pre-period is such that all observations in both groups are in control, whereas the post-period is such that observations from the treated group are in treatment and those from the comparison group are in control. The choice of the comparison group is the primary distinction between estimators in this category. To obtain higher-level averages such as the ATT, we then average over our estimates of $\tau_{l}$ (or $\tau_{g,l}$), typically employing appropriate, convex weights.

Three estimators in this category are appropriate only for the staggered setting.  \citet{sun2021-event} propose an interaction-weighted (\texttt{IW}) estimator, which is a weighted average of $\tau_{g,l}$ estimates obtained from a TWFE regression with cohort dummies fully interacted with indicators of relative time to the onset of treatment. They demonstrate that, in a balanced panel, each resulting estimate of $\tau_{g,l}$ can be characterized as a difference in the change in average outcome from a fixed pre-period $s < g$ to a post-period $l$ periods since $g$ between the treated cohort $g$ and the comparison cohort(s) in some set $\mc{C}$. The authors recommend using $\mc{C}={\sup_i E_i }$, which is either the never-treated cohort or, if no such cohort exists, the last-treated cohort. By default, \texttt{IW} uses $l=0$ as the reference period and can accommodate covariates in the TWFE regression. 

Employing the same general approach, \citet{callaway2021-did} propose two estimators, one of which uses never-treated units ($\hat\tau^{CS}_{nev}$) and the other not-yet-treated units ($\hat\tau^{CS}_{ny}$) as the comparison group. We label these estimators collectively as \texttt{CSDID}. Note that $\hat\tau^{CS}_{nev}$ uses the same comparison group as  \texttt{IW} when a never-treated cohort exists,%
\footnote{This equivalence holds when there are no missing data. Otherwise, \texttt{IW} from the saturated regression differs from one that directly estimates local DIDs, including the never-treated version of \texttt{CSDID}. These estimates are typically close but can differ substantially, as in \citet{Kuipers2023}.}
whereas $\hat\tau^{CS}_{ny}$ uses all untreated observations of not only never-treated units but also later adopters as controls for earlier adopters. Besides the choice of comparison cohort, \texttt{CSDID} estimators differ from \texttt{IW} in that they allow users to condition on pre-treatment covariates using both an explicit outcome model and inverse probability weighting simultaneously; consistency of the estimators requires at least one of these to be correct. By default, both \texttt{IW} and \texttt{CSDID} use the period immediately before the treatment's onset as the reference period for estimating the ATT.

Stacked DID, first formally introduced by \citet{cengiz2019effect}, is another related estimator sometimes used to address HTE concerns. As described by \cite{BLW2022}, it involves creating separate sub-datasets for each treated cohort by combining data from that cohort (around treatment adoption) and data from the never-treated cohort from the same periods. These cohort-specific datasets are then ``stacked'' to form a single dataset. An event-study regression akin to Equation~(\ref{eq:event-studies}) with the addition of sub-dataset specific unit and time dummies is then run. This method uses the same comparison group as \texttt{IW} and the never-treated version of \texttt{CSDID} without covariates, but stacked DID estimates a single dynamic treatment effect for a given relative period rather than separate estimates for each cohort. Essentially, stacked DID is a special case of \texttt{IW} that uses immutable weights selected by OLS. We denote the corresponding estimand $\tau_{l}^{vw}$ to reflect the fact that it is variance-weighted. These weights are generally neither proportional to cohort sizes nor guaranteed to sum to one \citep{wing2024stacked}. Thus, while stacked DID avoids the ``negative weighting'' problem and meets our criteria for HTE-robustness, its estimands are not the same as $\tau_{l}$ or the ATT. 


In settings with treatment reversals, separate groups of researchers have converged on the same strategy for choosing a comparison group: matching treated and control observations that belong to units with identical treatment histories. IKW~\citeyearpar{IKW2021} suggest one such estimator, \texttt{PanelMatch}, which begins by constructing a ``matched set'' for each observation $(i,t)$ such that unit $i$ transitions into treatment at time $t$. This matched set includes units that both (1) are not under treatment at time $t$ and (2) share the same treatment history as $i$ for a fixed number of periods leading up to the treatment onset. For each treated observation $(i,t)$ and for every post-period $(t+l-1)$ such that unit $i$ is still under treatment, it then estimates a local DID using the same pre-period $s<t$. The treatment ``group'' comprises solely of observation $(i,t)$, and the members of the matched set for $(i,t)$ that are still under control during $t+l-1$ serve as the comparison group. To obtain $\tau_{l}$ for a given $l$, it then averages over the corresponding local DID estimates from all treated observations. IKW~\citeyearpar{IKW2021} propose incorporating covariates by ``refining'' matched sets and use $l=0$ as the reference period.

Using a similar strategy, \citet{CDH2020} propose a ``multiple DID'' estimator, \texttt{DID\_multiple}. A notable difference is that they include local DIDs for units leaving the treatment and not just those joining the treatment; when there are no treatment reversals or covariates, \texttt{DID\_multiple} is a special case of \texttt{PanelMatch}. The original proposal for \texttt{DID\_multiple} also only considers the case where we match on a single period and where $l=1$, but since it has been extended \citep{de2024difference}. Consequently, the target estimand is not the ATT but rather an average of the contemporaneous effects of ``switching'' (i.e., the effect of joining or the negative of the effect of leaving at the time of doing so). In the staggered setting, the \texttt{PanelMatch} estimator aligns with the not-yet-treated version of \texttt{CSDID} (without covariate adjustment). We delve into details on the connections between these three estimators in the SM.

All DID extensions are built using local, $2\times 2$ DIDs, and their assumptions reflect this. Specifically, they each rely on a form of the PT assumption---that is, the expected changes in untreated potential outcomes from one period to the other are equal between the treated and the chosen comparison groups. We defer readers to the SM for a fuller account of each method's assumptions.

\subsection{The Imputation Method} Imputation estimators do not explicitly estimate local DIDs. Instead, they take the difference between the observed outcome and an imputed counterfactual outcome for each treated observation. The connection to the TWFE model is in the functional form assumption used to impute counterfactual outcomes. Specifically, an imputation estimator first fits a parametric model for the potential outcome under control $\Yit(0)$---in our case, $\Yit(0) = \Xit{'}\beta + \alpha_i + \xi_t + \epit$---using only control observations $\{(i,t):\Dit=0\}$. It is also through this outcome model that one can adjust for time-varying covariates. Then, it imputes $\Yit(0)$ for all treated observations $\{(i,t):\Dit=1\}$ using the estimated parameters. Finally, it estimates the individual treatment effect, $\tau_{i,t}$, for each treated observation $(i,t)$ by calculating the difference between the observation's observed outcome $\Yit=\Yit(1)$ and its imputed counterfactual outcome $\hat{Y}_{i,t}(0)$. Inference for the estimated $\hat\tau_{i,t}$ is possible, although uncertainty estimates need to be adjusted to account for the presence of idiosyncratic errors \citep[e.g.,][]{bai2021matrix}.
BJS \citeyearpar{borusyak2024revisiting} and LWX \citeyearpar{liu2024practical} each propose estimators in this category. Each paper proposes a more general framework that nests many models, including TWFE. The latter also introduces several specific imputation estimators, one of which uses the TWFE model, and the authors refer to the resulting estimator as the fixed effects counterfactual estimator, or \texttt{FEct}.

Although DID extensions and imputation methods rely on slightly different identification assumptions, these assumptions usually lead to similar observable implications. Researchers commonly use the presence or absence of pretrends to judge how plausible the PT assumption is. In the classic two-group setting, if there are data from multiple pre-treatment periods, researchers can plot the time series of average outcomes of each group and visually inspect whether they indeed trend together. The intuition is that if PT holds and the average outcome trends of the treated and control groups are indeed parallel in pre-treatment periods when $Y(0)$'s are observed for all units, then it is plausible that PT also holds in the post-treatment periods, when $Y(0)$'s are no longer observable for units in the treatment group. Conversely, differential trends in the pre-treatment periods should make us suspicious of PT. In more complex settings or where we wish to control for observed confounders, researchers often use dynamic estimates before and after the onset of treatment, $\tau_{l}$, to construct so-called ``event-study plots'' to judge the presence of pretrends. If PT holds, then pre-treatment dynamic estimates should be around zero. We provide a more thorough discussion and an example of the event-study plot in the next section when we introduce our procedure. 

\subsection{Choice of Estimators} 

In general, we believe the credibility of identifying assumptions is more important than the choice of estimator. After all, in the staggered setting, when assumptions hold, \texttt{IW}, \texttt{CSDID}, \texttt{DID\_multiple}, \texttt{PanelMatch}, and the imputation estimator all converge to the same or a similar estimand. However, there are a few reasons to favor the imputation estimator. First, it can handle complex settings, including those with treatment reversal---which account for over half of the studies in our sample---and can accommodate time-varying covariates, additional fixed effects, and unit- or group-specific time trends commonly seen in social science research. The imputation estimator connects to TWFE through a shared outcome model, and thus any of the aforementioned modifications to the outcome model can be directly mirrored.  DID extensions relates to TWFE through their shared connection to DID in the two-group, two period setting. Classic DID's inability to naturally accommodate these complexities limits DID extensions on this front. Just like TWFE, the imputation estimator risks misspecification bias, and adding more redundant terms may significantly increase variance. However, we still consider the added flexibility to be an advantage. Second, imputation estimators are the most efficient under homoskedastic errors (BJS, \citeyear{borusyak2024revisiting}).\footnote{In Section A.3 of the SM, we demonstrate that the imputation estimator tends to yield larger $z$-scores, based on data from our sample. In the majority of cases, the imputation estimator has a smaller standard error, and the difference can be especially dramatic for \texttt{IW} and the never-treated version of \texttt{CSDID}, which often discard the vast majority of untreated observations.} Moreover, by using the average of all pre-treatment periods as the reference point rather than a single pre-period, as the default in DID extensions, they provide greater power in hypothesis testing for pretrends. The main drawback of imputation estimators is that their current implementations (either \texttt{FEct} or \texttt{DID\_impute}) do not allow for automated adjustment of time-invariant covariates, an advantage offered by \texttt{CSDID} and \texttt{PanelMatch}. Adjusting for pre-treatment characteristics can improve credibility of research, as conditional PT may be more plausible than the unconditional one \citep{SantAnna2018-sj}.  

\FloatBarrier


\section{Data and Procedure}\label{sc:data}

Next, we assess the robustness of empirical findings from causal panel analyses in
political science and compare results obtained using the different methods we have discussed. We will explain our sample selection rules, describe standard practices in the field, and outline our reanalysis approach. Readers can find a more detailed explanation of our sample selection criteria and replication and reanalysis procedure in Section A.2 of the SM.

\subsection{Data} Our replication sample comprises studies from three leading political science journals, {\it APSR}, {\it AJPS}, and {\it JOP}, published over a recent 7-year span from 2017 to 2023. We initially include all studies, including both long and short articles, that employ panel data analyses with a binary treatment as a key component of their causal argument, resulting in a total of 102 studies. After a careful review, as explained in footnote 1, we find that 64 studies employ a TWFE model similar to Equation~(\ref{eq:twfe}). We then attempt to replicate the main results of these 64 studies and are successful in 49 cases (76.6\%). Though a significant proportion of studies failed to replicate, we note that the success rate is still higher than that of \citet{hainmueller2019much} at 55\% and is comparable to that of \citet{lal2021much} at 67\%. We credit this to the new replicability standards set by journals. A detailed explanation of how we select the ``main model'' is provided in Section A.3 of the SM. Table~\ref{tb:replicability} depicts the distribution of successful replications, along with reasons for replication failures, across the various journals.

\begin{table}[!htbp]
  \centering\small
   \caption{Sample Selection and Replicability of Qualified Studies}
    \label{tb:replicability}
   \begin{tabular}{C{1.8cm}C{1.7cm}C{2.2cm}cccc}\hline\hline\small
      &      & TWFE & Incomplete & Replication &  &  Success\\
    Journal  & All &  (attempted)  &  data    &  error  &   Replicable & rate\% \\\hline
    {\it APSR} &  22 & 13    & 2     & 1     & 10     & 76.9 \\
    {\it AJPS} &  31 & 21    & 3     & 3     & 15     & 71.4 \\
    {\it JOP}  &  49 & 30    & 6     & 0     & 24    & 80.0 \\ \hline
    Total      & 102 & 64    & 11    & 4     & 49    & 76.6 \\\hline
    \end{tabular}%
\end{table}\vspace*{-1em}

\subsection{Settings and Common Practices} 

Table~\ref{tb:practice} presents an overview of the standard practices and settings in the studies that we successfully replicated. The majority of studies in our sample (67.3\%) use the DID design/method/approach to justify the use of the TWFE model, while the remaining studies advocate for the model's ability to exploit ``within'' variations in the data. Out of the 49 studies, nine (18.4\%) employ a classic block DID setting, which includes two-group, two-period designs (three studies) and multi-period block DID designs (six studies). Thirteen studies (26.5\%) use a staggered DID design, while the remaining 27 studies (55.1\%) fall into the ``general'' category, meaning they allow for treatment reversals. Except for five, all studies feature a continuous outcome of interest. Most use cluster-robust SEs or panel-corrected SEs \citep{Beck1995-am}, and eight studies employ bootstrap procedures for estimating uncertainties. A subset of studies explore alternative model specifications by adding lagged dependent variables (eight studies), unit-specific linear time trends (fifteen studies), and higher-than-unit-level time trends (one study). Notably, 32 studies use some type of visual inspection---either average outcomes over time, event-study plots, or both---to evaluate the plausibility of PT.  Four studies published in 2023 (33\%) employ HTE-robust estimators, compared to none before 2023, indicating rapid adoption of these methods. Of these, two use \texttt{CSDID}, one \texttt{PanelMatch}, and one the imputation estimator. 


\begin{table}[!htbp]
  \centering\small
  \caption{Settings and Common Practice}\label{tb:practice}%
    \begin{tabular}{lccclcc}\hline\hline
        \textit{Motivations for TWFE} &       &       &  &  \textit{Variance Estimator}   &  &        \\ 
        ``Difference-in-differences" & 33     & 67.3\% &  &     Cluster-robust SE or PCSE & 48    & 98.0\%  \\
    ``Within" variations & 16     & 32.7\% &  &     Cluster-bootstrapped procedures & 8     & 16.3\%    \\
          &       &       &  &  \\
    \textit{Treatment setting} &       &       &  &  \textit{Variants in specifications}   &  &        \\ 
    Classic 2$\times$2 DID & 3     & 6.1\% &       &  Lagged dependent variables  & 8     & 16.3\%  \\
    Multi-period block DID & 6     & 12.2\% &       & Higher-than-unit-level time trends & 1 & 2.0\% \\
    Staggered DID & 13    & 26.5\% &       & Unit-specific linear time trends & 15     & 30.6\%  \\
    General & 27    & 55.1\% &       &    \\ 
          &       &       &  &                       \textit{Data visualization}   \\
    \textit{Outcome variable} & &    &   & Group average outcomes & 19    & 38.8\%     \\ 
    Continuous  &  44   &  89.8\%     &    &  Event-study plots & 23    & 46.9\%   \\ 
    Binary  &   5    &   10.2\%    &         & Neither   &  17  & 34.7\% \\ \hline
\end{tabular}%
\end{table}\vspace{-1em}%


\subsection{Procedure} We use data from \citet{Grumbach2020} to illustrate our process for replication and reanalysis. The authors investigate coethnic mobilization by minority candidates during U.S. congressional elections. To simplify our analysis, we focus on the impact of the presence of an Asian candidate on the proportion of general election contributions from Asian donors. To begin, we aim to understand the research setting and data structure. We visualize the patterns of treatment and outcome variables using plots, which are shown in the SM. In this application, treatment reversals clearly take place. Some data are missing (due to redistricting), but the issue does not seem to be severe. We record important details such as the number of observations, units, and time periods, the type of variance estimator, and other specifics of the main model. Next, we replicate the main finding, employing both the original variance estimator and a cluster-bootstrap procedure.

\begin{figure}[!ht]
  \caption{Reanalysis of \citet{Grumbach2020}}\label{fg:gs2020}
  \centering
  \begin{minipage}{1\linewidth}{
  \begin{center}
  \hspace{-1em}
  \subfigure[Treatment effect estimates]{\includegraphics[width = 0.28\textwidth]{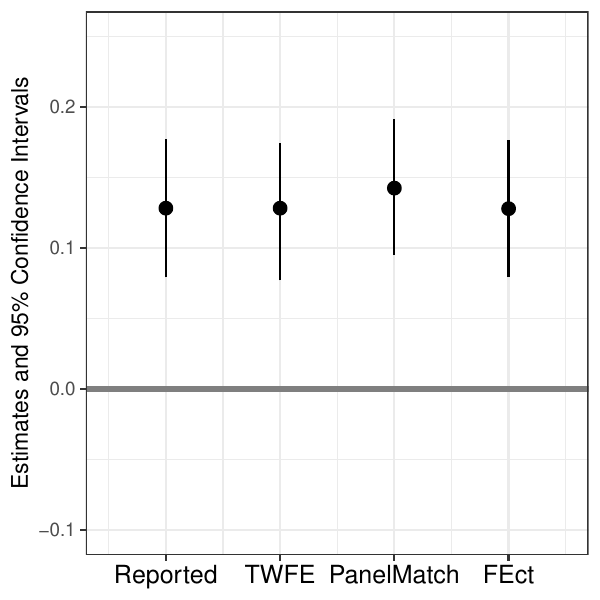}}\hspace{1em}
  \subfigure[Dynamic: TWFE]{\includegraphics[width = 0.28\textwidth]{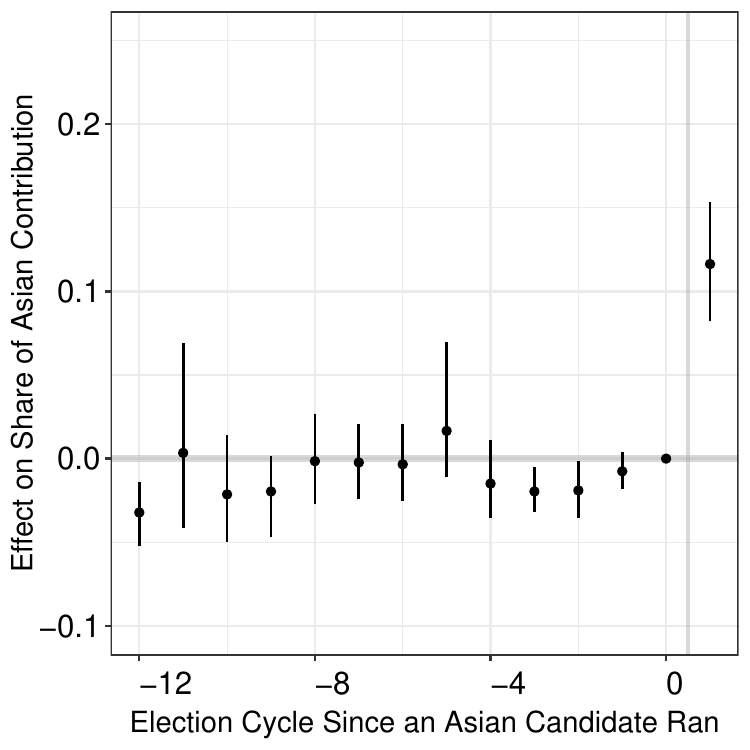}}\hspace{1em}
  \subfigure[Dynamic: PanelMatch]{\includegraphics[width = 0.28\textwidth]{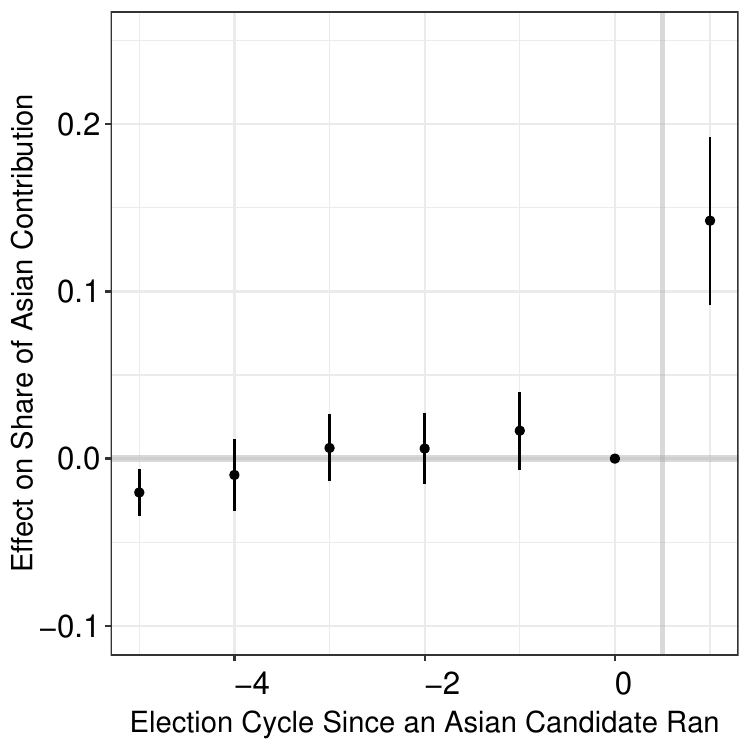}}\\
  \subfigure[Dynamic: Imputation]{\includegraphics[width = 0.28\textwidth]{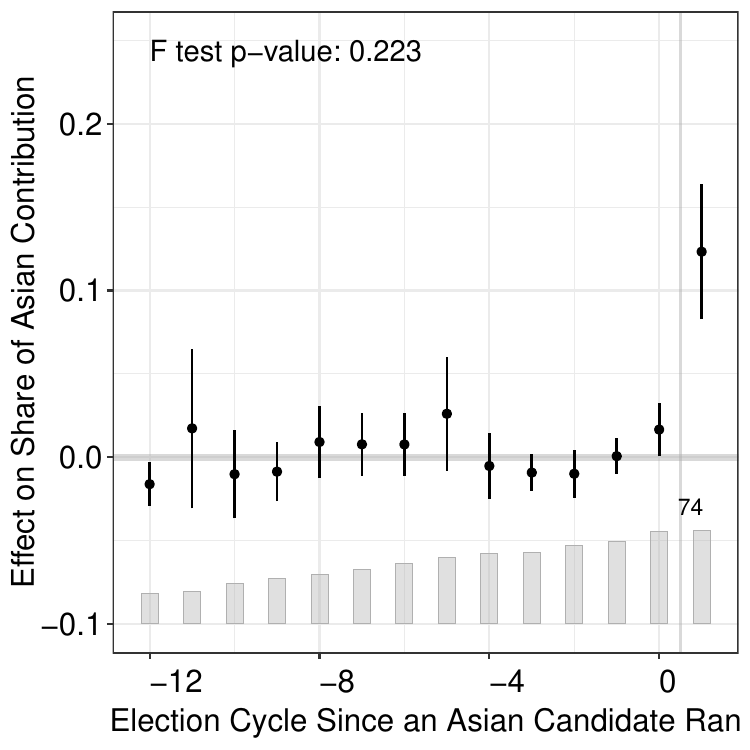}}\hspace{1em}
  \subfigure[Placebo \& sensitivity tests]{\includegraphics[width = 0.28\textwidth]{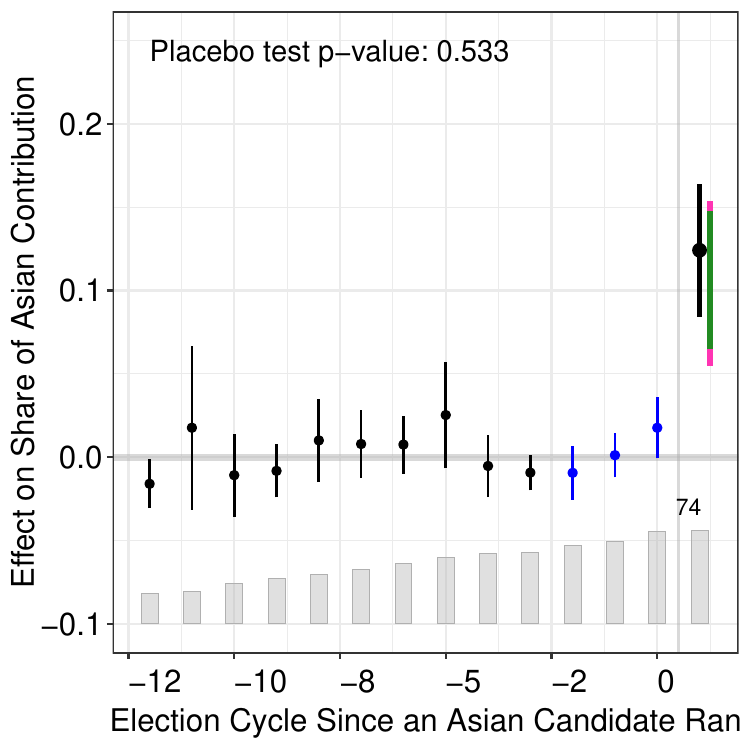}}\hspace{1em}
  \subfigure[Test for carryover effects]{\includegraphics[width = 0.28\textwidth]{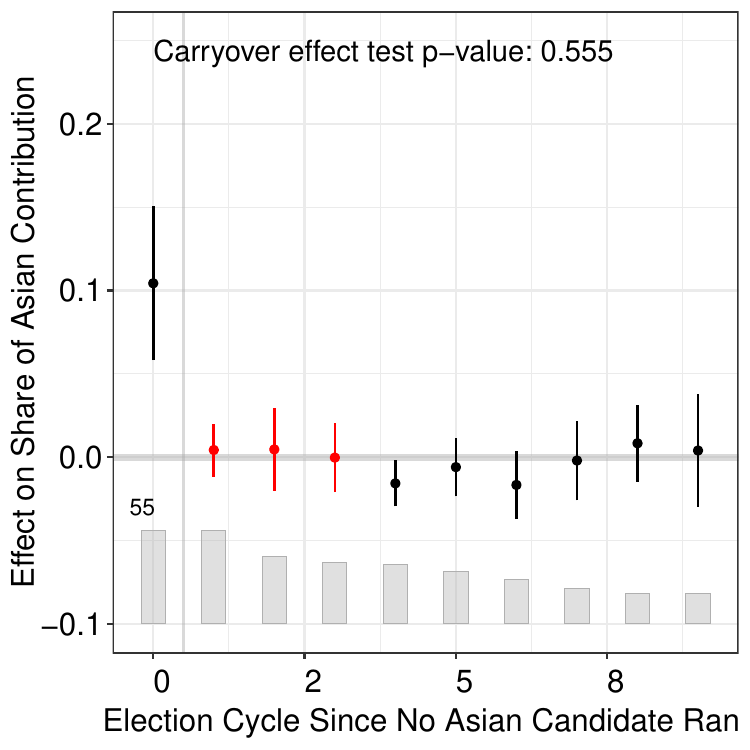}}
  \end{center}
{\footnotesize\textbf{Note:} Reanalysis of data from \citet{Grumbach2020}. Subfigures (a): Treatment effect or ATT estimates from multiple methods; subfigures (b)-(d): event-study plots using TWFE, \texttt{PanelMatch}, and the imputation estimator (\texttt{FEct}); Subfigure (e)-(f): results from the placebo test (and robust CS) and test for carryover effects using \texttt{FEct}---the blue points in (e) and red points in (f) represent the holdout periods in the respective tests. In (e), the green and pink bars represent the 95\% robust CSs when $\bar{M} = 0$ and $\bar{M} = 0.5$, respectively. CIs in all subfigures---excepted for the reported estimate in (a)---are produced by bootstrap percentile methods.}} \end{minipage}\vspace{-0.5em}
  \end{figure}

We then re-estimate the ATT and dynamic treatment effects using estimators discussed in the previous section. For staggered adoption treatment cases, we apply seven estimators: TWFE (with always treated units removed for easier comparisons with other estimators), the imputation estimator (\texttt{FEct}), \texttt{PanelMatch}, \texttt{DID\_multiple}, \texttt{StackedDID}, \texttt{IW}, and \texttt{CSDID} (both not-yet-treated and never-treated versions). For applications with treatment reversals like \citet{Grumbach2020}, we use the first three estimators only.\footnote{ A recent development of \texttt{DID\_multiple}, DIDmultiplegtDYN, allows for the estimation of dynamic and long-term effects \citep{de2024difference}. It defines an estimand that accounts for carryover effects and targets the first onset of the treatment. To remain consistent with the imputation framework, which allows multiple treatment onsets, we apply \texttt{DID\_multiple} only to cases featuring staggered treatment timing.} The comparison between the TWFE estimate and the other estimates sheds light on whether original findings are sensitive to relaxing the constant treatment effect assumption. Figures~\ref{fg:gs2020}(a)-(d) show the results from this example. The similarity between estimates for the ATT in panel (a) suggests that the original finding is robust to the choice of estimators. The event-study plots from HTE-robust estimators in panels (c) and (d) are broadly consistent with the event-study plot from TWFE in panel (b).

Next, we conduct diagnostic tests based on the imputation estimator, including the $F$ test and the placebo test, to further assess the plausibility of PT and, in applications with treatment reversal, the no-carryover-effect assumption. We use the imputation estimator because it is applicable across all studies in our replication sample, can incorporate time-varying covariates, and remains highly efficient. Figures~\ref{fg:gs2020}(d)--(f) show the results from the $F$ test, placebo test, and test for no carryover effects on our running example, respectively. Both a visual inspection and the formal tests suggest that PT and no-carryover-effect assumptions are quite plausible. 

Finally, we compute the robust confidence sets proposed by \citet{rambachan2023more}, which account for potential PT violations  when testing the null hypothesis of no post-treatment effect. Specifically, we employ the relative magnitude restriction, with two modifications to accommodate the imputation method. First, we use estimates from the placebo test to ensure that benchmark pre-treatment estimates are obtained using the same approach as post-treatment ATT estimates. This alignment prevents potential asymmetry in testing and treatment effect estimation \citep{Roth2024interpret}. Second, since the imputation method does not rely on a single reference period, we explicitly incorporate the placebo estimate from the last pre-treatment period ($\hat\delta_{0}$) to account for deviations of post-treatment estimates from earlier reference periods. Mathematically, we decompose each dynamic estimate, $\mu_t$, into the true treatment effect, $\tau_t$, and a trend (bias) component, $\delta_t$, such that: $\mu_t = \tau_t + \delta_t$. Our modified relative magnitude restriction then requires that, for all $t \geq 0$, 
\begin{equation}\label{eq:rm}
\bigl|\delta_{t+1} - \delta_t\bigr| \leq \bar{M} \cdot \max_{\,s\in \mathcal{P}\setminus\{0\}} \bigl|\delta_{s+1} - \delta_{s}\bigr|,
\end{equation}
where $\mathcal{P}$ is the set of placebo periods. In our application, we set $\mathcal{P} = \{-2, -1, 0\}$, so the maximum violation among placebo periods is:
$\max\!\Bigl\{\bigl|\delta_{0}-\delta_{-1}\bigr|,\;\bigl|\delta_{-1}-\delta_{-2}\bigr|\Bigr\}.$ 

When $\bar{M} = 0$, the relative magnitude restriction reported in Equation ~(\ref{eq:rm}) implies that $\delta_{t} = \delta_{0}$ for all $t > 0$, meaning that the PT violation remains fixed at the same level as in the last pre-treatment placebo period. In this case, the robust CS obtained at $\bar{M} = 0$ acts as a debiased confidence interval, using the placebo estimate from the last pre-treatment period as the benchmark for bias. Allowing $\bar{M} > 0$ permits PT violations to vary over time, but constrains the change in magnitude of violations between consecutive post-treatment periods to remain within $\bar{M}$ times the largest consecutive discrepancy observed during the placebo periods.

\citet[][2653]{rambachan2023more} suggest using $\bar{M} = 1$ as a ``natural benchmark'' when the number of placebo periods is roughly equal to the number of post-treatment periods, treating any potential PT violations as no worse than those already observed.\footnote{In our setting, the number of post-treatment periods typically exceeds the number of placebo periods, which means the criterion is even more lenient than the authors have suggested.} In our reanalysis, we first construct 95\% robust confidence sets for each post-treatment dynamic effect and the ATT at $\bar{M} = 0$ and $\bar{M} = 0.5$. Figure~\ref{fg:gs2020}(e) illustrates these robust confidence sets for the estimated ATT using the imputation method. The center of the robust confidence sets is smaller than the point estimate because $\hat{\delta}_{0} > 0$. If the confidence sets for $\bar{M} = 0$ does not include zero, as in this case, we conduct a sensitivity analysis by varying $\bar{M}$ over a wider range to determine the ``breakdown value'' $\tilde{M}$, which is the smallest value of $\bar{M}$ at which the robust confidence set first includes zero. In \citet{Grumbach2020}, the breakdown value is $\tilde{M} = 2.5$, which means that the estimated coethnic mobilization effect remains statistically distinguishable from zero unless PT violations are more than 2.5 times the largest discrepancy observed during the placebo periods.

Overall, the results from \citet{Grumbach2020} appear highly robust, regardless of the choice of point and variance estimators. The PT and no-carryover-effect assumptions seem plausible. The study also has sufficient power to distinguish the ATT from zero, even under potential, realistic PT violations.

\FloatBarrier


\section{Systematic Assessment}\label{sc:results}

We perform the replication and reanalysis procedure described above for all 49 studies in our sample. This section offers a summary of our findings, with complete results for each article available in the SM. We organize our results around two main questions: (1) Are existing empirical findings based on TWFE models robust to HTE-robust estimators? (2) Is the PT assumption plausible, and do original findings remain robust to mild PT violations informed by pretrends? We also discuss other issues observed in the replicated studies, including the presence of carryover effects and sensitivity to model specifications.

                             

\begin{figure}[!ht]
  \caption{TWFE vs. The Imputation Estimator: All Cases}\label{fg:all}
  \centering
  \vspace{-0.5em}
  \begin{minipage}{1\linewidth}{
  \begin{center}  
  \includegraphics[width = 1\textwidth]{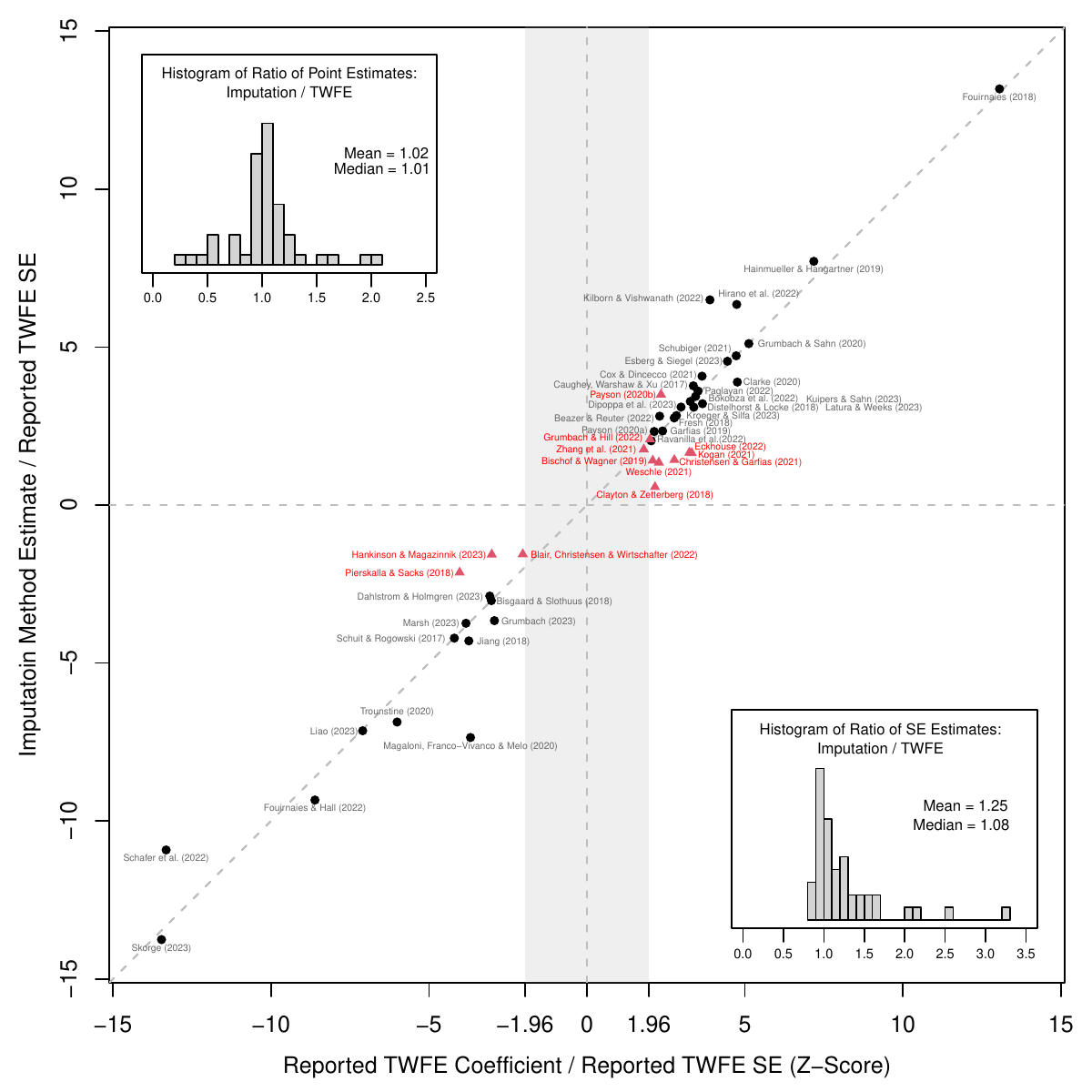}
  \end{center}\vspace{-1em}
  {\footnotesize\textbf{Note:} The above figure compares reported TWFE coefficients with imputation method (\texttt{FEct}) estimates. Both estimates for each application are normalized by the same reported TWFE SE. \citet{fh2018} and \citet{Hall2022} are close to the 45-degree line but are not included in the figure as their TWFE $z$-scores exceed 15. Black circles (red triangles) represent studies whose imputation method estimates for the ATT are statistically significant (insignificant) at the 5\% level, based on cluster-bootstrapped SEs. The top-left (bottom-right) corners display histograms of the ratio of point (SE) estimates based on the imputation method and TWFE. These plots show that changes in point estimates, combined with the efficiency loss from using the imputation method, contribute to the loss of statistical significance in some studies.}}
  \end{minipage}\vspace{-1em}
\end{figure}

\paragraph*{HTE-robust estimators yield qualitatively similar but highly variable estimates.} To examine the impact of the weighting problem caused by HTE associated with TWFE models, we first compare the estimates obtained from the imputation estimator, \texttt{FEct}, for all studies to those originally reported. We choose the imputation estimator for the reason mentioned earlier. Most importantly, it is applicable to all studies in our sample, including those with treatment reversals and those with additional time trends. Figure~\ref{fg:all} plots the comparison. The horizontal axis represents the originally reported TWFE estimates, and the vertical axis represents \texttt{FEct} estimates, both normalized using the same originally reported SEs. If the point estimates are identical, then the corresponding point should lie exactly on the 45-degree line. Red triangles represent studies where the imputation estimates are statistically insignificant at the 5\% level, based on cluster-bootstrapped SEs. 

We observe several patterns. First, TWFE coefficients are statistically significant at the 5\% level in all but one study, and the absolute values of $z$-scores for a significant minority of studies cluster around $1.96$, indicating possibly a file-drawer problem and potential publication bias. Second, the points largely follow the 45-degree line, with the imputation estimates \emph{always} having the same sign as the original estimates. This suggests that while scenarios where accounting for HTE completely reverses the empirical findings are theoretically possible, they are rare.

However, results sometimes deviate significantly. In the top left corner of Figure~\ref{fg:all}, we plot the histogram of the ratio of imputation to TWFE estimates. Although the mean and median of the ratio are close to one, at the extremes, we observe imputation estimates as small as one-fourth or as large as more than double the TWFE estimates. The most consequential deviations occur in studies that were originally near the margins of statistical significance. Additionally, we plot the ratio of SE estimates from the imputation method to TWFE in the bottom right corner of Figure~\ref{fg:all}. The median is 1.08, meaning that in the majority of cases the SE estimate from the imputation method is at least 8\% larger than the SE from TWFE. The mean is 1.25, and the distribution is right-skewed; in the extreme, the ratio was almost as high as three. Combined, drops in point estimates and increases in uncertainty lead to the third pattern: When we switch from TWFE to the imputation estimator, the number of studies that are statistically insignificant at the 5\% level increases from one to twelve (24\%).

\begin{figure}[!th]
  \caption{Comparison of Estimates: The Staggered Setting}\label{fg:st}
  \centering
  \begin{minipage}{1\linewidth}{
  \begin{center}  
  \includegraphics[height = 0.5\textwidth]{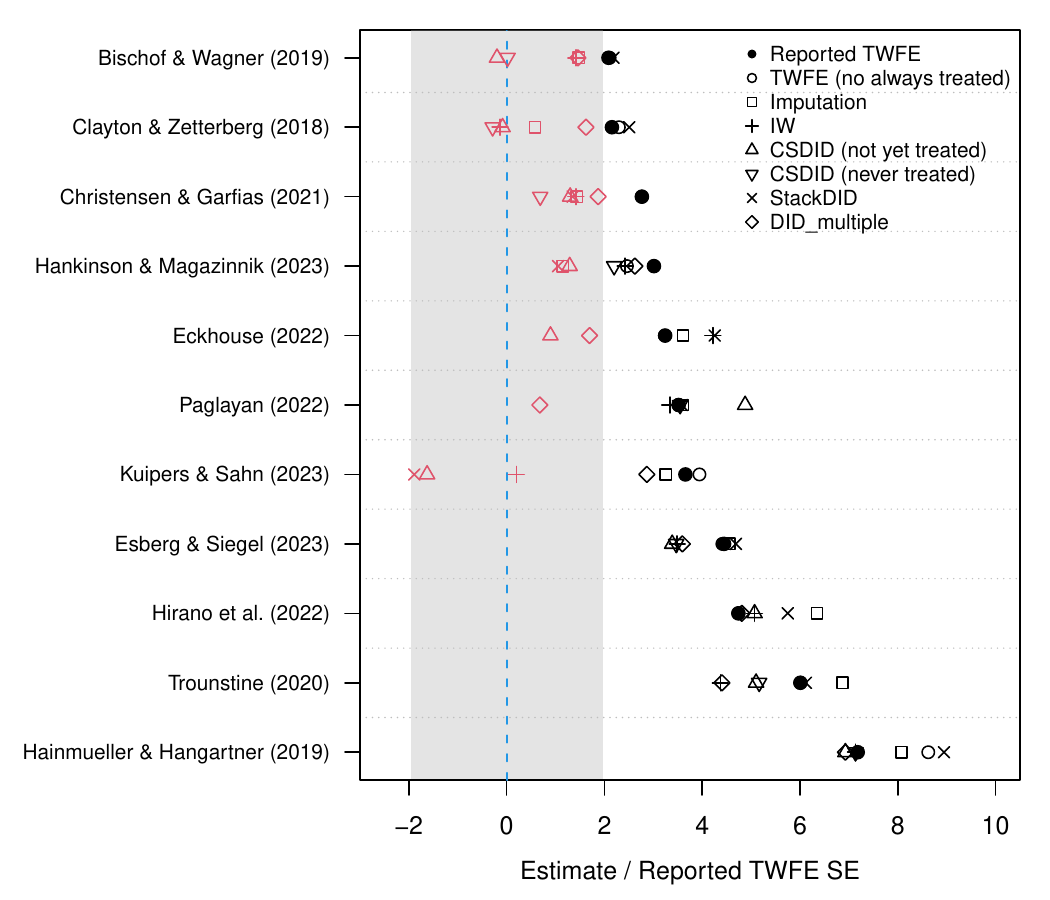}\hspace{-0.7em}
  \includegraphics[height = 0.5\textwidth]{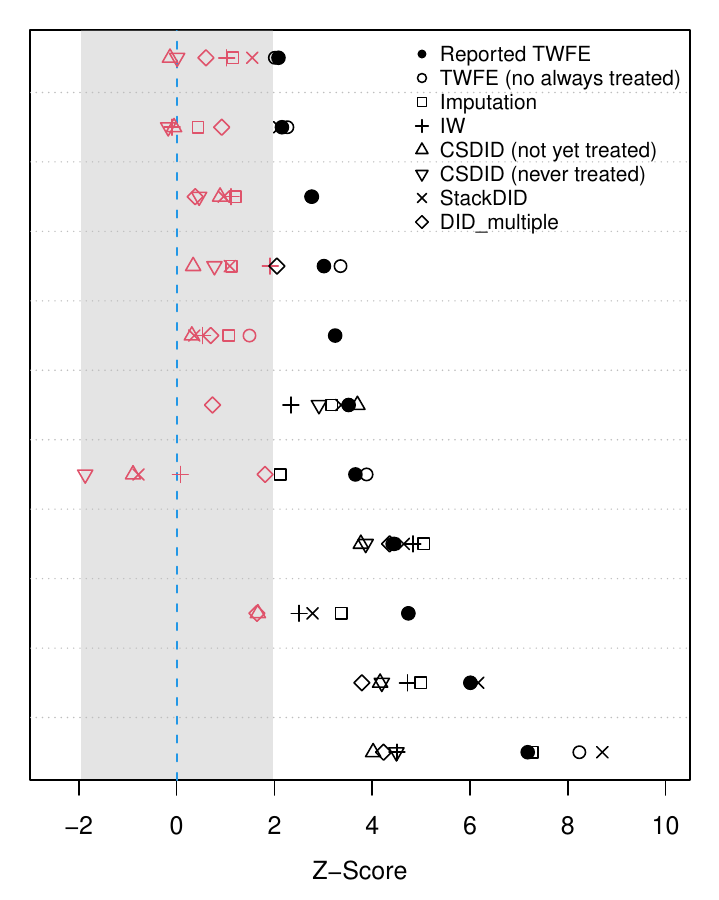}\vspace{-1em}
   \end{center}
  {\footnotesize\textbf{Note:} The above figures compare reported TWFE coefficients with estimates from various alternative estimators. In the left panel, all eight estimates for each application are normalized by the same reported SE to highlight changes resulting from the use of alternative estimators. In the right panel, the estimates are divided by their respective bootstrapped SEs. To facilitate visualization, we multiply all estimates by the sign of the reported coefficient. In both figures, black (red) symbols represent estimates that would be statistically significant (insignificant) at the 5\% level, assuming they were treated as $z$-scores. The normalized \texttt{CSDID} (never treated) estimate for \citet{Kuipers2023}, -5.36, falls out of plotting area and is therefore not shown in the left panel. \citet{Kogan2021} and \citet{magaloni2020killing} are excluded because the authors' original TWFE specifications include unit-specific linear time trends, which are not supported by most HTE-robust estimators except the imputation estimator. Some  estimates are missing because of too few never-treated units. \texttt{PanelMatch} is excluded because it targets a different estimand.}}
  \end{minipage}
\end{figure}

If we restrict our attention to eleven studies with staggered treatments, we can broaden our comparison set to include more HTE-robust estimators discussed earlier.\footnote{\citet{Kogan2021} and \citet{magaloni2020killing} are excluded because the original specifications include additional time trends, which are not supported by HTE-robust estimators except the imputation estimator.} Figure~\ref{fg:st} visually compares the points estimates (left panel) and $z$-scores (right panel). In the left panel, all point estimates are normalized by the same reported TWFE SEs for each study. In all but three studies, the estimates from all HTE-robust estimators share the same sign, though there is a noticeable amount of variation in the estimated effect size.\footnote{Although a sign change is observed in \citet{Clayton2018} when using \texttt{CSDID}, \texttt{IW}, and \texttt{DID\_multiple}, these estimates are negligibly small and statistically insignificant. Similarly, the not-yet-treated version of \texttt{CSDID} is the opposite sign but miniscule and statistically insignificant in \citet{Bischof2019}. The estimates from \texttt{IW} and \texttt{CSDID} are also of the opposite sign for \citet{Kuipers2023}, and they are of a much larger magnitude. The never-treated version of \texttt{CSDID} is also statistically significant.}  

As in Figure~\ref{fg:all}, TWFE does not appear to be systematically upward or downward biased compared to HTE-robust estimators. Another observation that carries over is that HTE-robust estimators generally require more power to reject the null hypothesis of no effect. In five of the eleven studies, at least four HTE-robust estimates per study are statistically insignificant. The left panel shows that the changes in point estimate alone are often sufficient to render the results statistically insignificant. The comparison of $z$-scores in the right panel highlights that increased uncertainties can be substantial. Combined with earlier evidence, these findings from the staggered cases suggest that the HTE issue regarding TWFE is empirically significant and warrants careful consideration by researchers.

It is worth noting that while only a small fraction of studies in our sample (eight studies, 16.3\%) employ a bootstrap procedure to estimate SEs or CIs, the widely practiced cluster-robust SE typically performs adequately. This is because the number of units (clusters) is generally large, with a median of 317. However, ten studies have fewer than 50 units; among them, two studies that were significant at the 5\% level using cluster-robust SEs fell below this threshold when using cluster-bootstrapped SEs, both of which were already marginally significant. We provide comparisons of reported, cluster-robust, and cluster-bootstrapped SEs in the SM. 
\FloatBarrier


\begin{figure}[!th]
\caption{Dynamic Treatment Effects w/ Imputation Estimator}\label{fg:dynamic}
\vspace{0.5em}
\centering\scriptsize
\begin{minipage}{1\linewidth}{
\centering
\hspace{0em}
\resizebox{1\textwidth}{!}{
\begin{tabular}{C{3.8cm}C{3.8cm}C{3.8cm}C{3.8cm}C{3.8cm}}
   \citet{Beazer2022} \newline ATT: 24.38 (8.79)&  
   \citet{Bischof2019} \newline ATT:  0.08 (0.07)&  
   \citet{Bisgaard2018} \newline ATT: -0.05 (0.02)& 
   \citet{Blair2022} \newline ATT: -0.50 (0.40)& 
   \citet{Bokobza2022} \newline ATT: 0.11 (0.03)\\
   \hspace{-2em} \includegraphics[width = 0.22\textwidth]{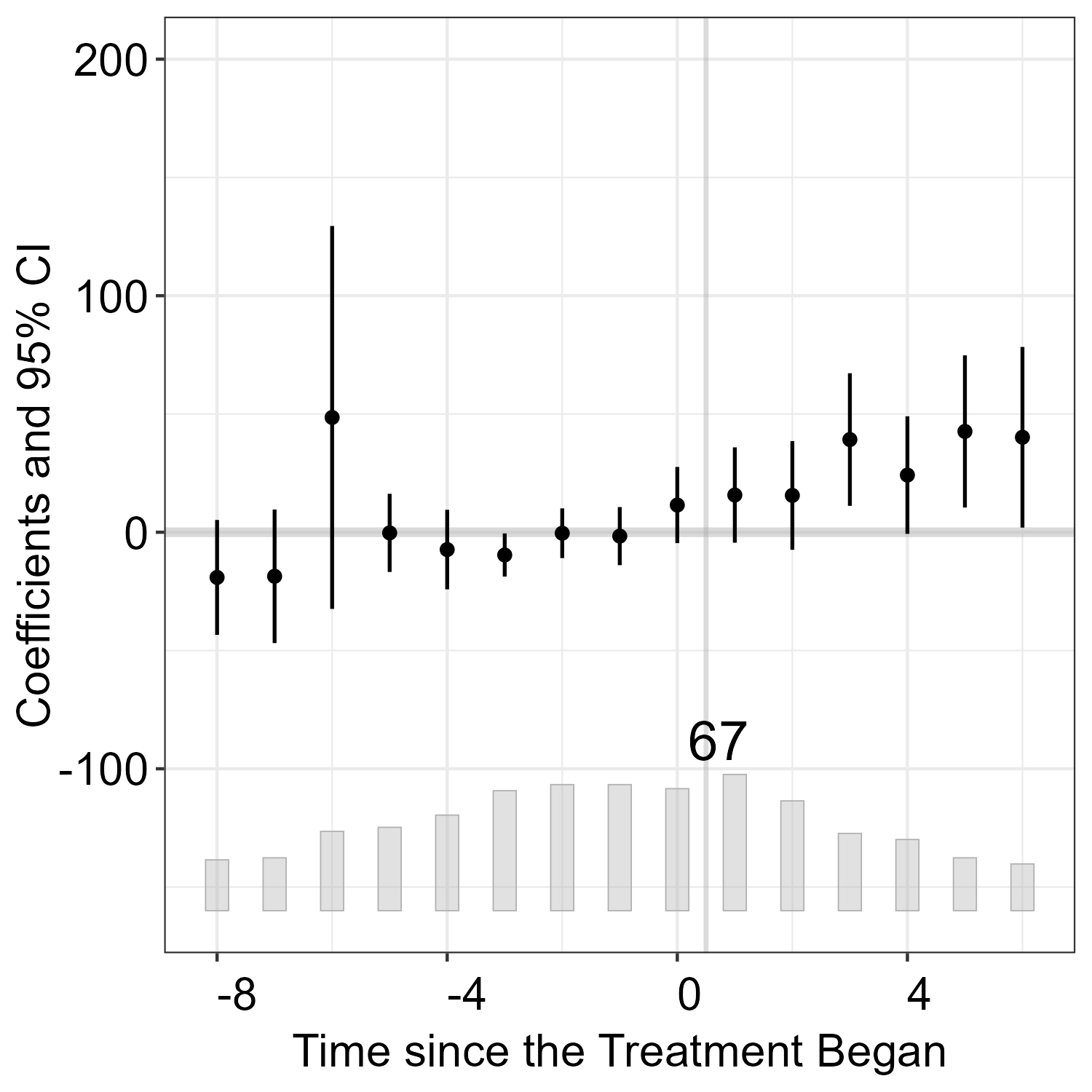}  & 
   \hspace{-2em} \includegraphics[width = 0.22\textwidth]{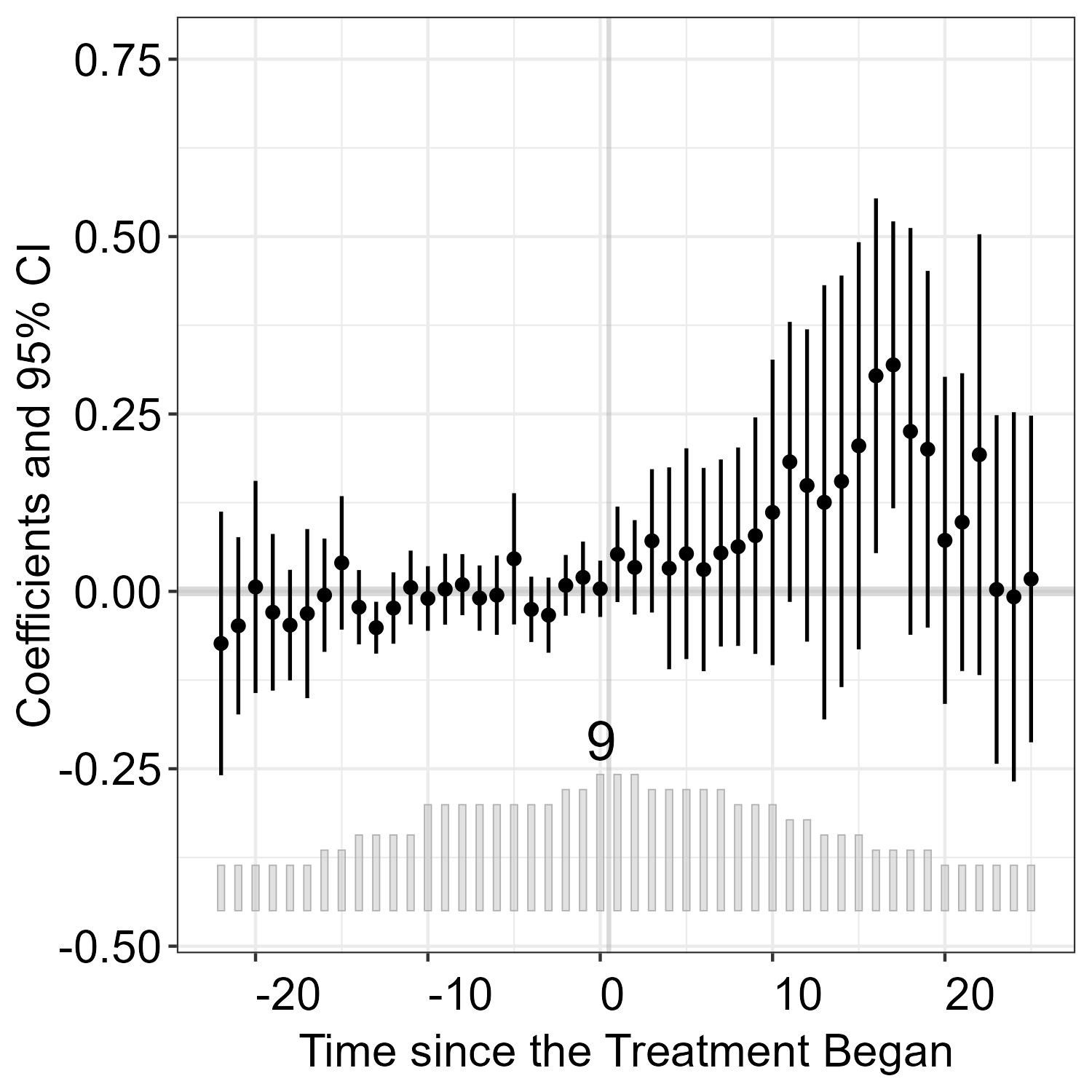}  &
   \hspace{-2em} \includegraphics[width = 0.22\textwidth]{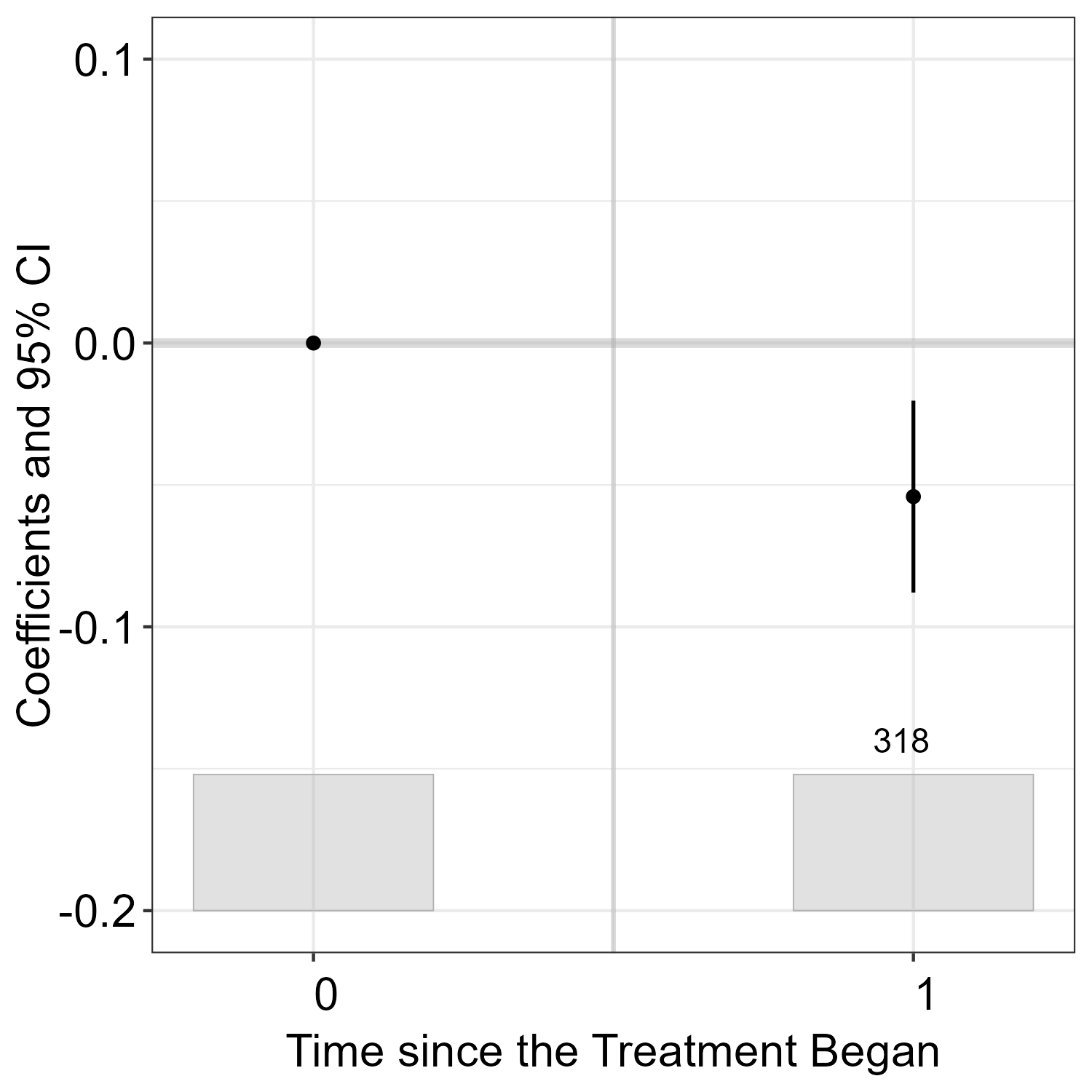}  &
   \hspace{-2em} \includegraphics[width = 0.22\textwidth]{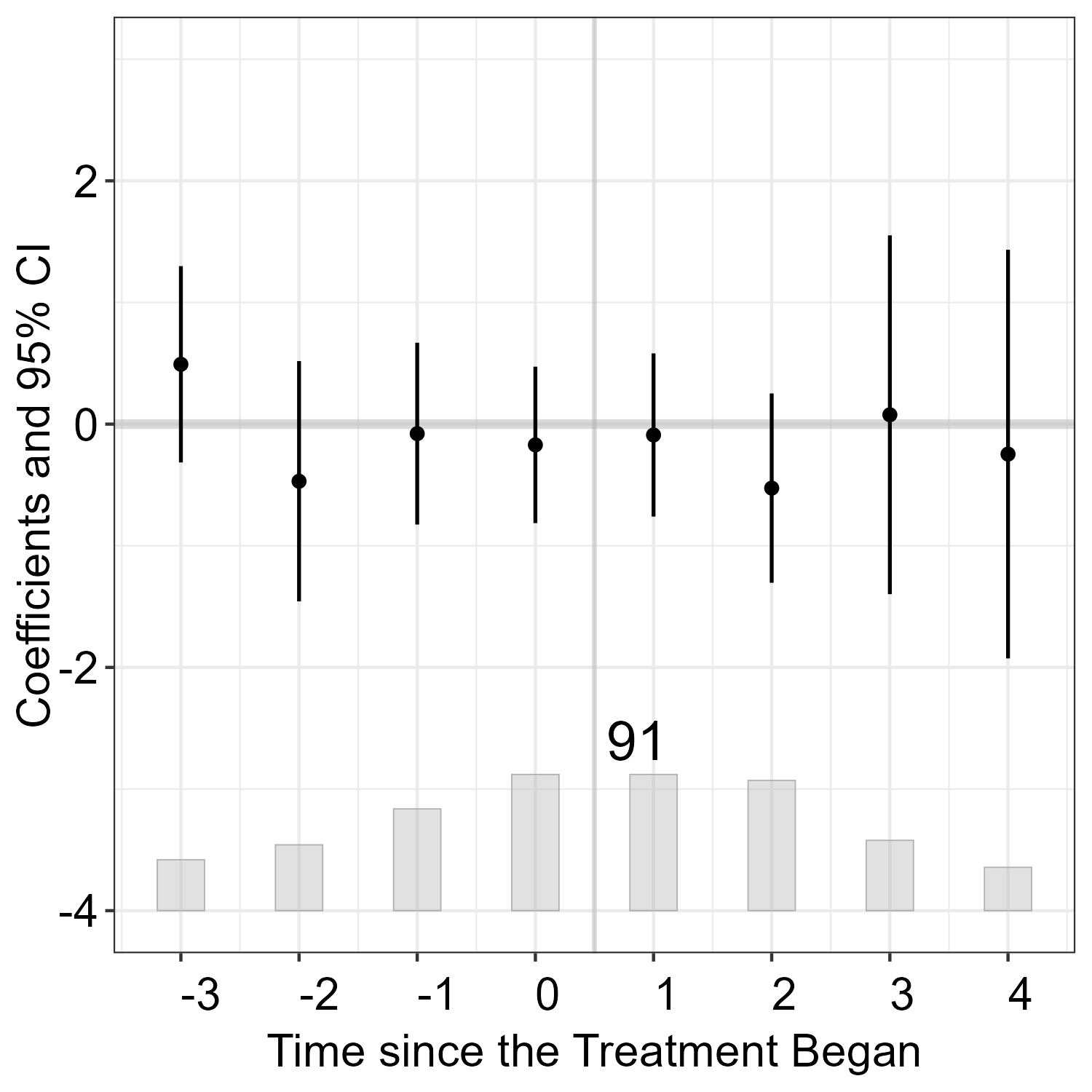}  &
   \hspace{-2em}  \includegraphics[width = 0.22\textwidth]{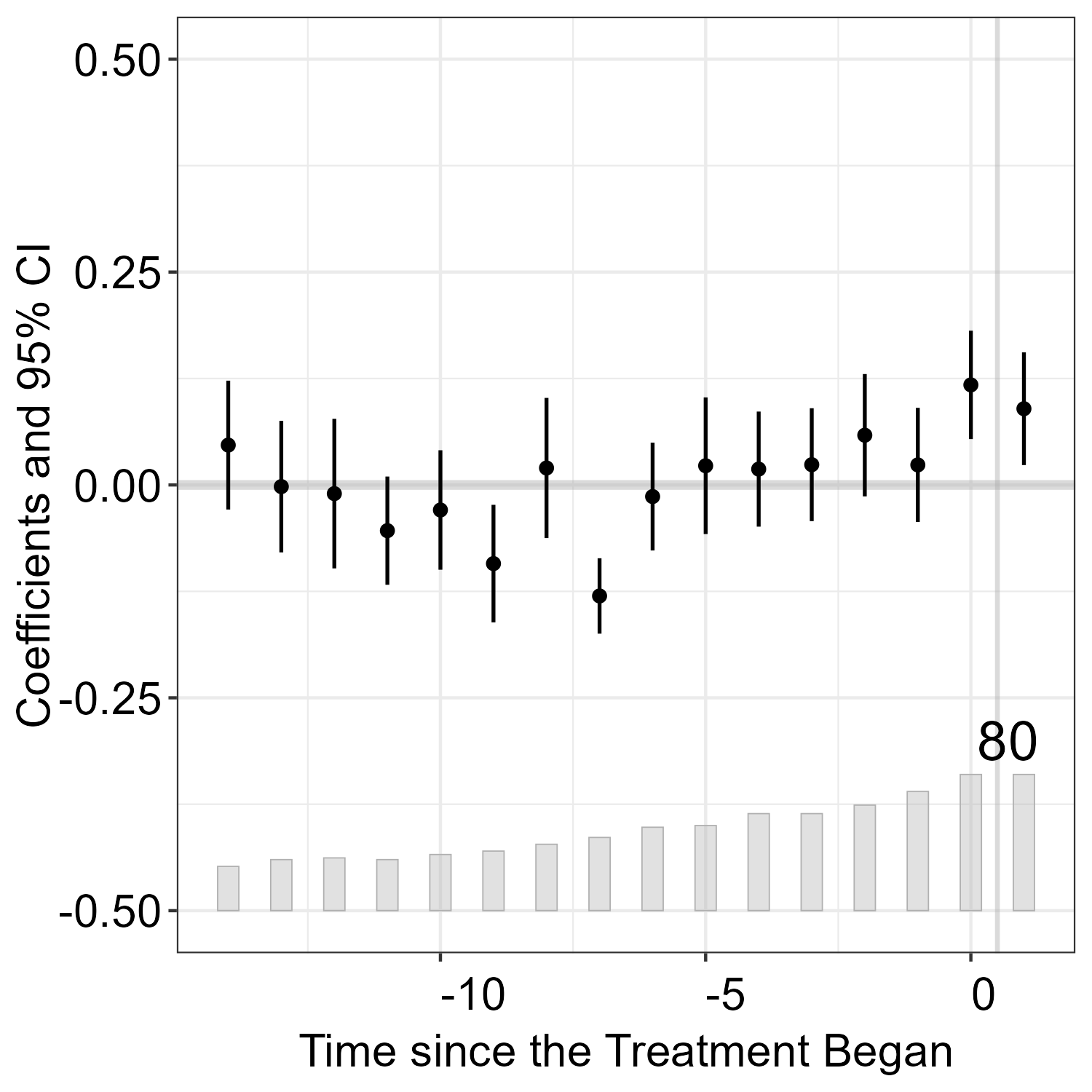} \\ \\
   \citet{Caughey2017} \newline ATT: 0.014 (0.005)&  
   \citet{Christensen2021}  \newline ATT: 0.04 (0.04)& 
   \citet{Clarke2020}  \newline ATT: 0.08 (0.03)& 
   \citet{Clayton2018} \newline  ATT: 0.19 (0.44)&
   \citet{cox2021budgetary} \newline  ATT: 0.44 (0.11)\\
   \hspace{-2em} \includegraphics[width = 0.22\textwidth]{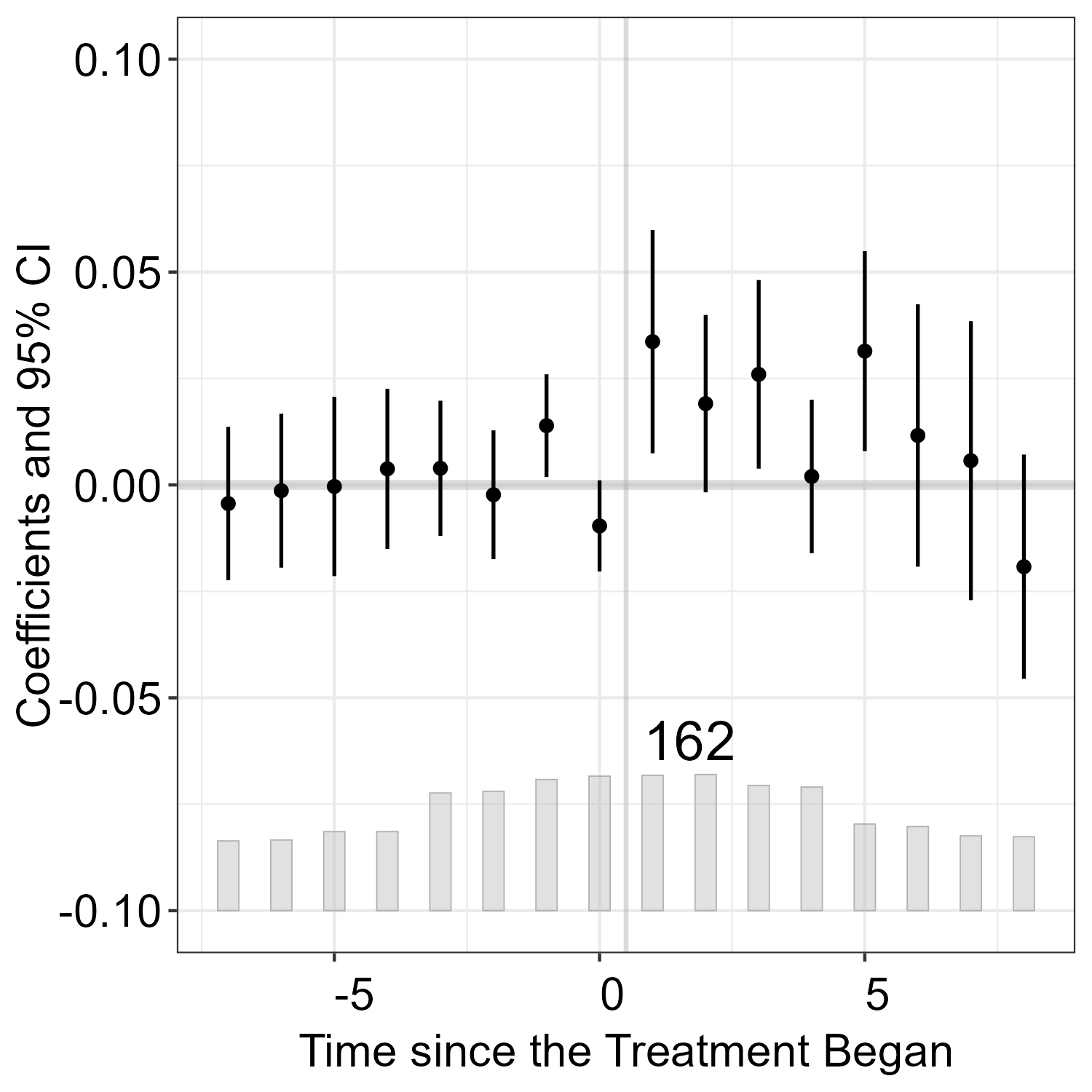}  &
   \hspace{-2em} \includegraphics[width = 0.22\textwidth]{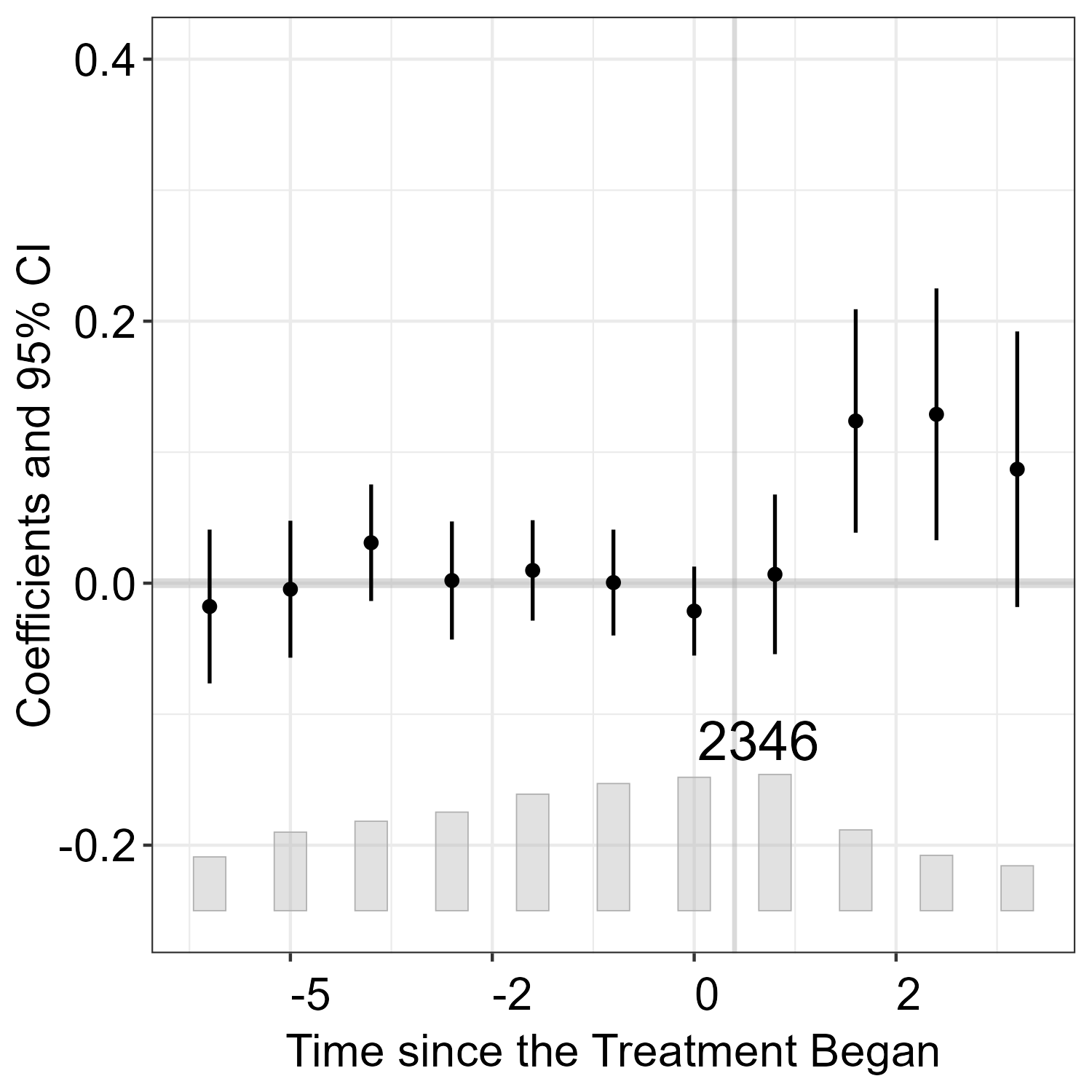}  & 
   \hspace{-2em}  \includegraphics[width = 0.22\textwidth]{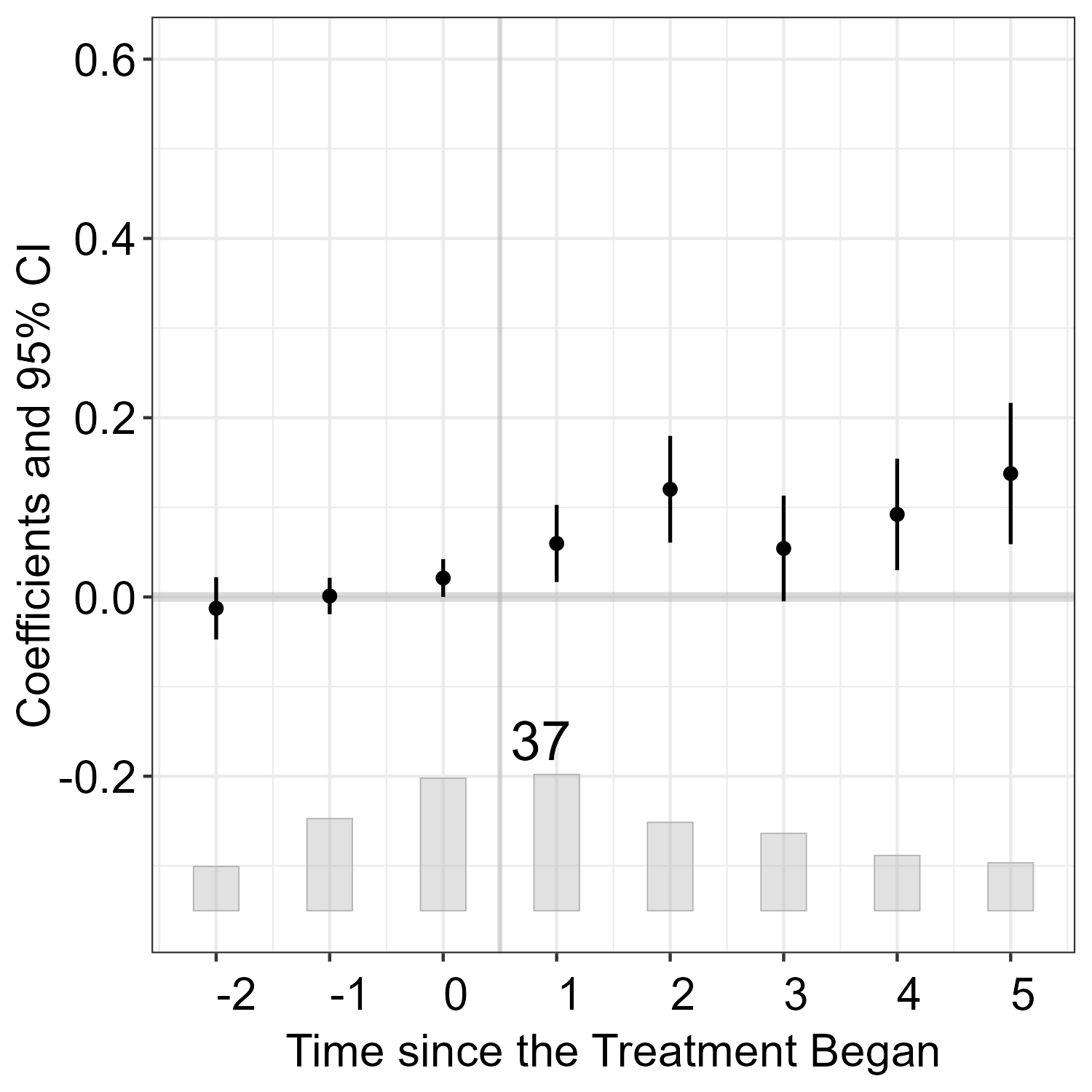}  & 
   \hspace{-2em}\includegraphics[width = 0.22\textwidth]{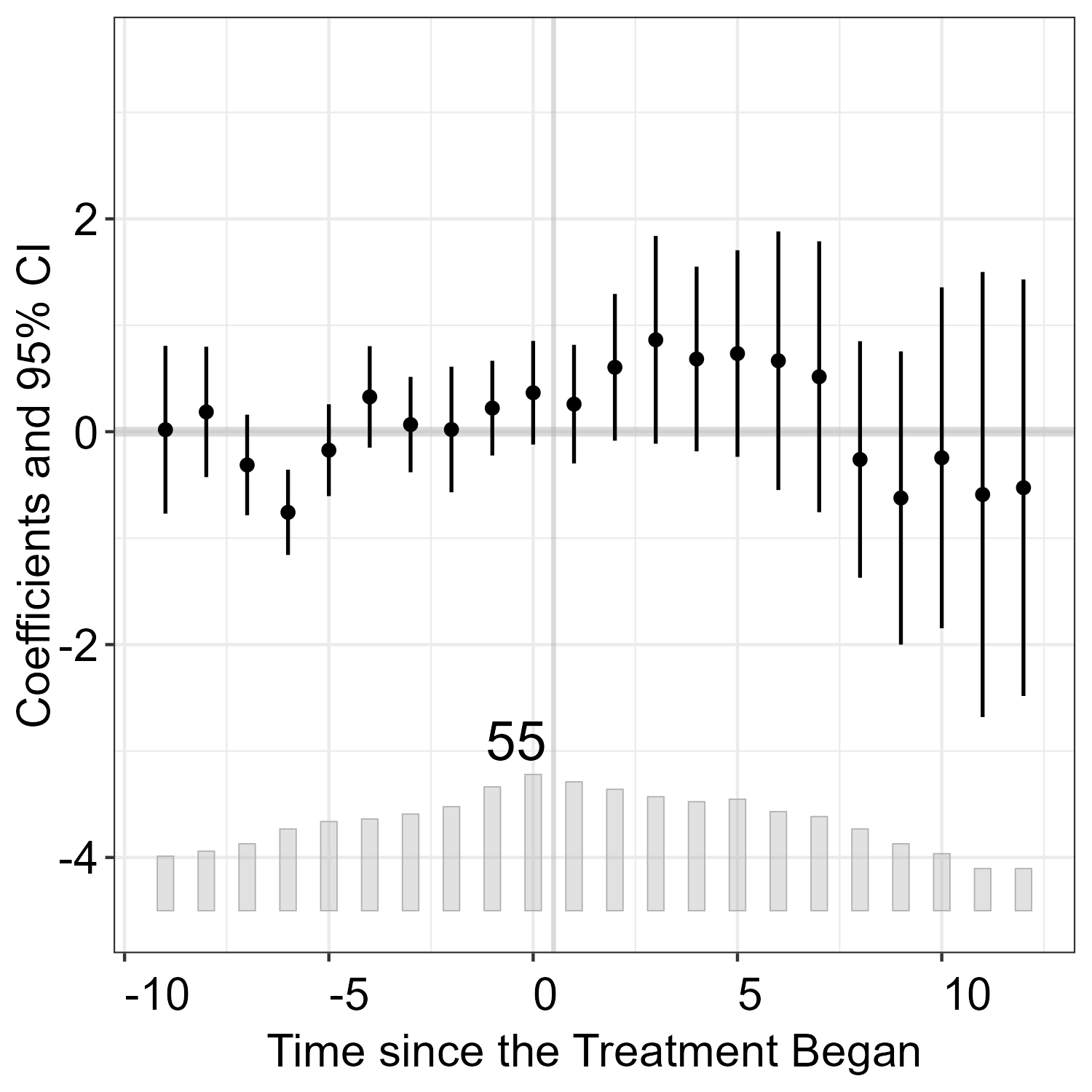} & 
   \hspace{-2em} \includegraphics[width = 0.22\textwidth]{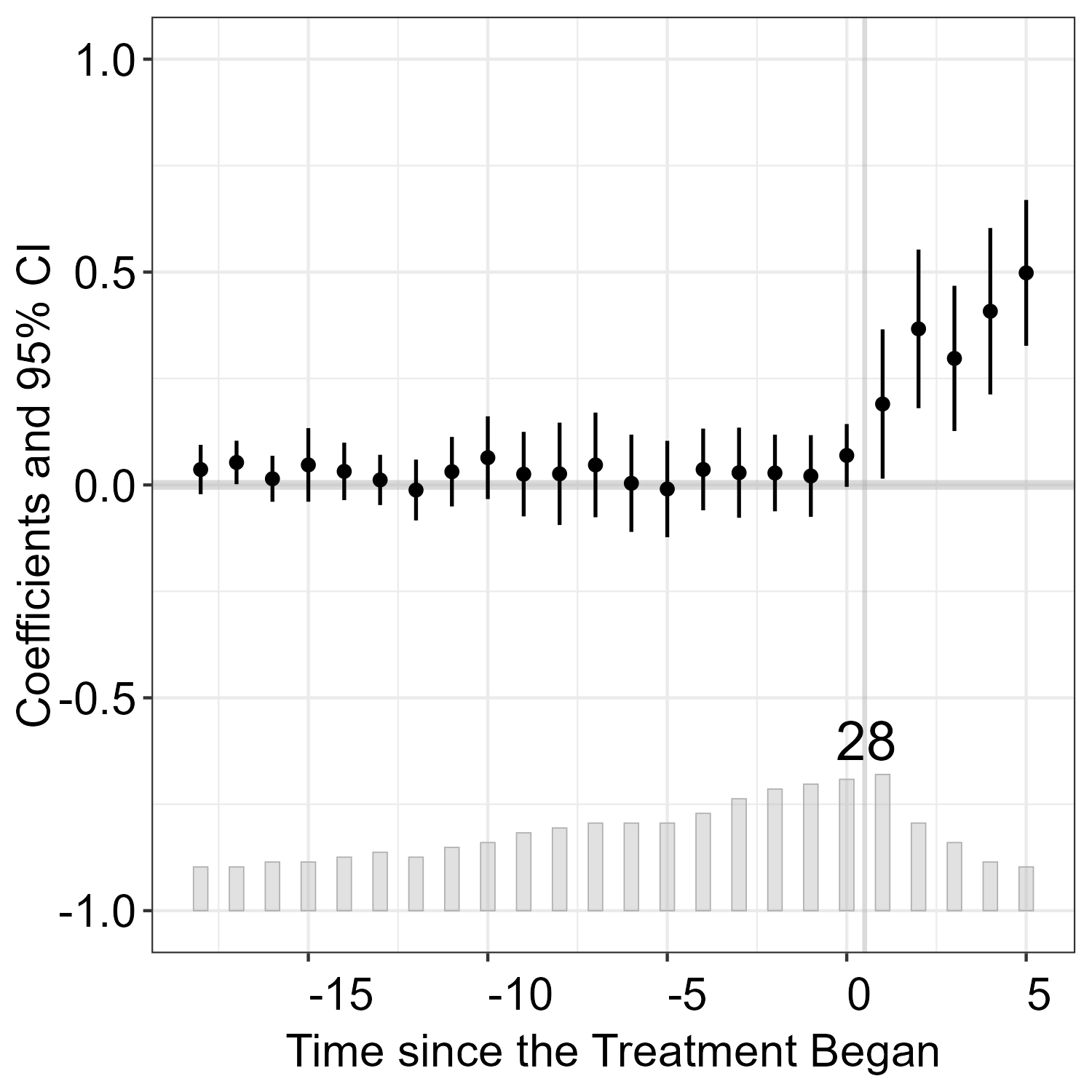} \\ \\      
    \citet{Dahlstrom2023} \newline ATT: -1.67 (0.65) & 
     \citet{Dipoppa2023} \newline ATT: 0.0012 (0.0004) & 
     \citet{Distelhorst2018} \newline ATT: 0.15 (0.05)& 
      \citet{eckhouse2022metrics} \newline ATT: 0.015 (0.014)& 
   \citet{Esberg2023} \newline ATT: 2.89 (0.57)\\
   \hspace{-2em}\includegraphics[width = 0.22\textwidth]{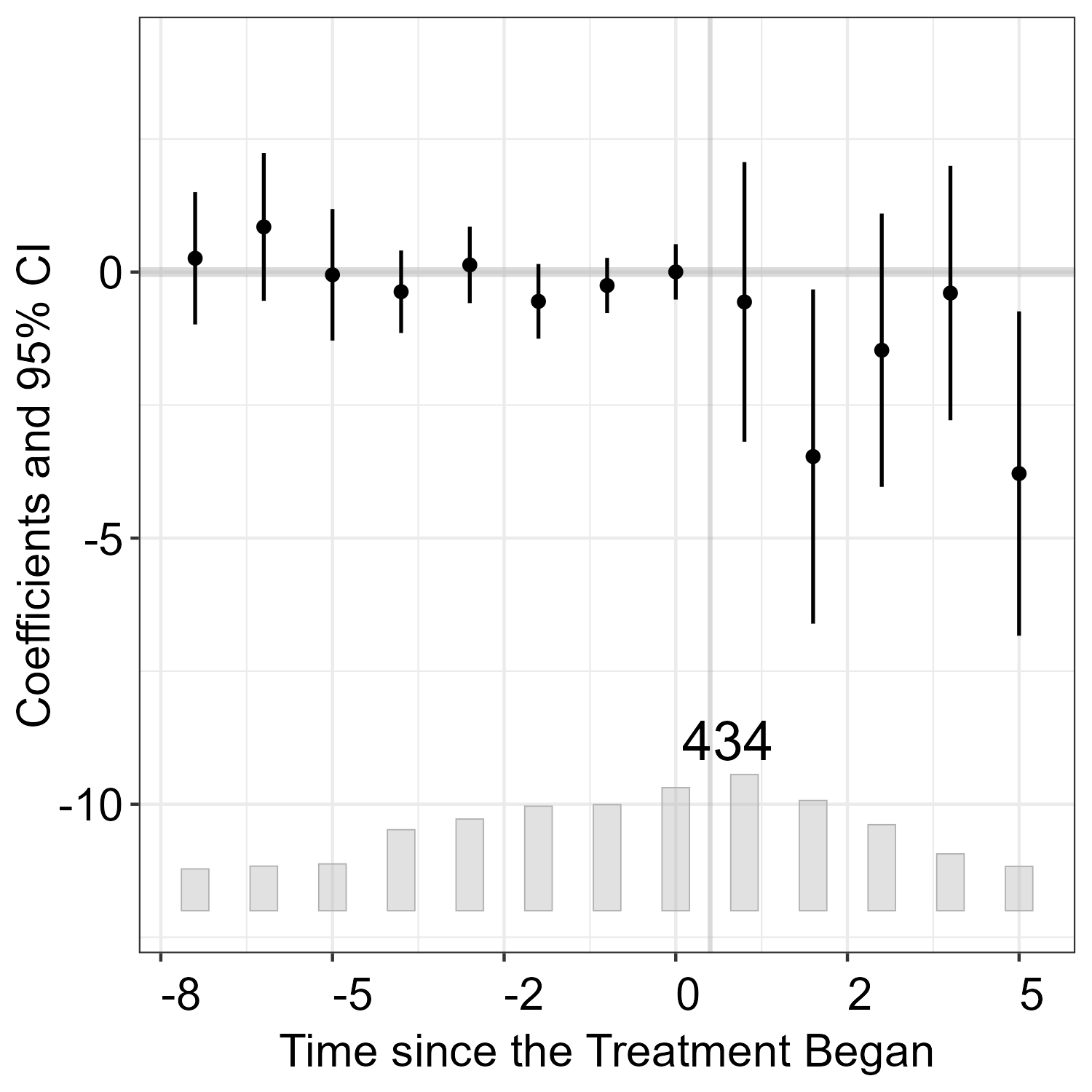} &
    \hspace{-2em} \includegraphics[width = 0.22\textwidth]{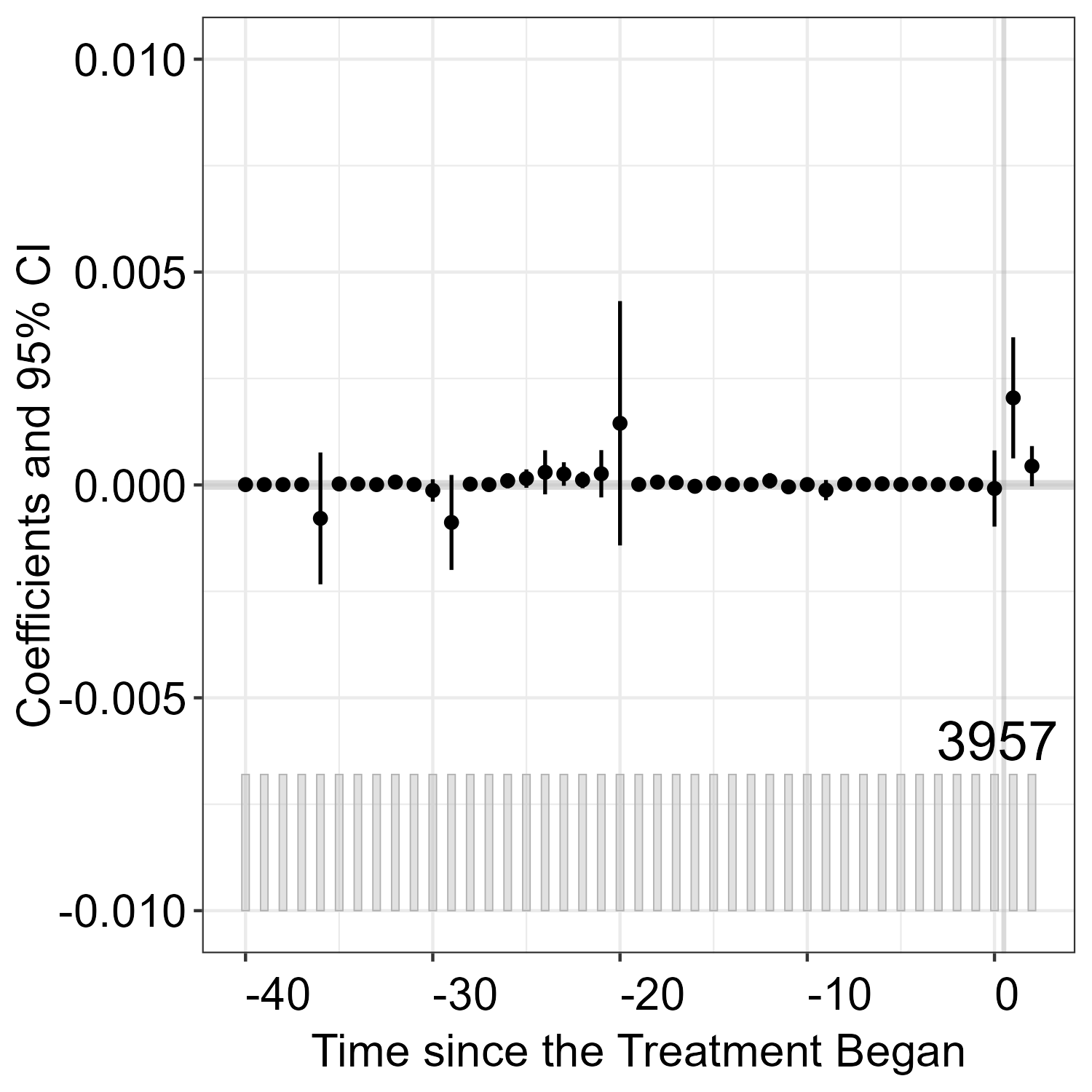} & 
    \hspace{-2em}\includegraphics[width = 0.22\textwidth]{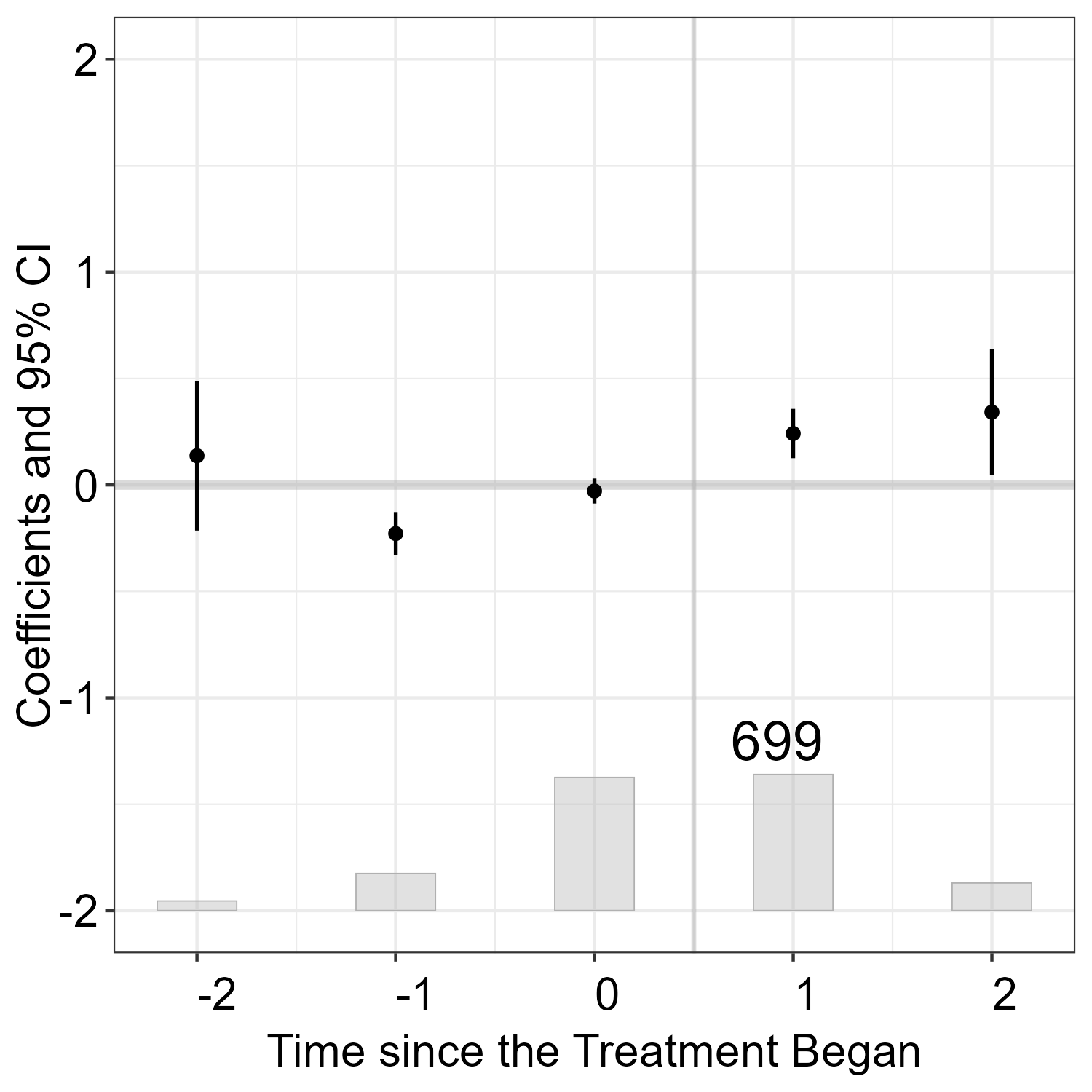} &
   \hspace{-2em} \includegraphics[width = 0.22\textwidth]{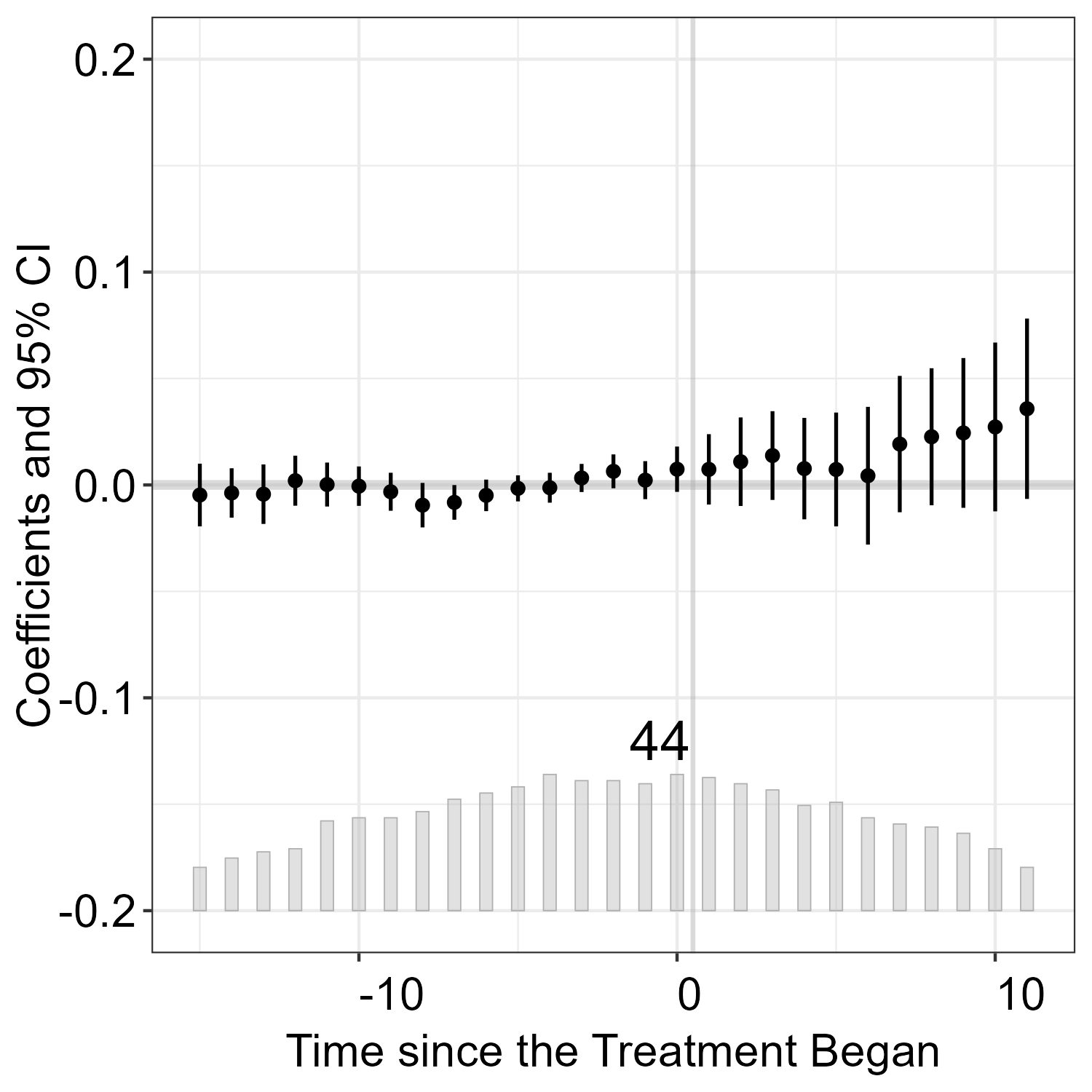} &
   \hspace{-2em}\includegraphics[width = 0.22\textwidth]{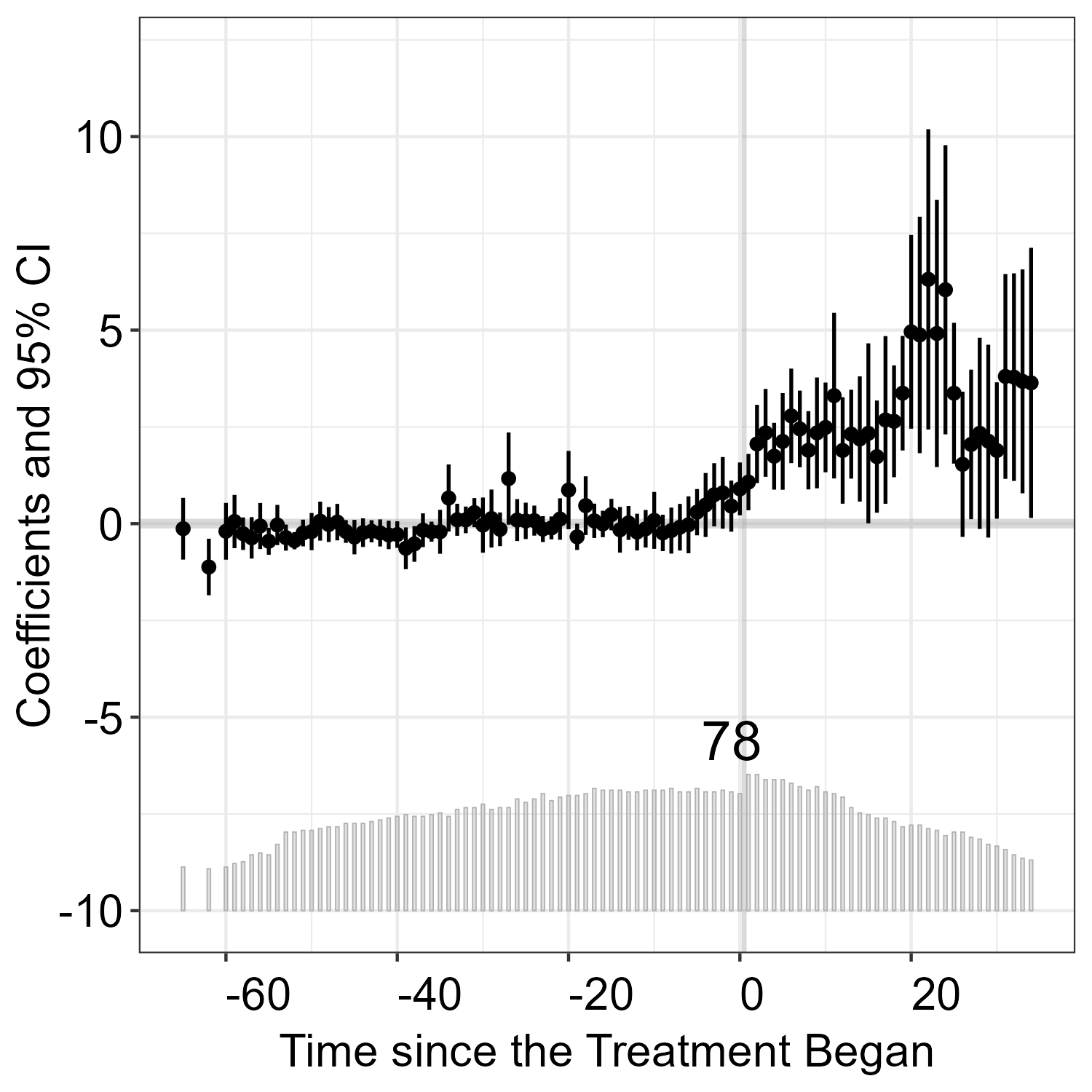}\\ \\  
   \citet{Fouirnaies2018ajps} \newline ATT: 0.87 (0.11)&
   \citet{fh2018} \newline ATT:  0.29 (0.02)&
   \citet{Fouirnaies2022} \newline ATT: -0.26 (0.03)& 
    \citet{Fresh2018} \newline  ATT: 0.17 (0.06)& 
    \citet{Garfias2019jop} \newline  ATT: 0.07 (0.03)\\   
    \hspace{-2em} \includegraphics[width = 0.22\textwidth]{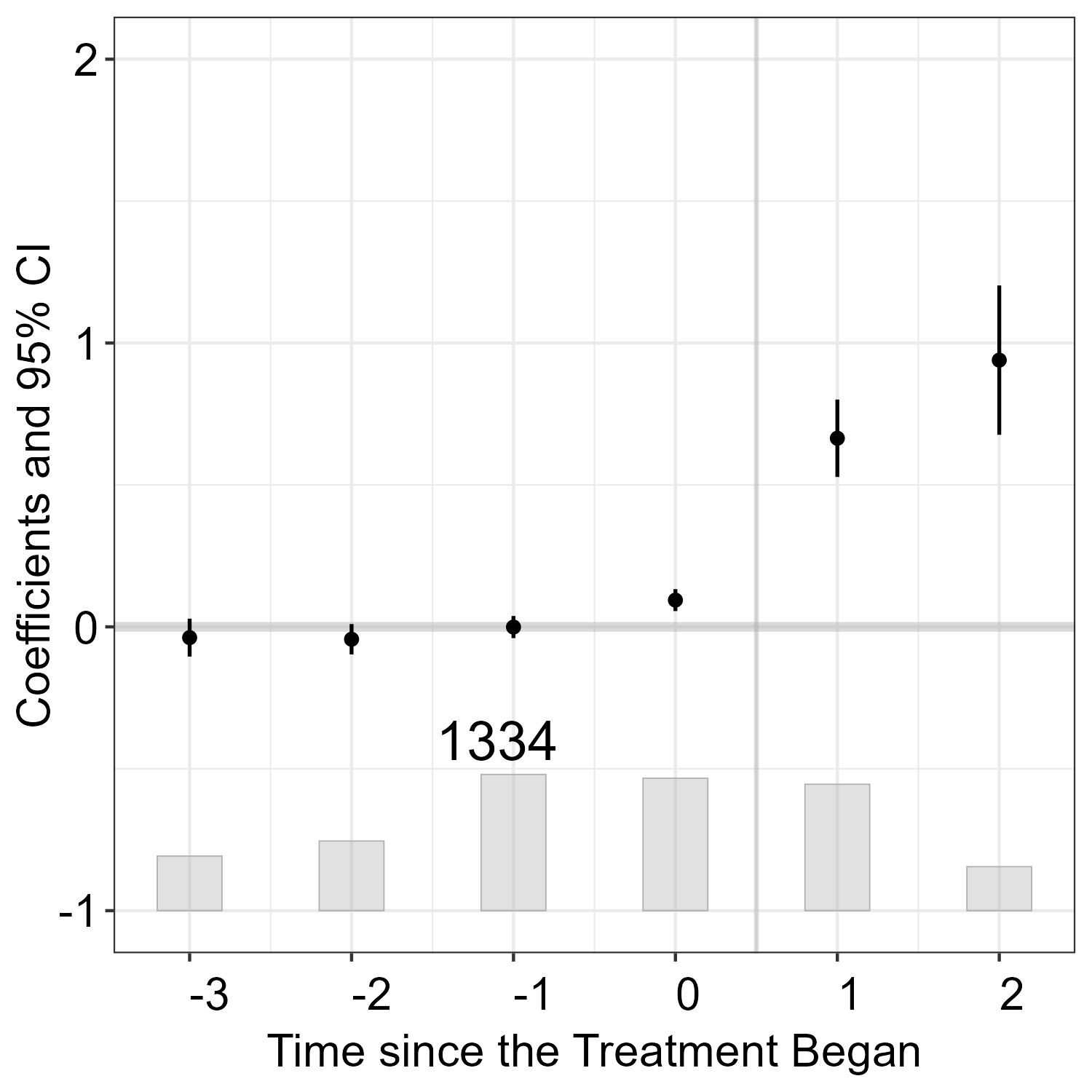} &
   \hspace{-2em}  \includegraphics[width = 0.22\textwidth]{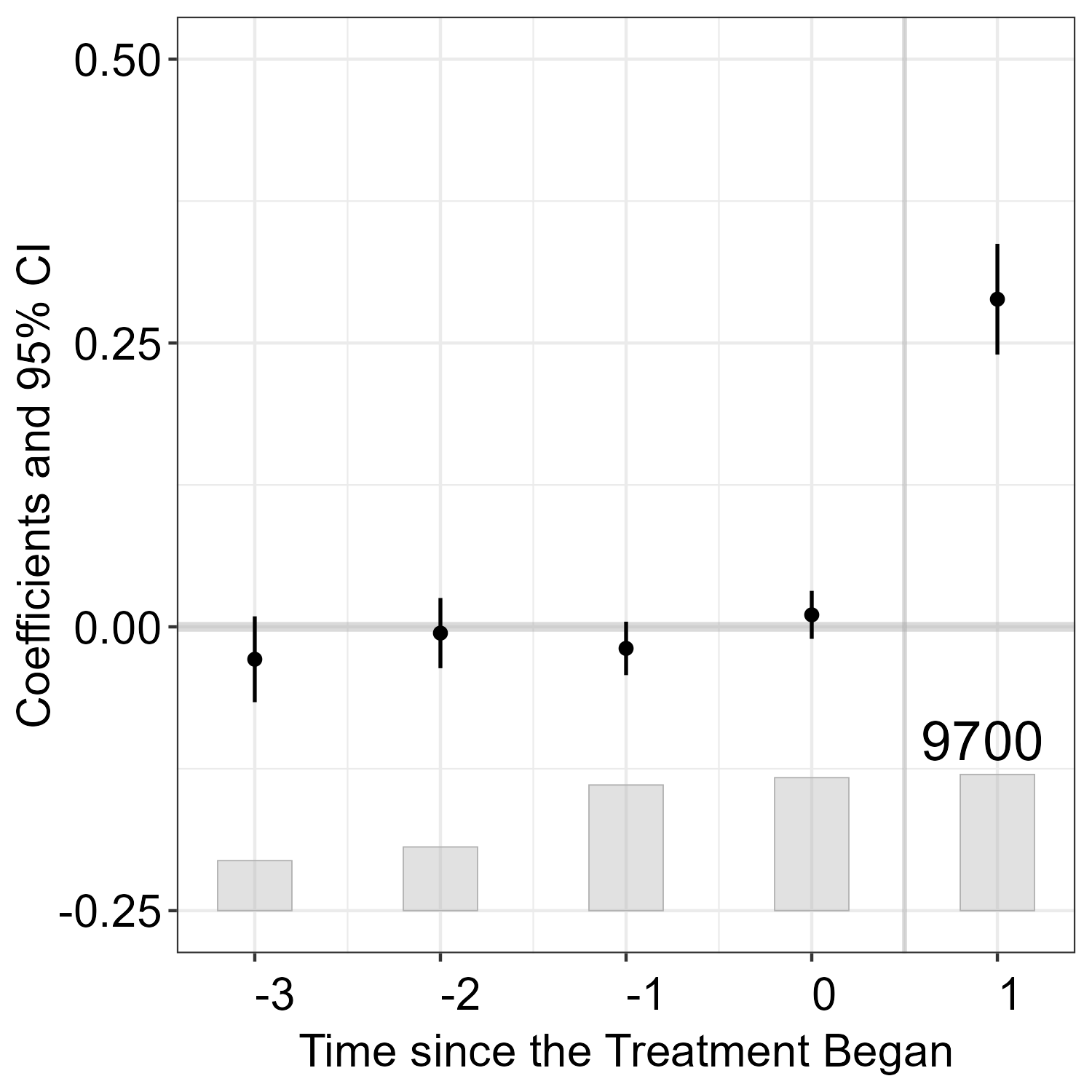} & 
   \hspace{-2em} \includegraphics[width = 0.22\textwidth]{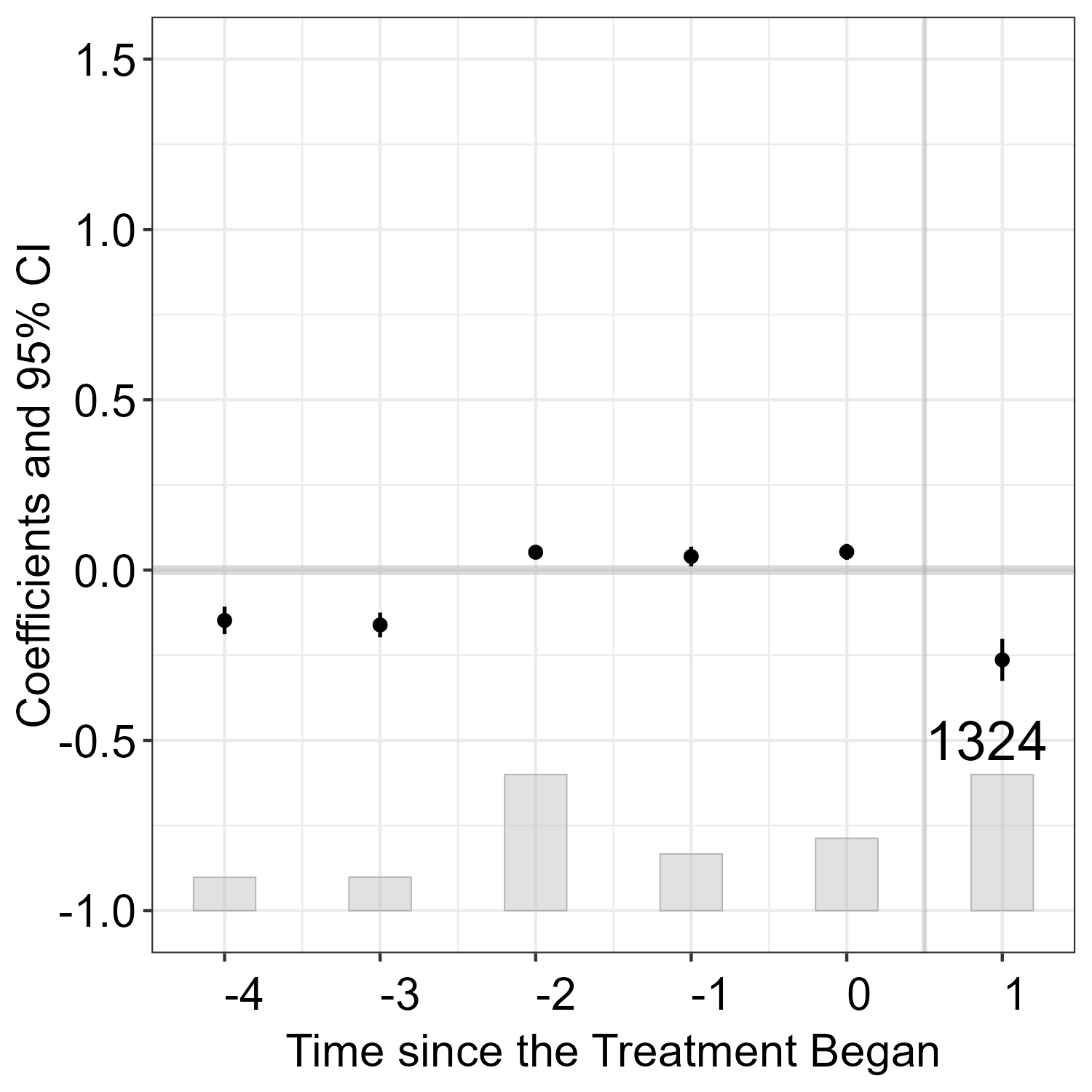}  &
    \hspace{-2em}  \includegraphics[width = 0.22\textwidth]{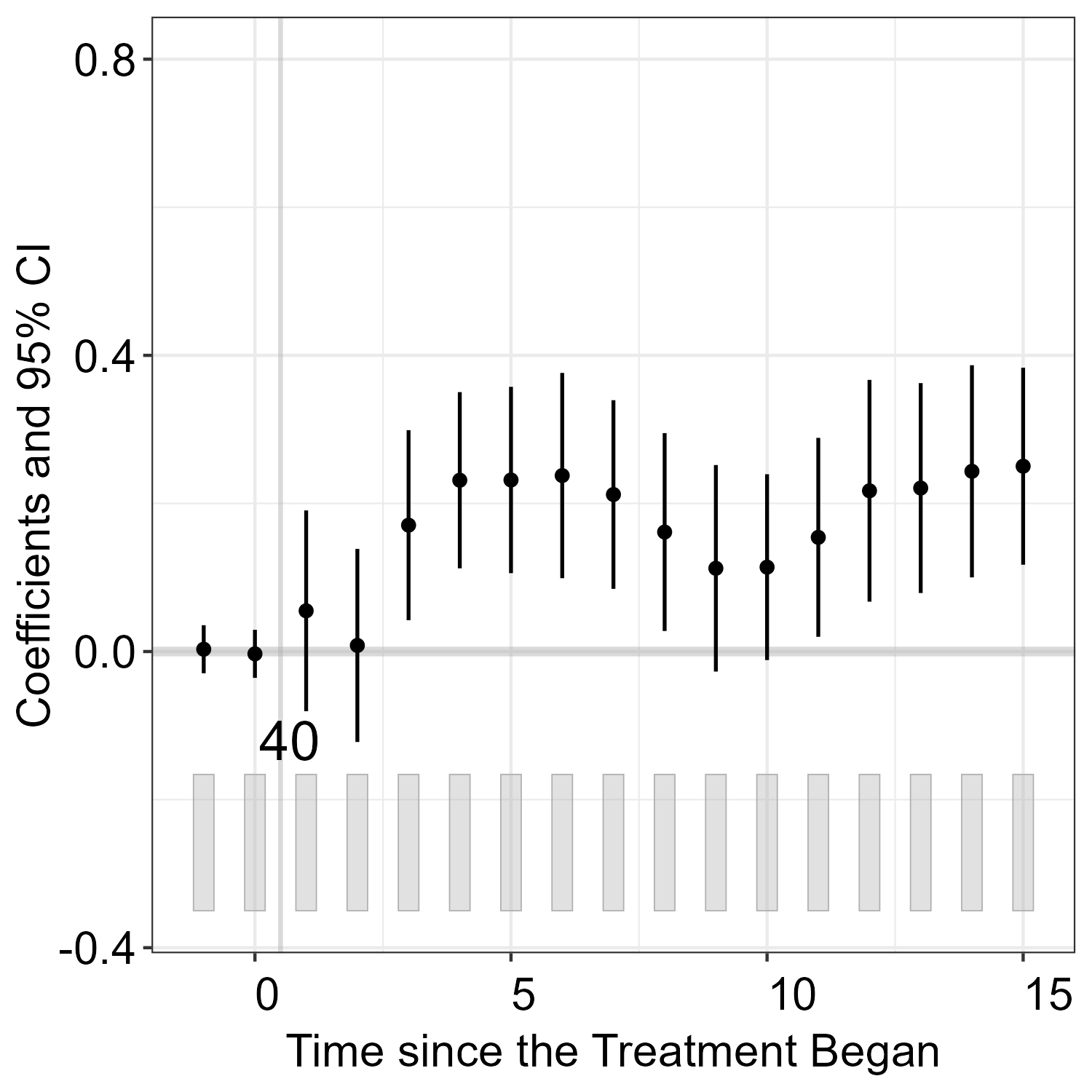} &
    \hspace{-2em}  \includegraphics[width = 0.22\textwidth]{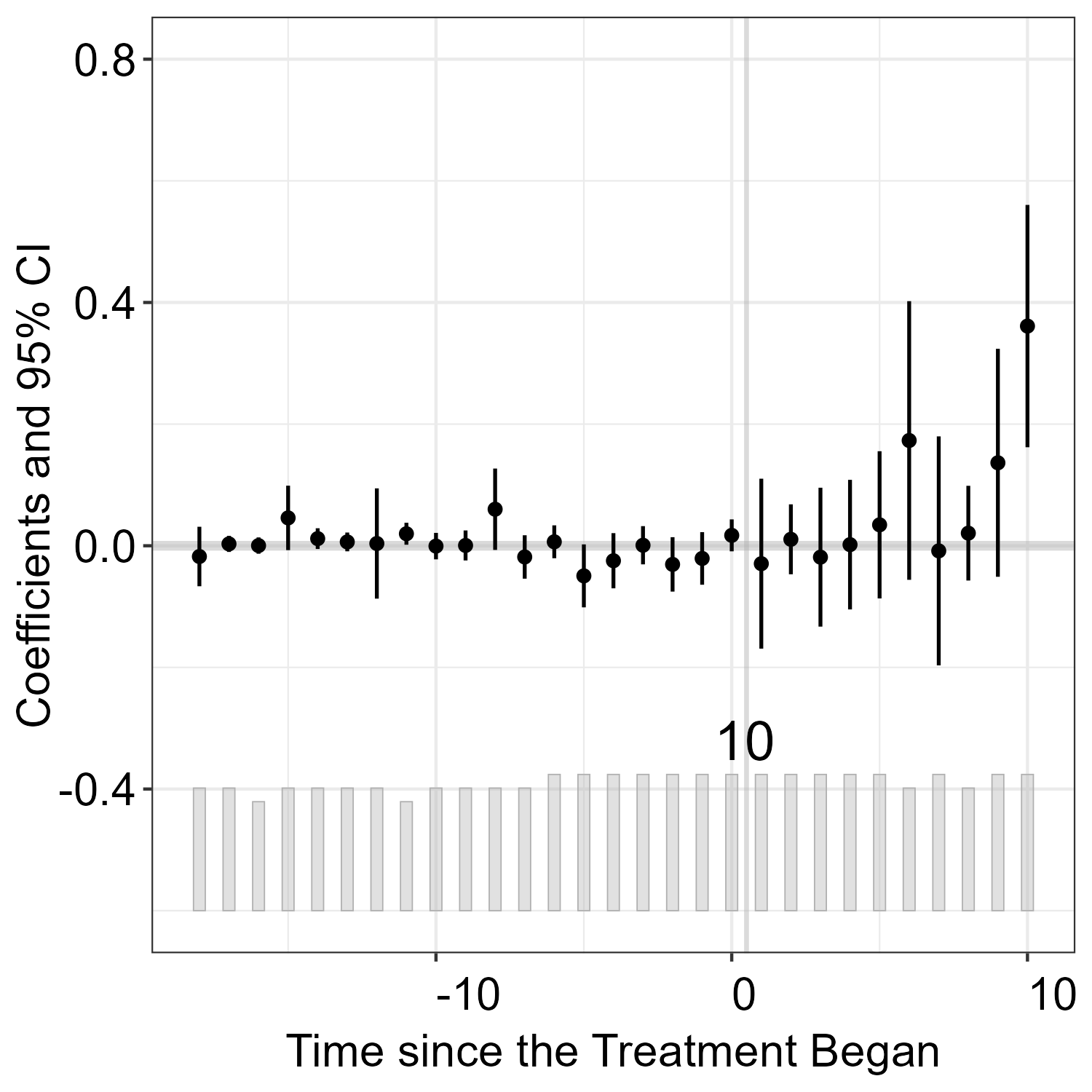}\\ \\ 
   \\
       \citet{Grumbach2023}\newline  ATT: -0.57 (0.15) &
   \citet{Grumbach2020}\newline  ATT: 0.13 (0.03)&
   \citet{Grumbach2022} \newline  ATT: 0.033 (0.019) &
      \citet{hainmueller2019does} \newline  ATT: 1.51 (0.21) & 
    \citet{Hankinson2023} \newline  ATT: -0.51 (0.45)\\   
     \hspace{-2em}  \includegraphics[width = 0.22\textwidth]{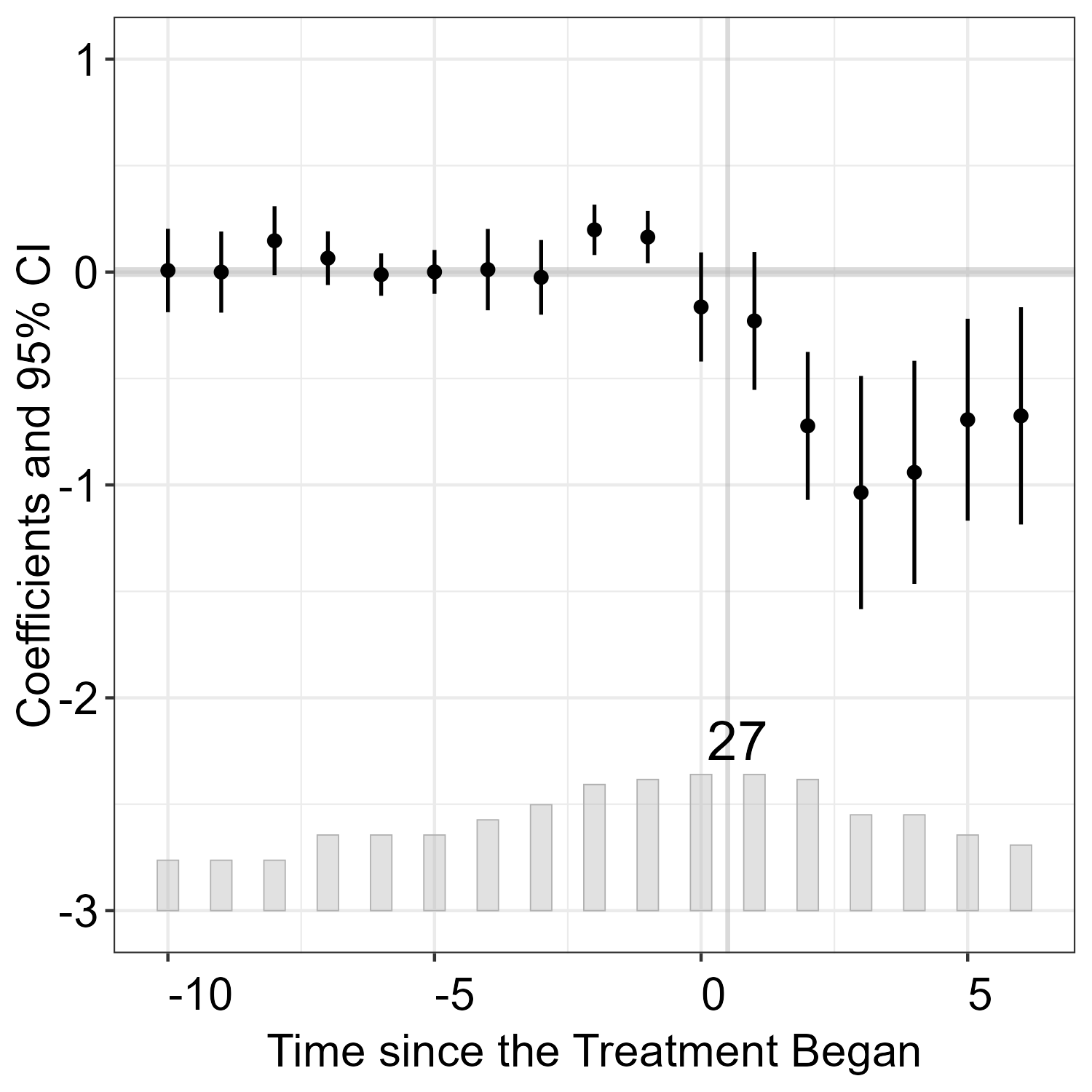} &
   \hspace{-2em}  \includegraphics[width = 0.22\textwidth]{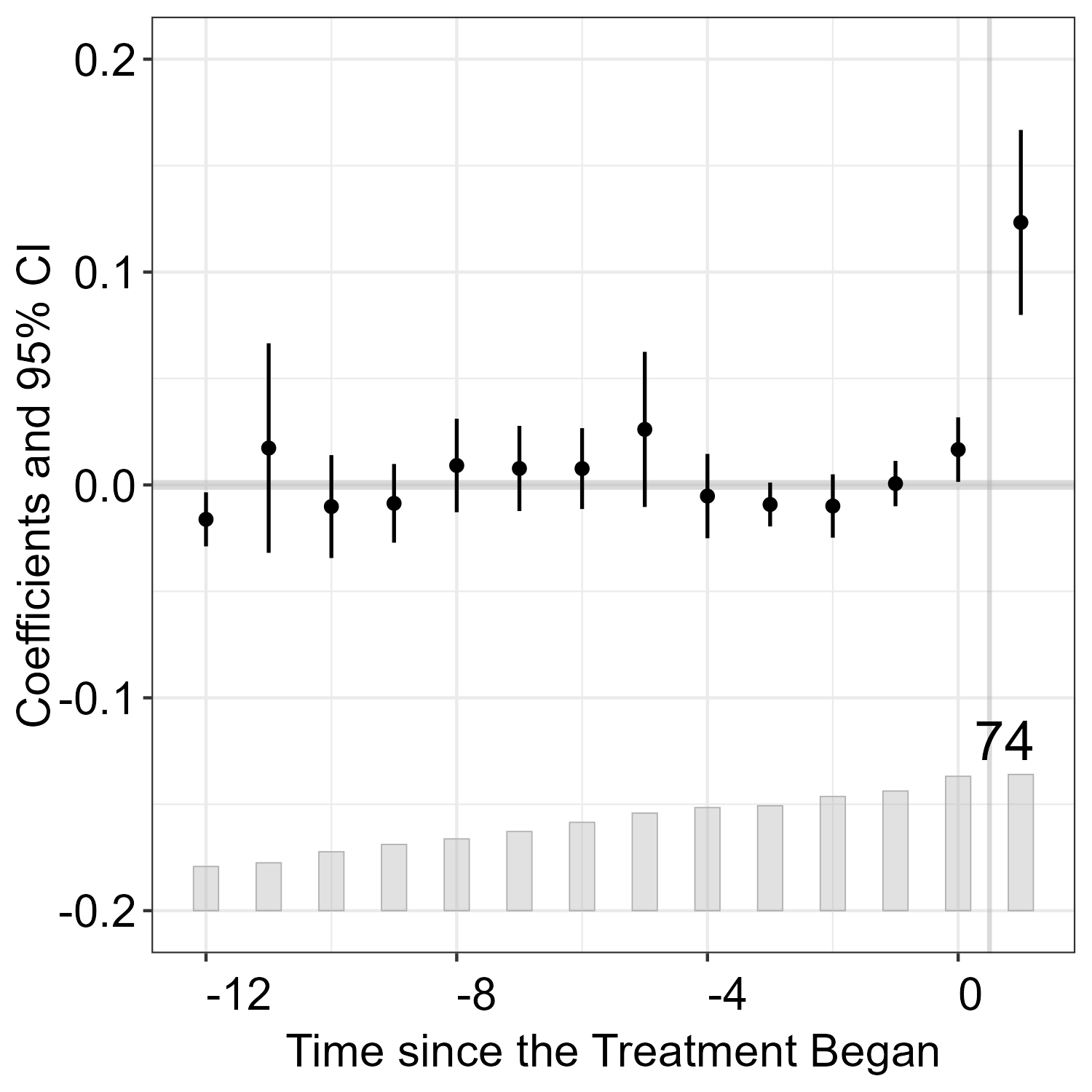} &
   \hspace{-2em}  \includegraphics[width = 0.22\textwidth]{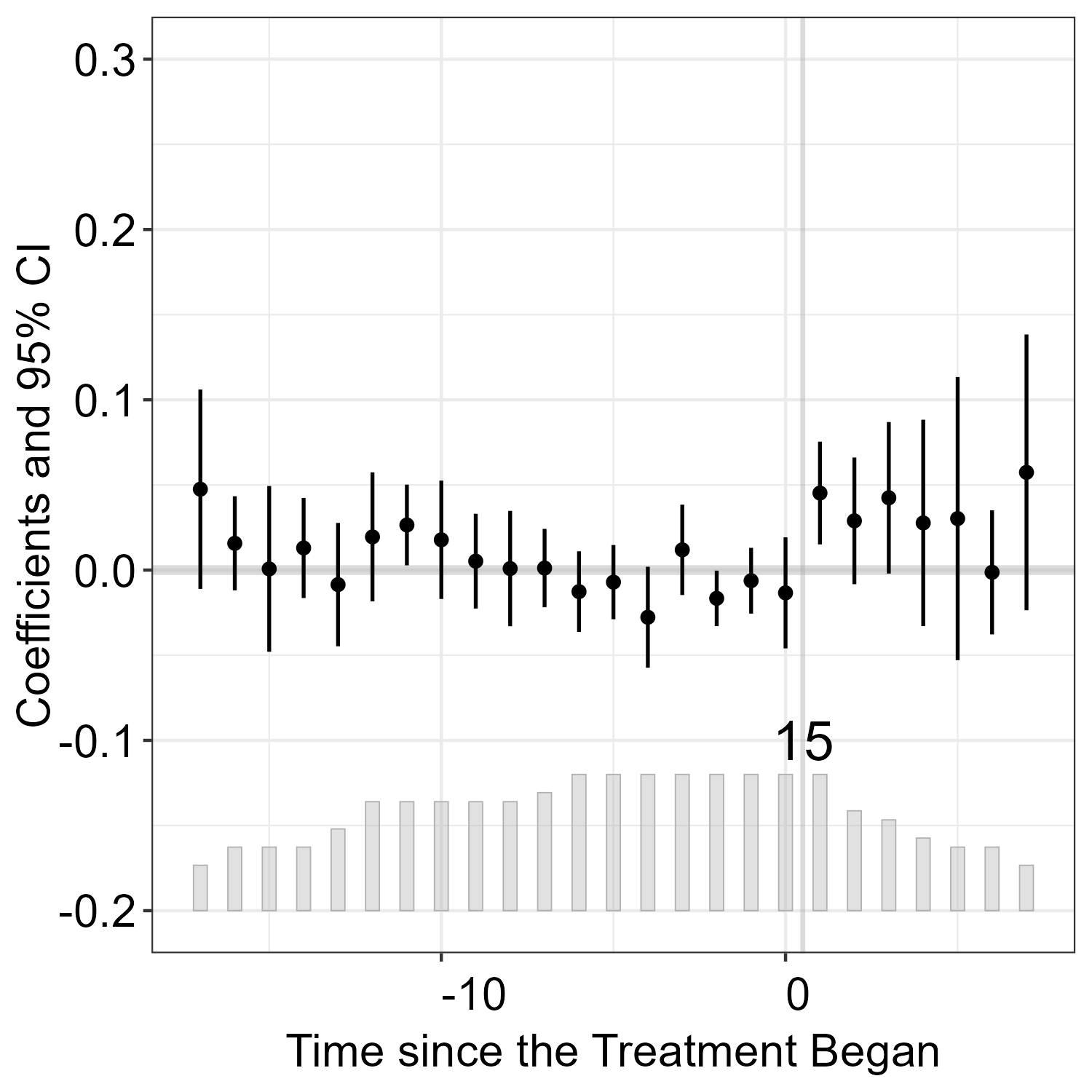}  &
   \hspace{-2em}  \includegraphics[width = 0.22\textwidth]{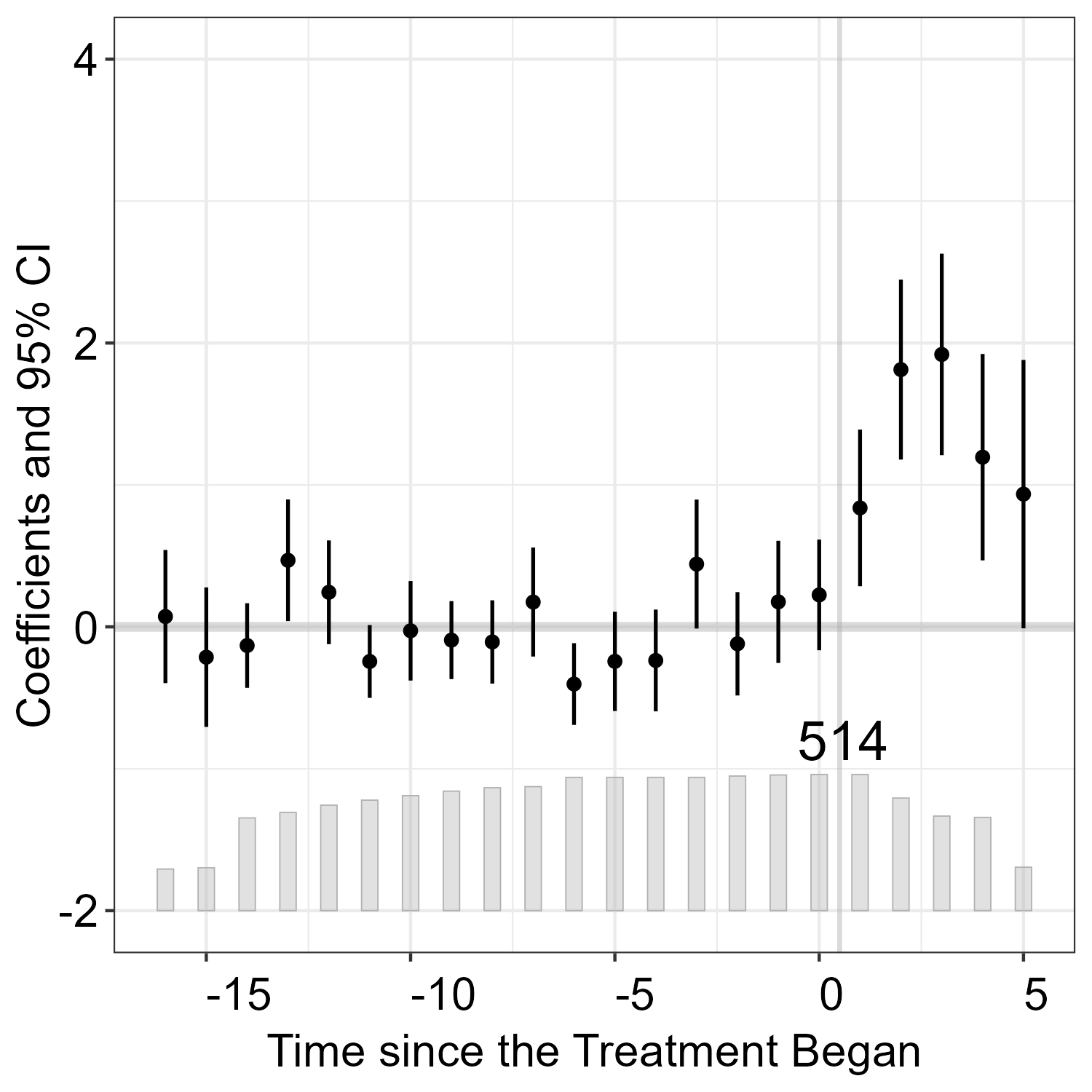} &
    \hspace{-2em}  \includegraphics[width = 0.22\textwidth]{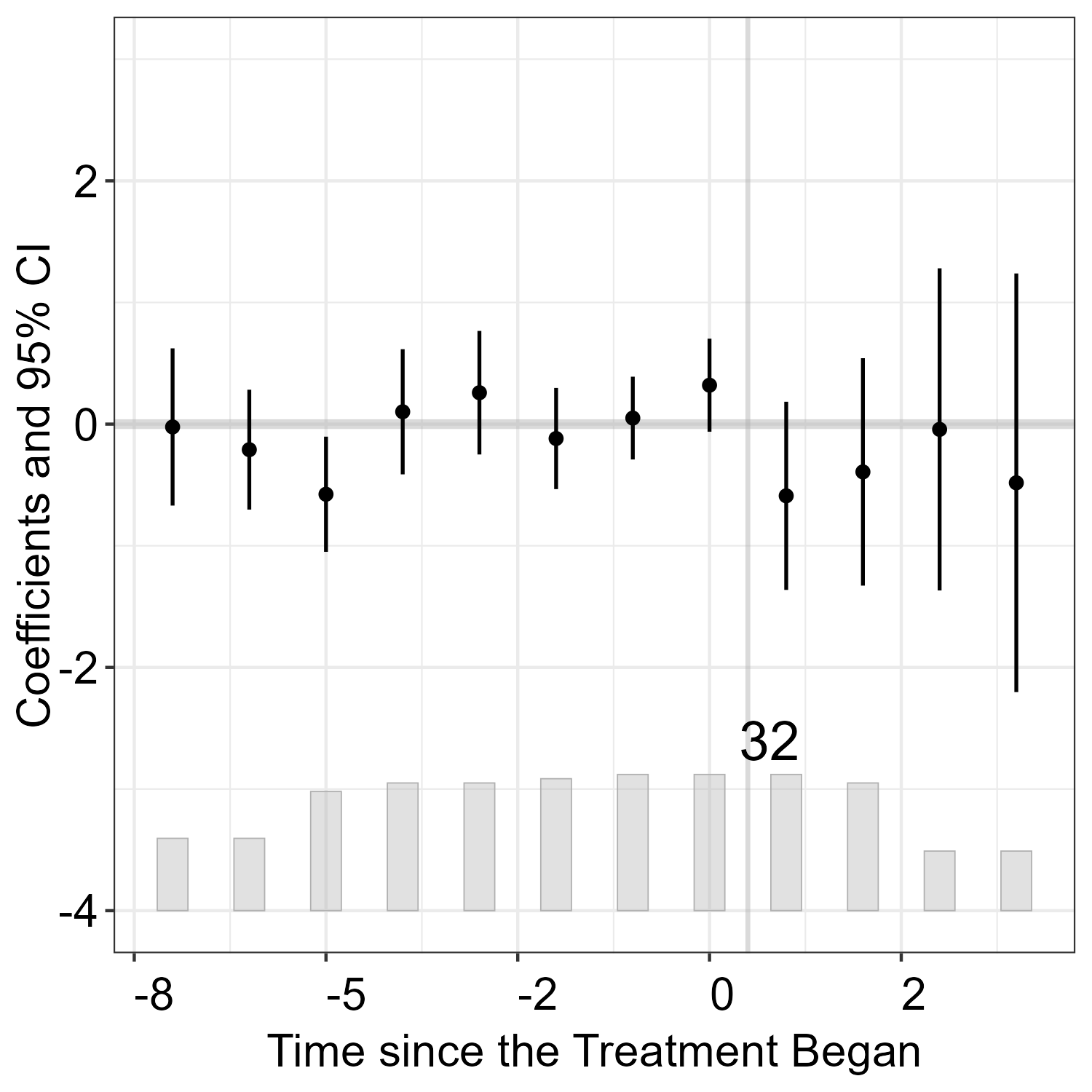} \\ \\ 
   \\
\end{tabular}}
}
\end{minipage}\vspace{-0.5em}
\addtocounter{figure}{-1}
\end{figure}

\begin{figure}[!ht]
\caption{Dynamic Treatment Effects w/ Imputation Estimator (Cont.)}
\vspace{0.5em}
\centering\scriptsize
\begin{minipage}{1\linewidth}{
\centering
\hspace{0em}
\resizebox{1\textwidth}{!}{
\begin{tabular}{C{3.8cm}C{3.8cm}C{3.8cm}C{3.8cm}C{3.8cm}}    
   \citet{Hall2022} \newline  ATT: 0.058 (0.003)& 
   \citet{Hirano2022} \newline  ATT: 0.56 (0.17)&
   \citet{Jiang2018} \newline  ATT: -0.87 (0.20)& 
   \citet{kilborn2022public} \newline ATT: 0.12 (0.04) &
   \citet{Kogan2021} \newline  ATT: 0.45 (0.43)\\ 
   \hspace{-2em}  \includegraphics[width = 0.22\textwidth]{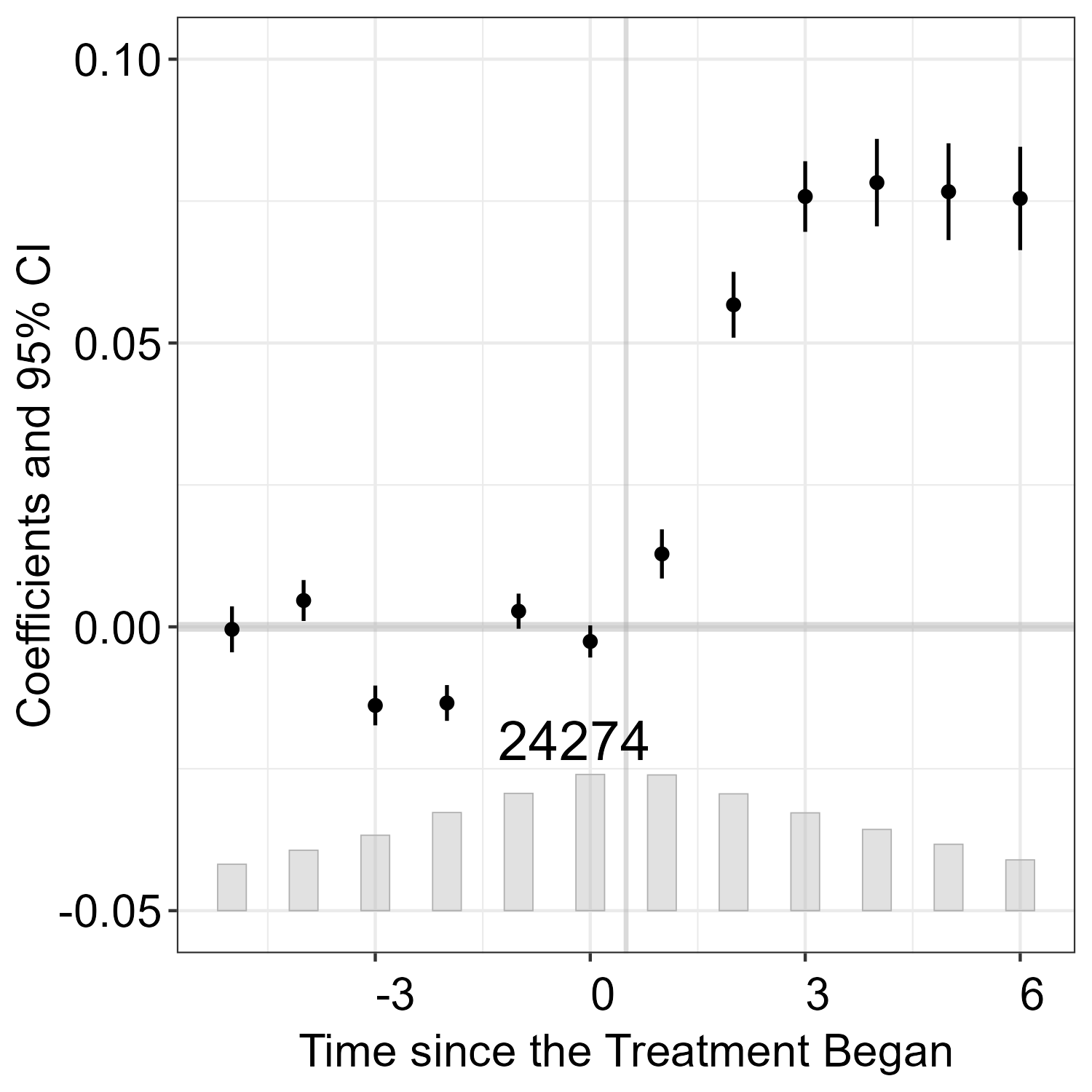}  &
   \hspace{-2em} \includegraphics[width = 0.22\textwidth]{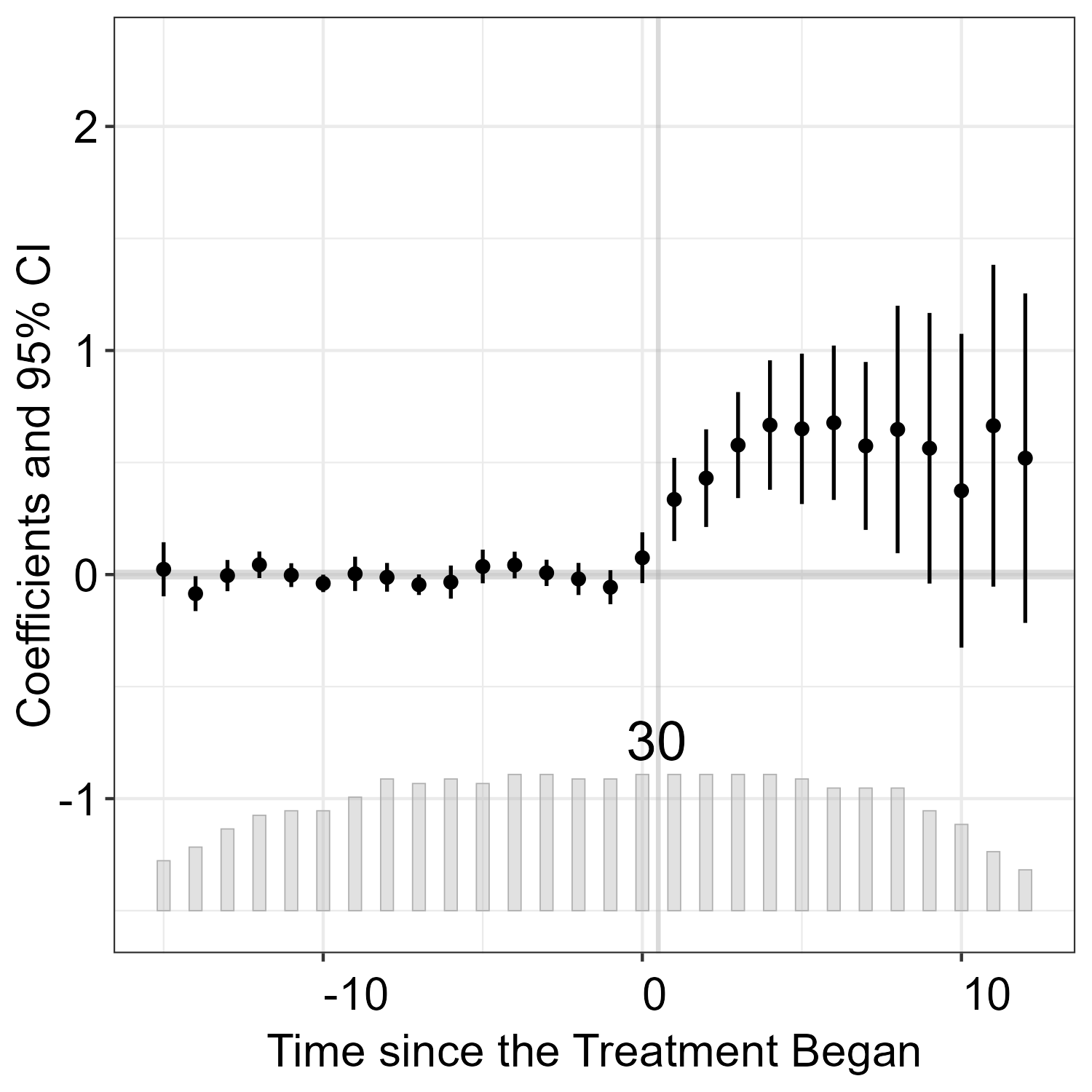}  &
   \hspace{-2em} \includegraphics[width = 0.22\textwidth]{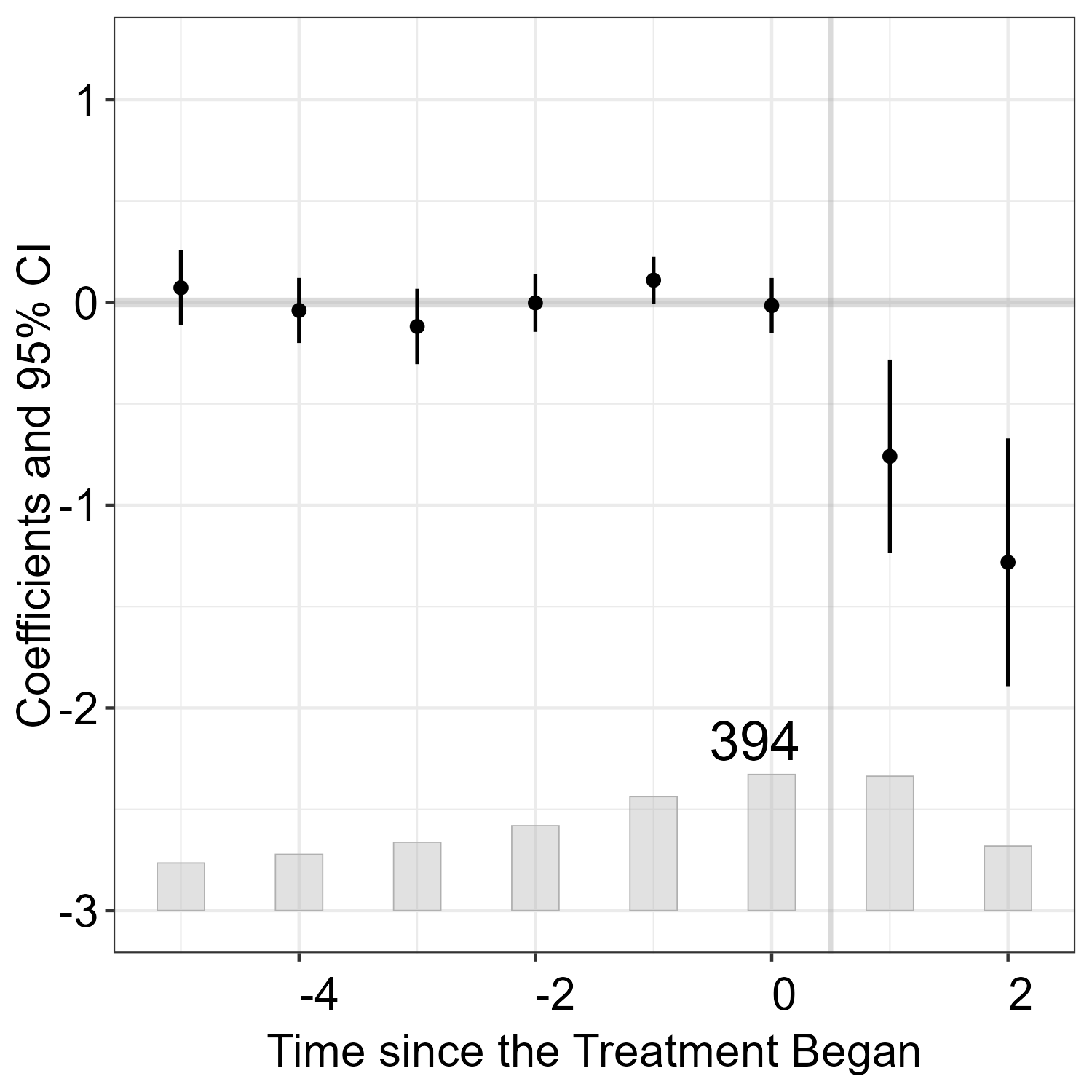}  & 
   \hspace{-2em} \includegraphics[width = 0.22\textwidth]{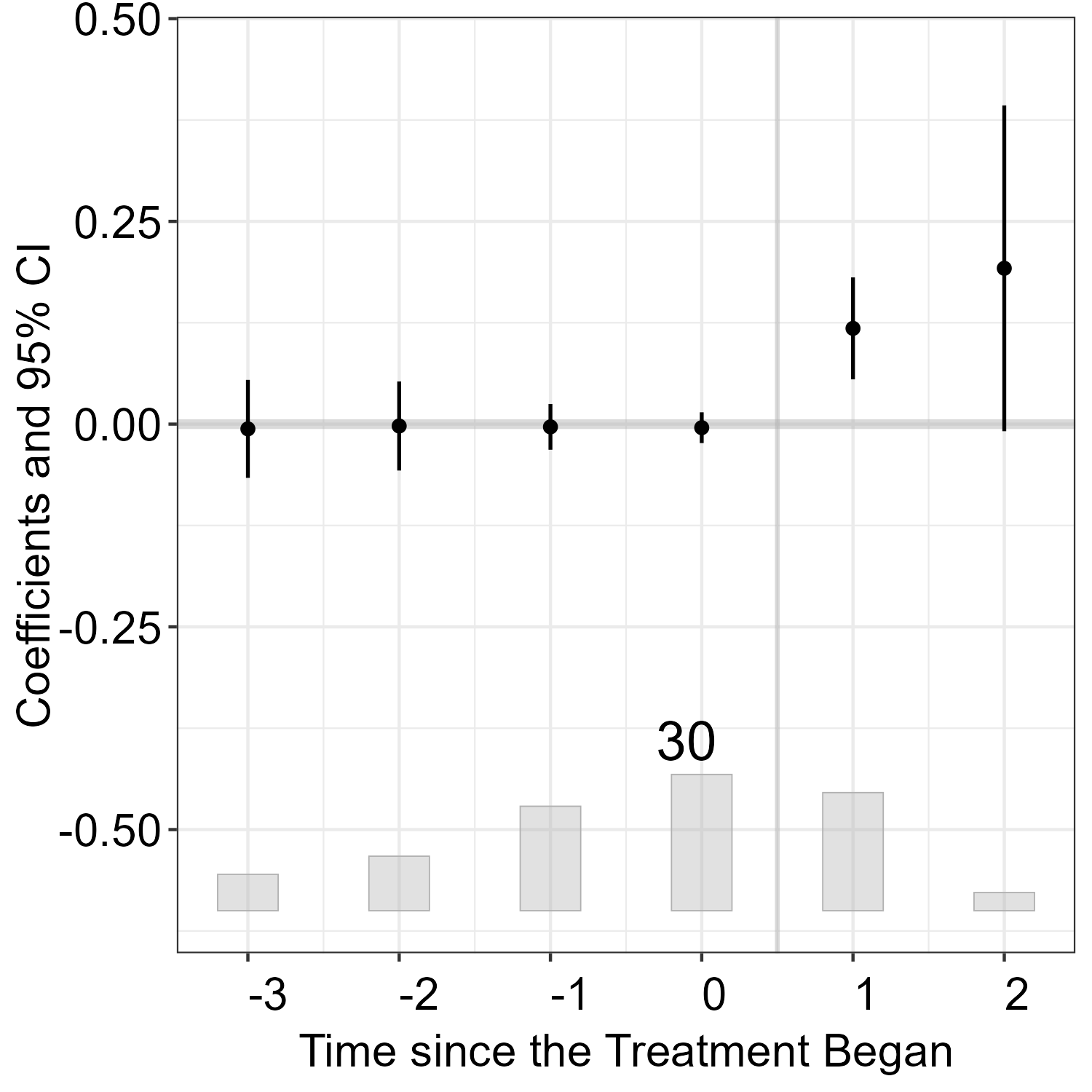}  & 
   \hspace{-2em}  \includegraphics[width = 0.22\textwidth]{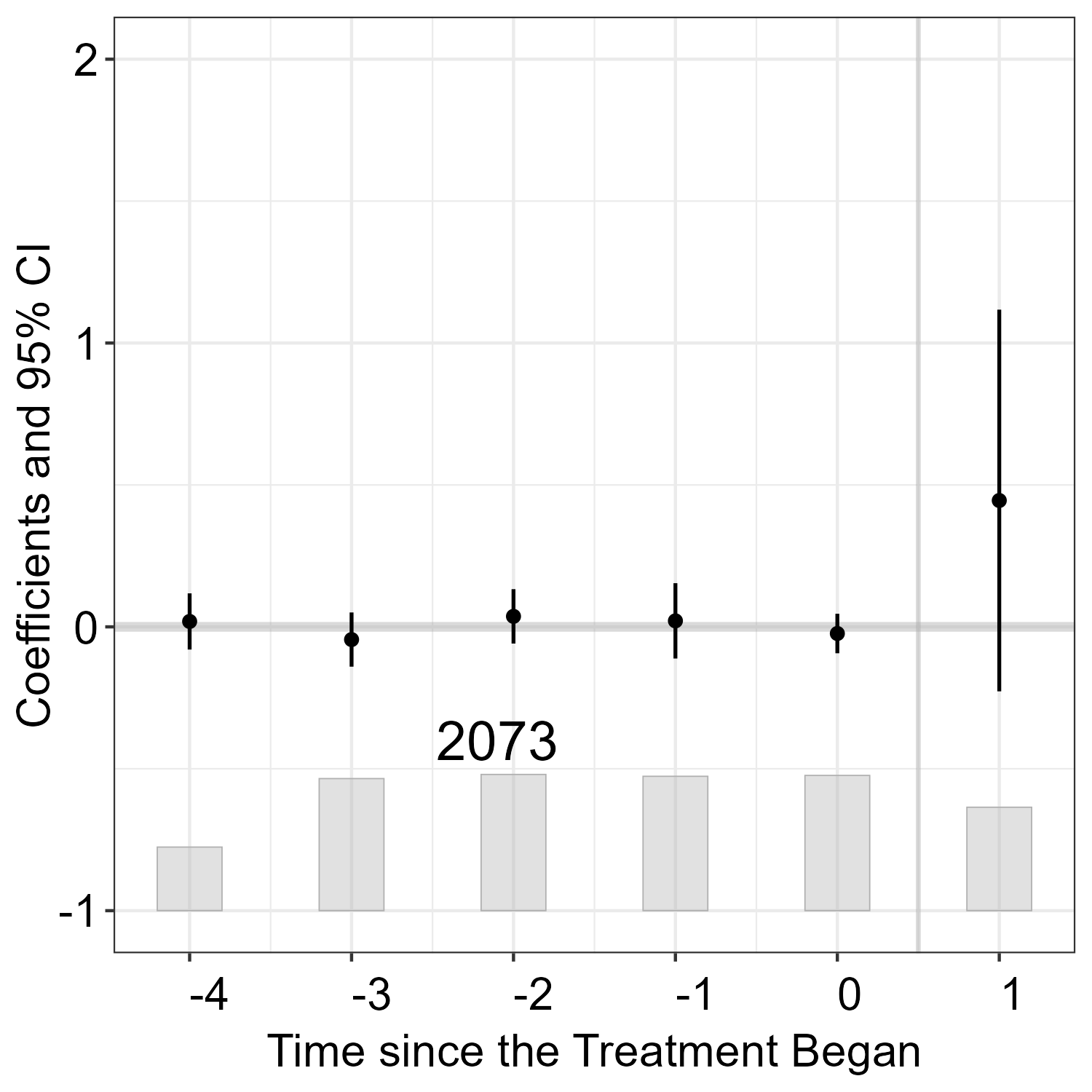} \\ \\ 
   \citet{Kroeger2023} \newline  ATT: 0.03 (0.01)& 
   \citet{Kuipers2023} \newline  ATT: 0.05 (0.02) &  
   \citet{latura2023corporate} \newline  ATT: 0.03 (0.01) &
   \citet{Liao2023} \newline  ATT: -0.045 (0.006)& 
   \citet{magaloni2020killing} \newline  ATT: -2.28 (0.96) \\    
   \hspace{-2em}  \includegraphics[width = 0.22\textwidth]{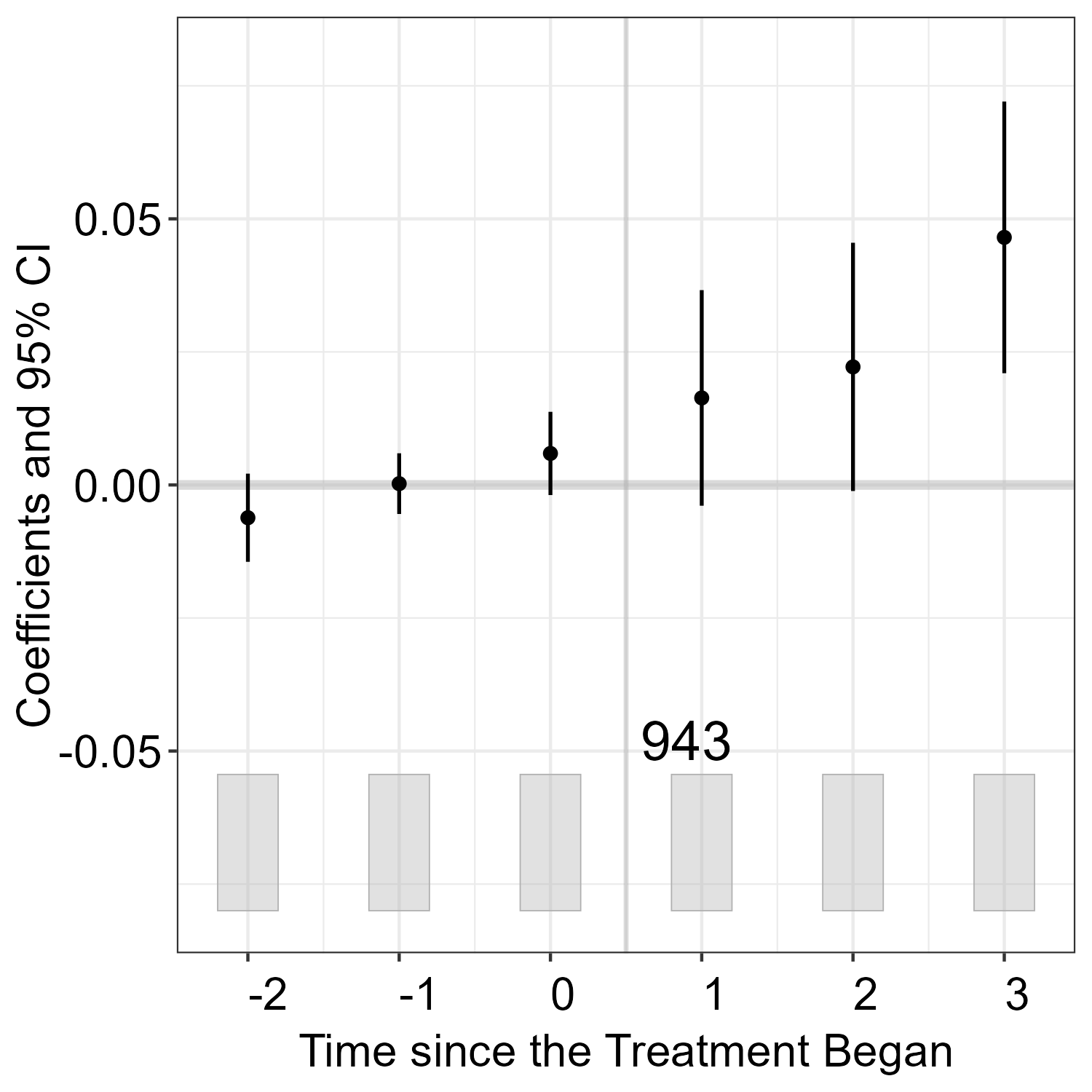} &
   \hspace{-2em} \includegraphics[width = 0.22\textwidth]{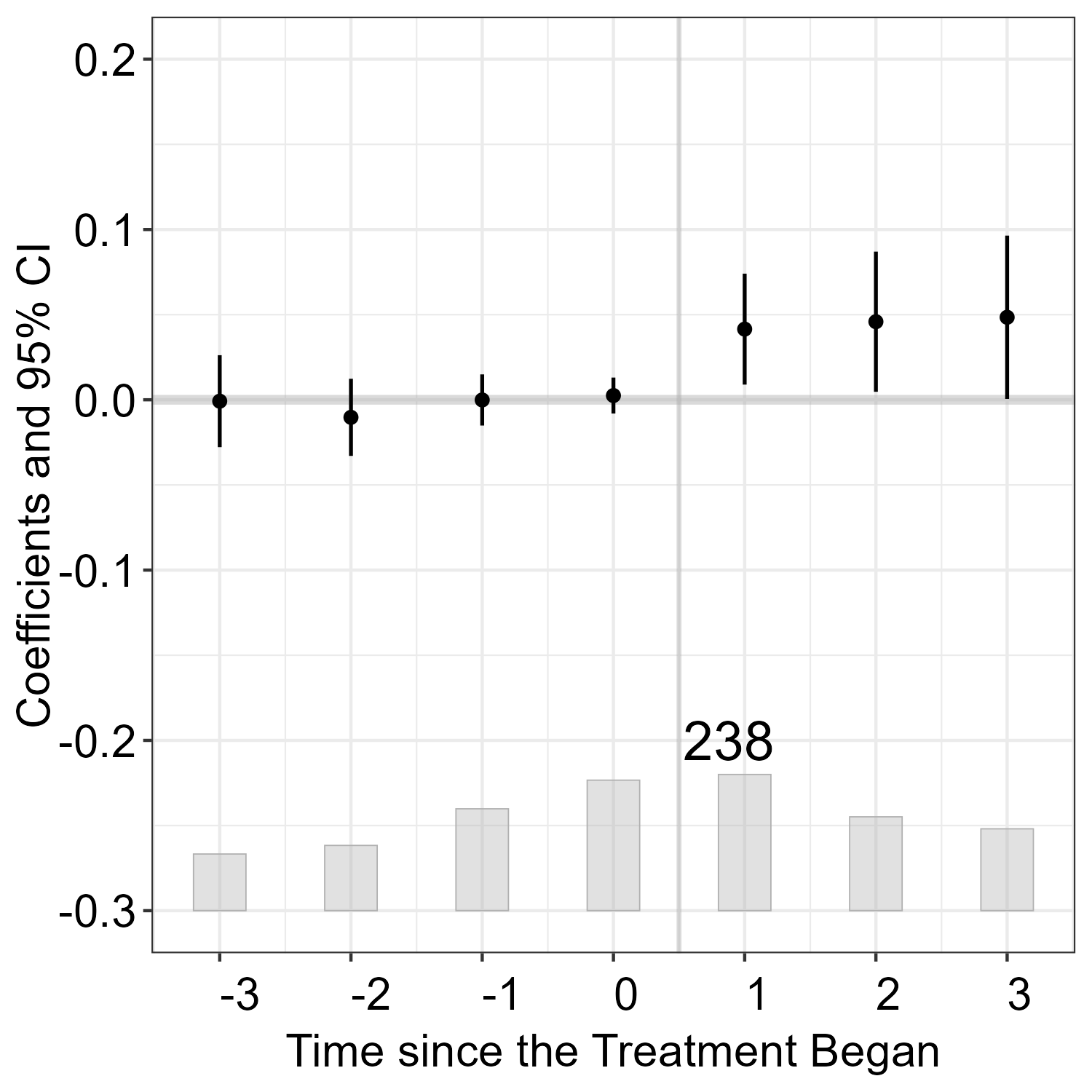} &
   \hspace{-2em}  \includegraphics[width = 0.22\textwidth]{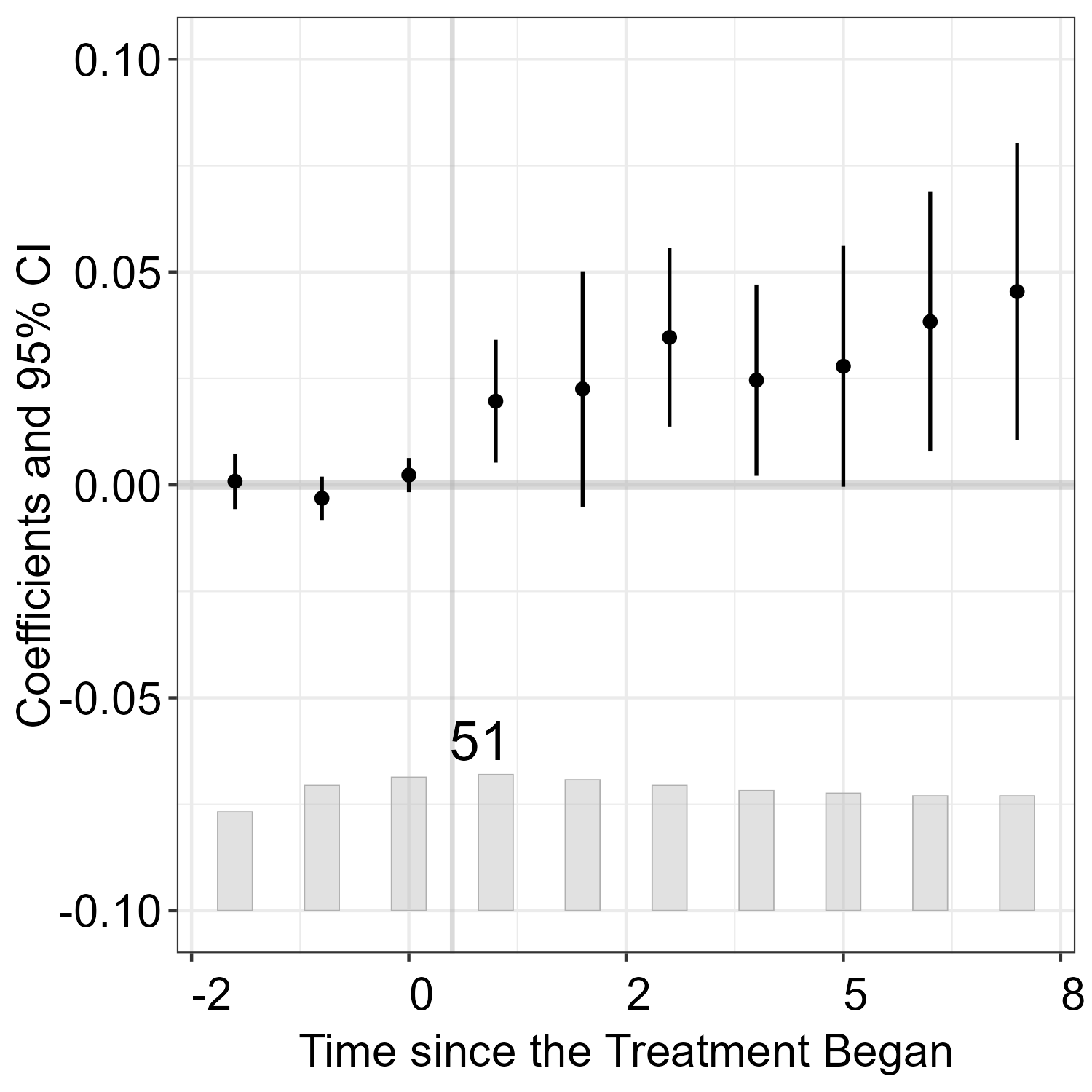} &
   \hspace{-2em}  \includegraphics[width = 0.22\textwidth]{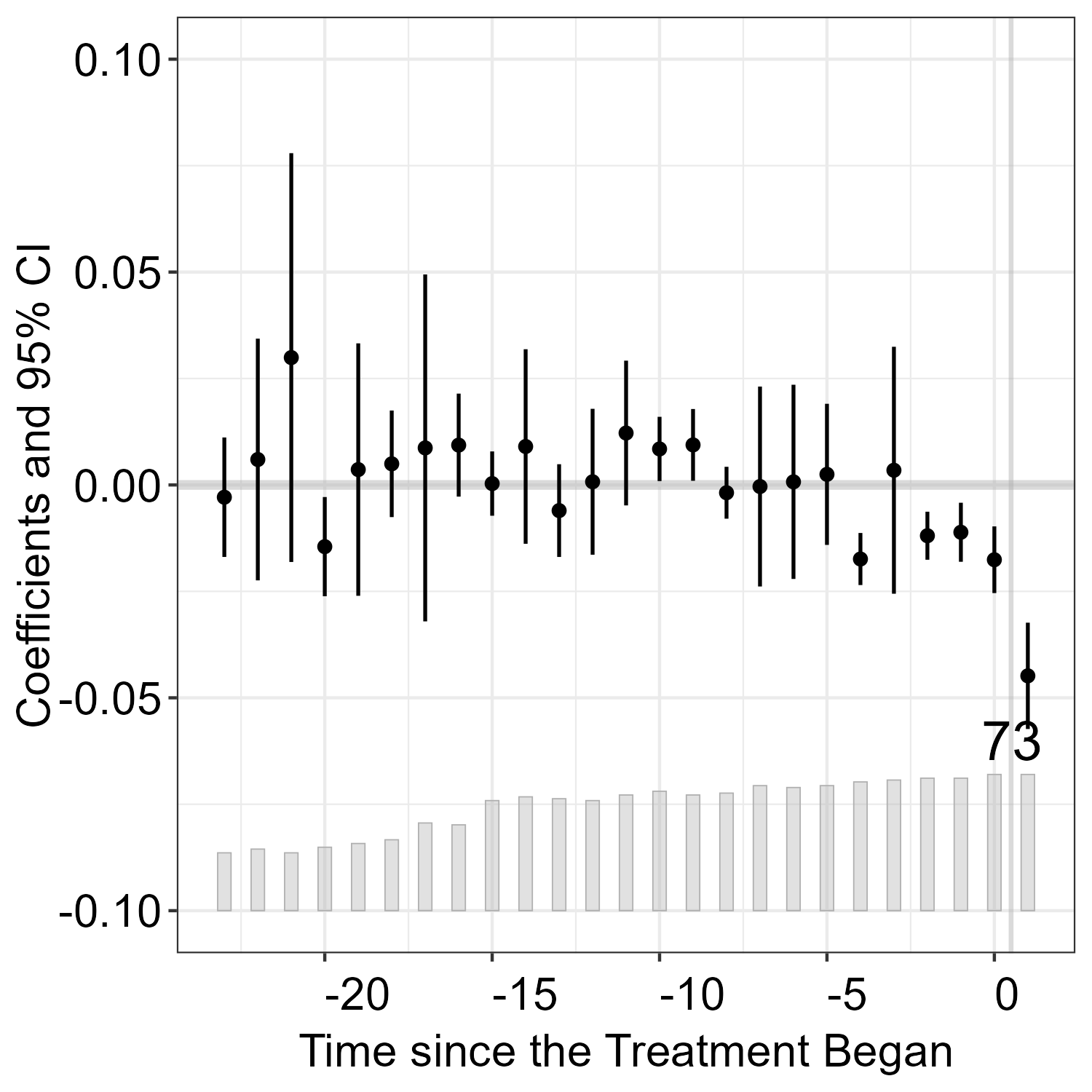} &
   \hspace{-2em}  \includegraphics[width = 0.22\textwidth]{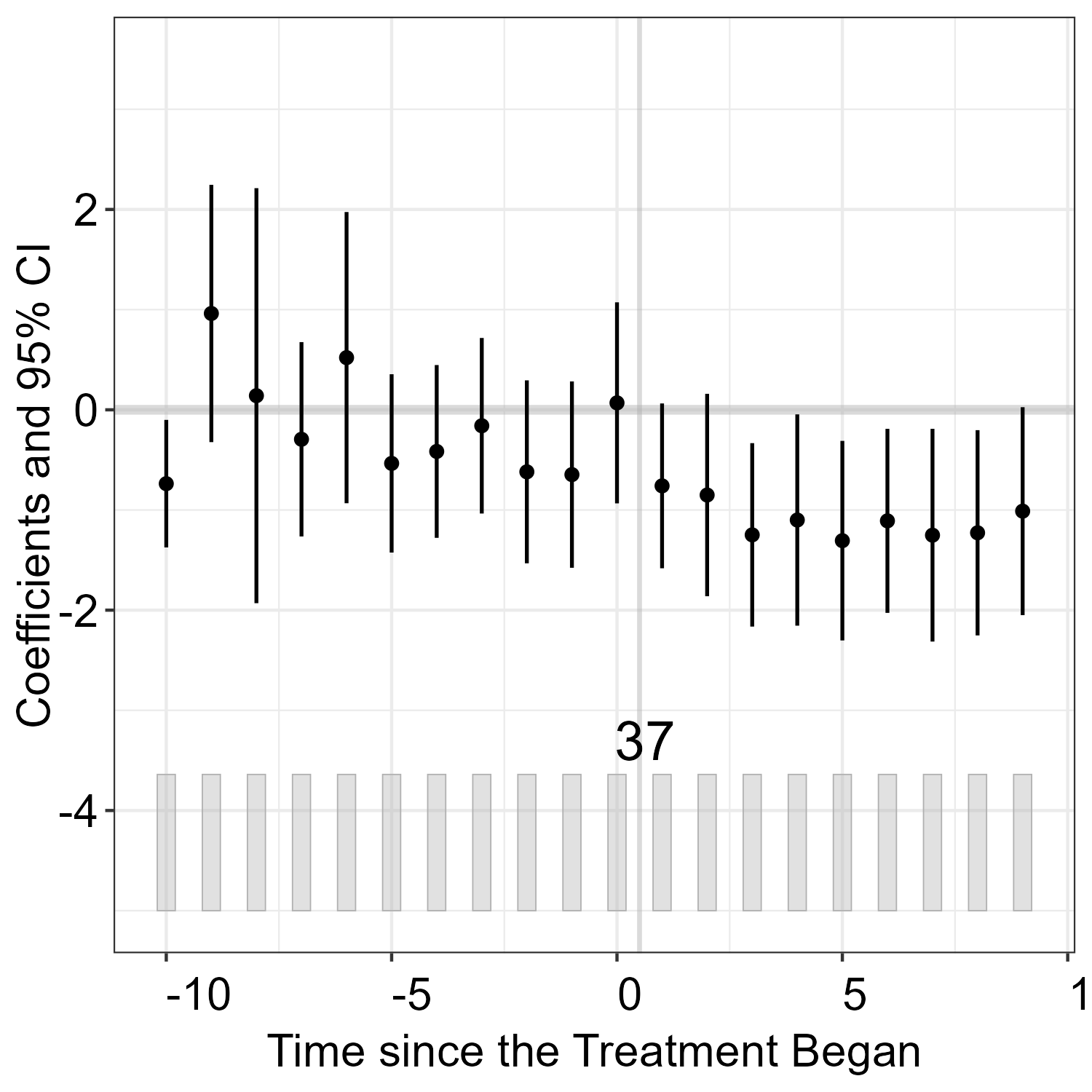} \\ \\  
   \citet{Marsh2023} \newline  ATT: -0.012 (0.003) &
   \citet{Paglayan2022} \newline  ATT: 6.45 (2.03) &
    \citet{Payson2020apsr} \newline  ATT: 0.04 (0.02)&
    \citet{Payson2020jop} \newline  ATT: 13.04 (7.46)& 
     \citet{Pierskalla2018} \newline ATT: -0.07 (0.04)
   \\
    \hspace{-2em} \includegraphics[width = 0.22\textwidth]{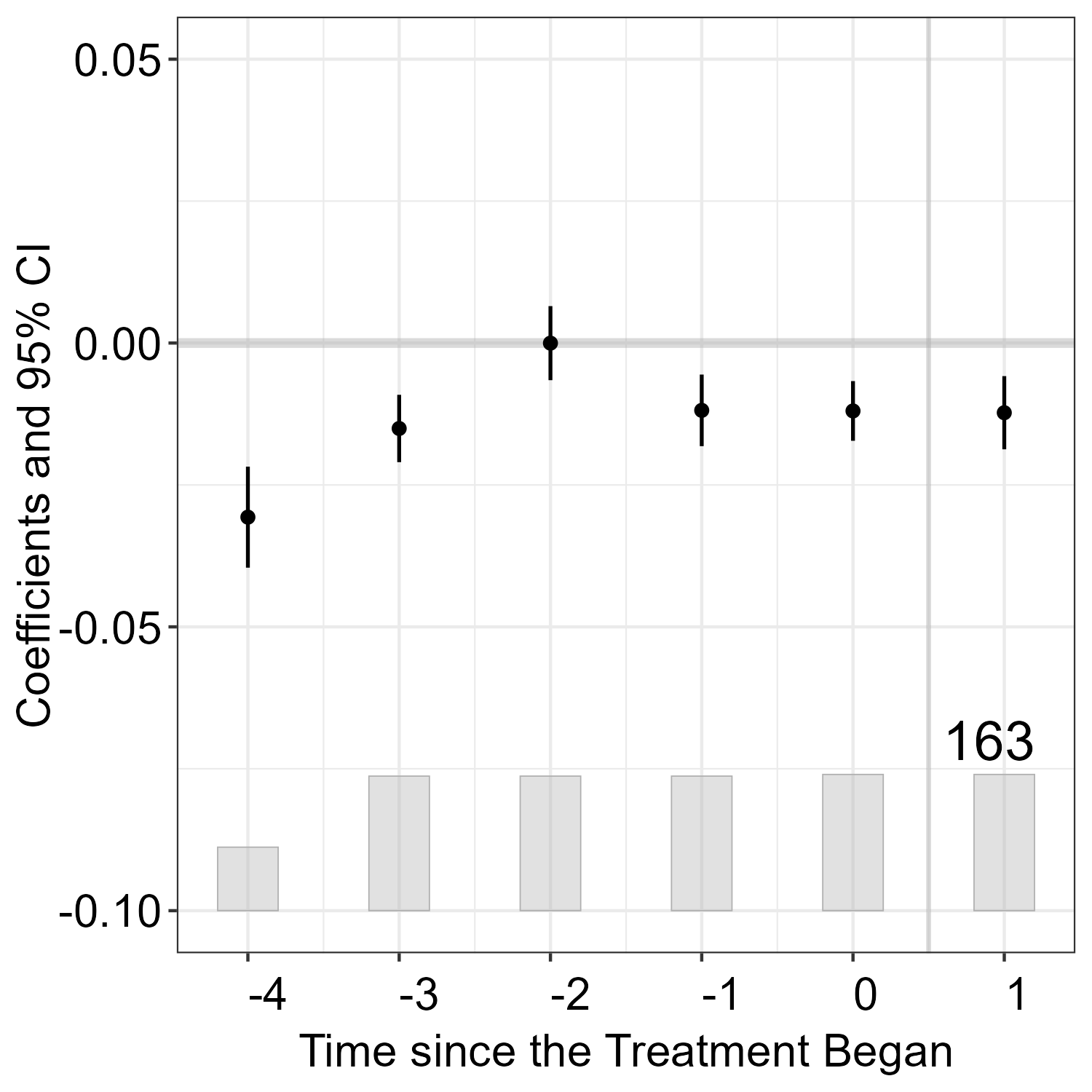} &
   \hspace{-2em}  \includegraphics[width = 0.22\textwidth]{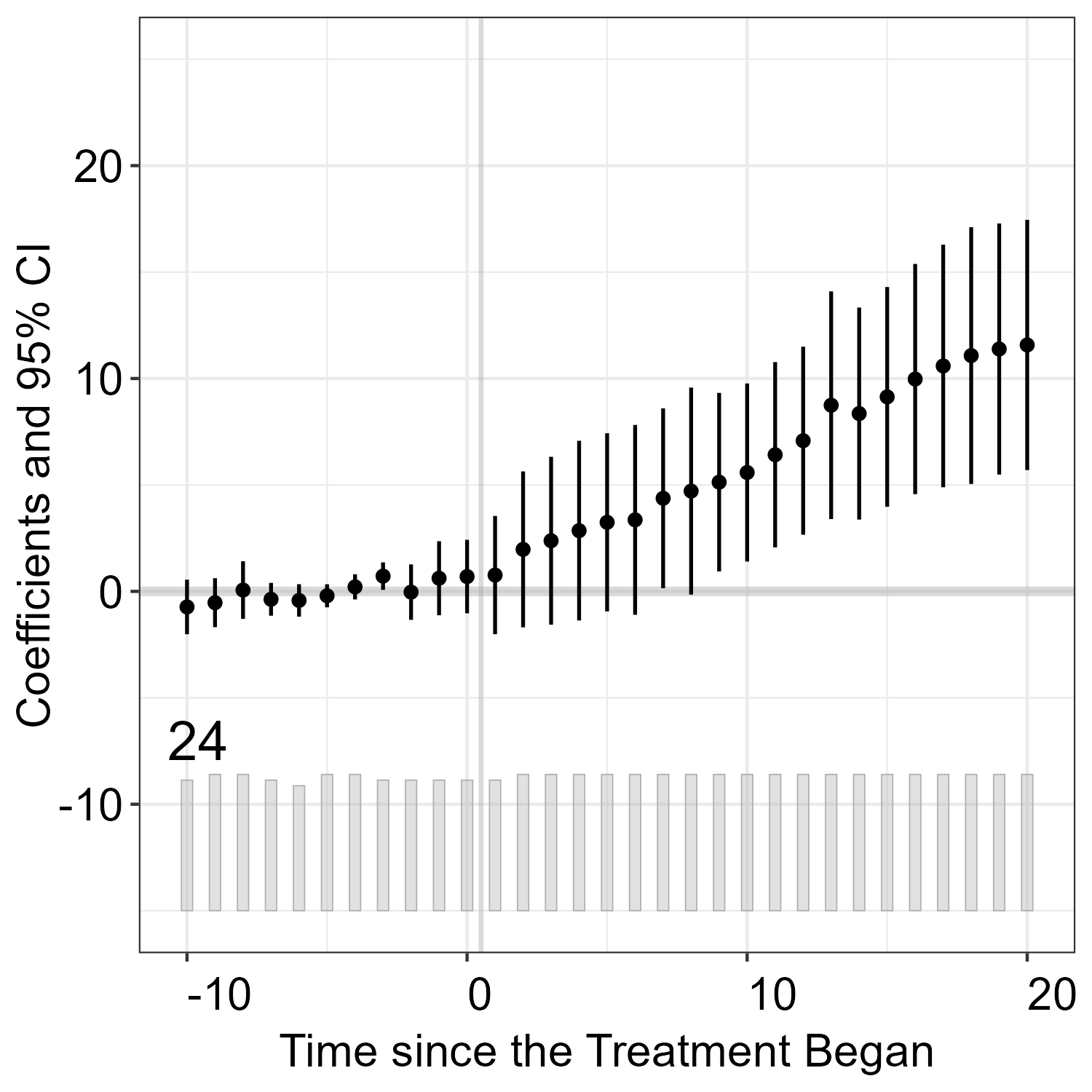} &
      \hspace{-2em} \includegraphics[width = 0.22\textwidth]{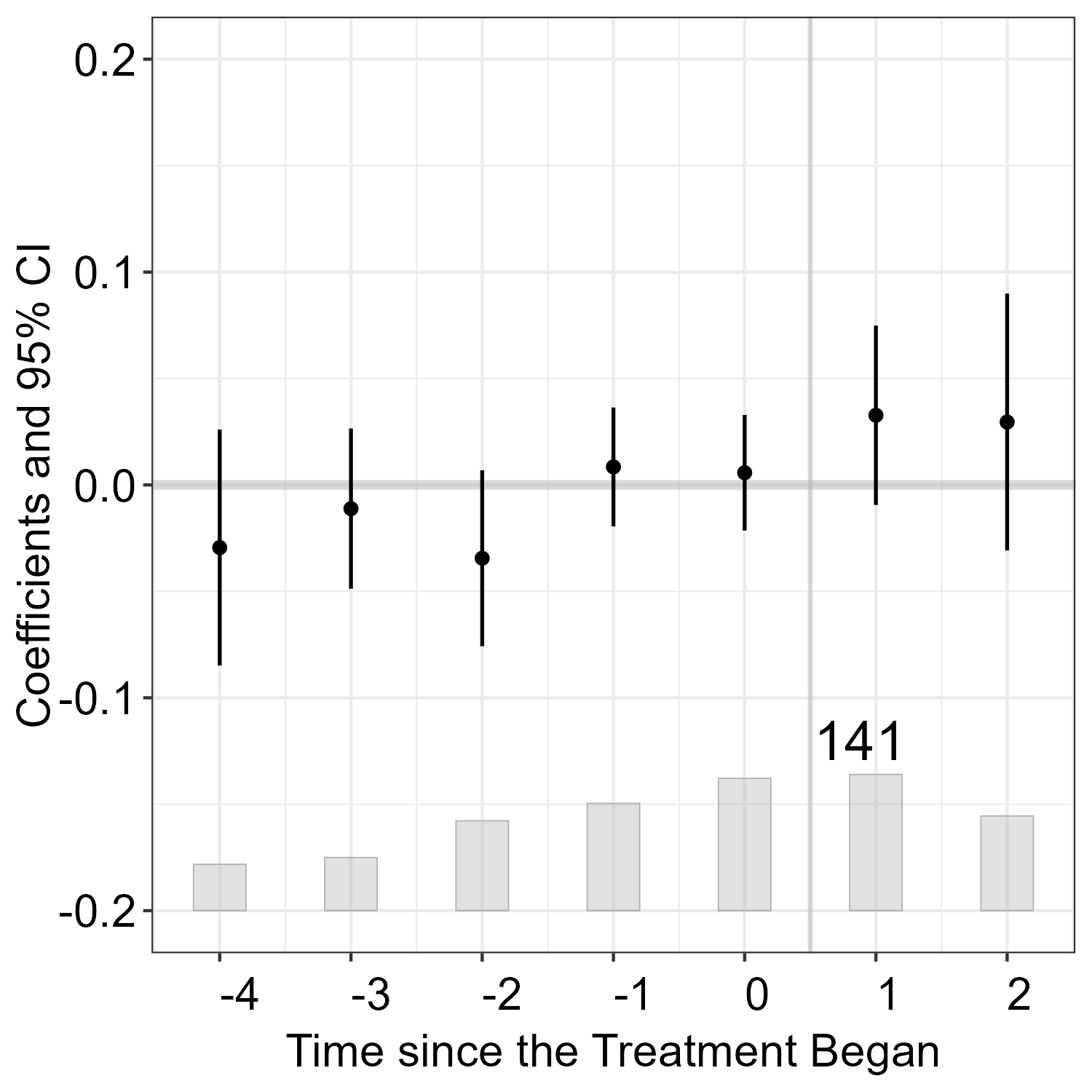} & \hspace{-2em}  \includegraphics[width = 0.22\textwidth]{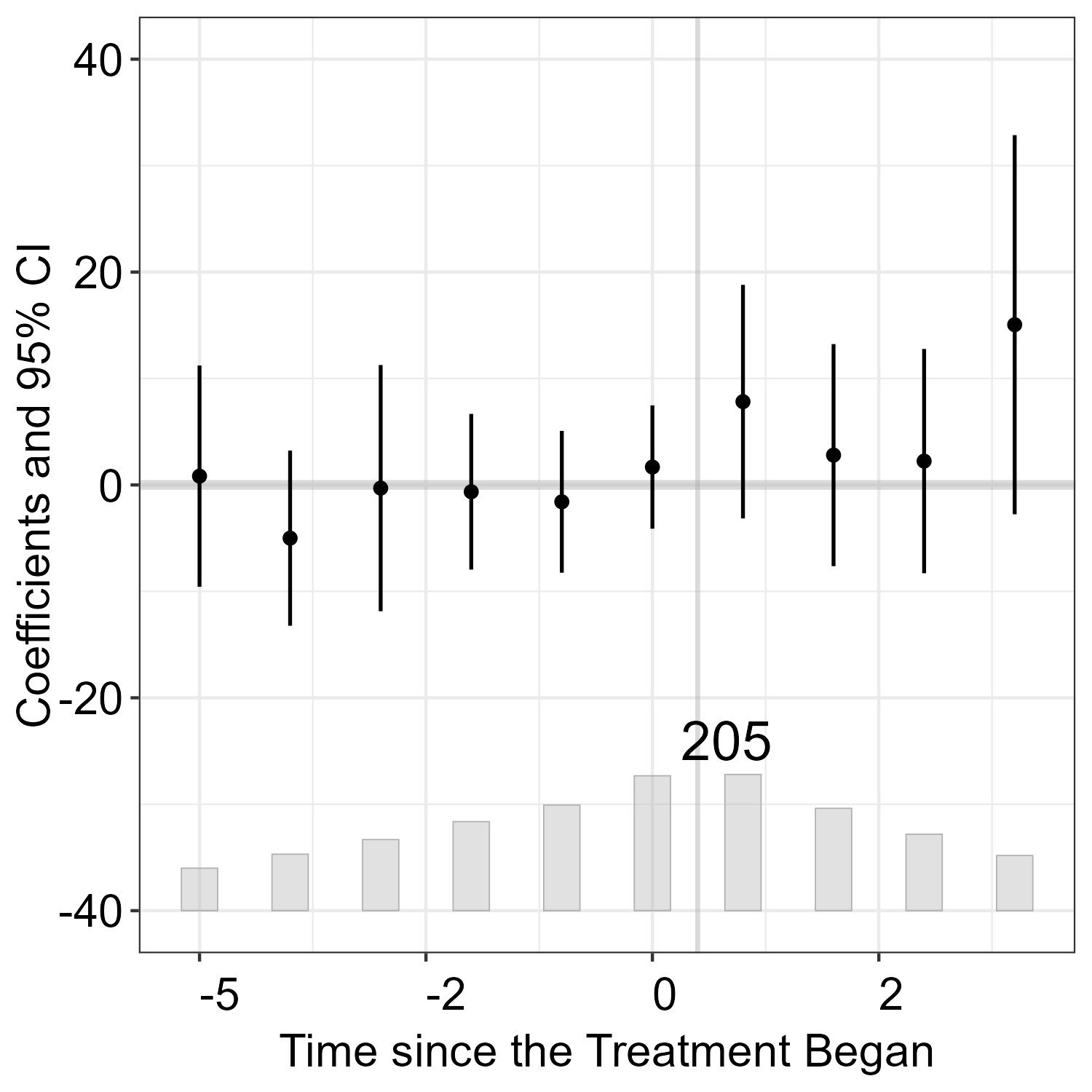} & \hspace{-2em}  \includegraphics[width = 0.22\textwidth]{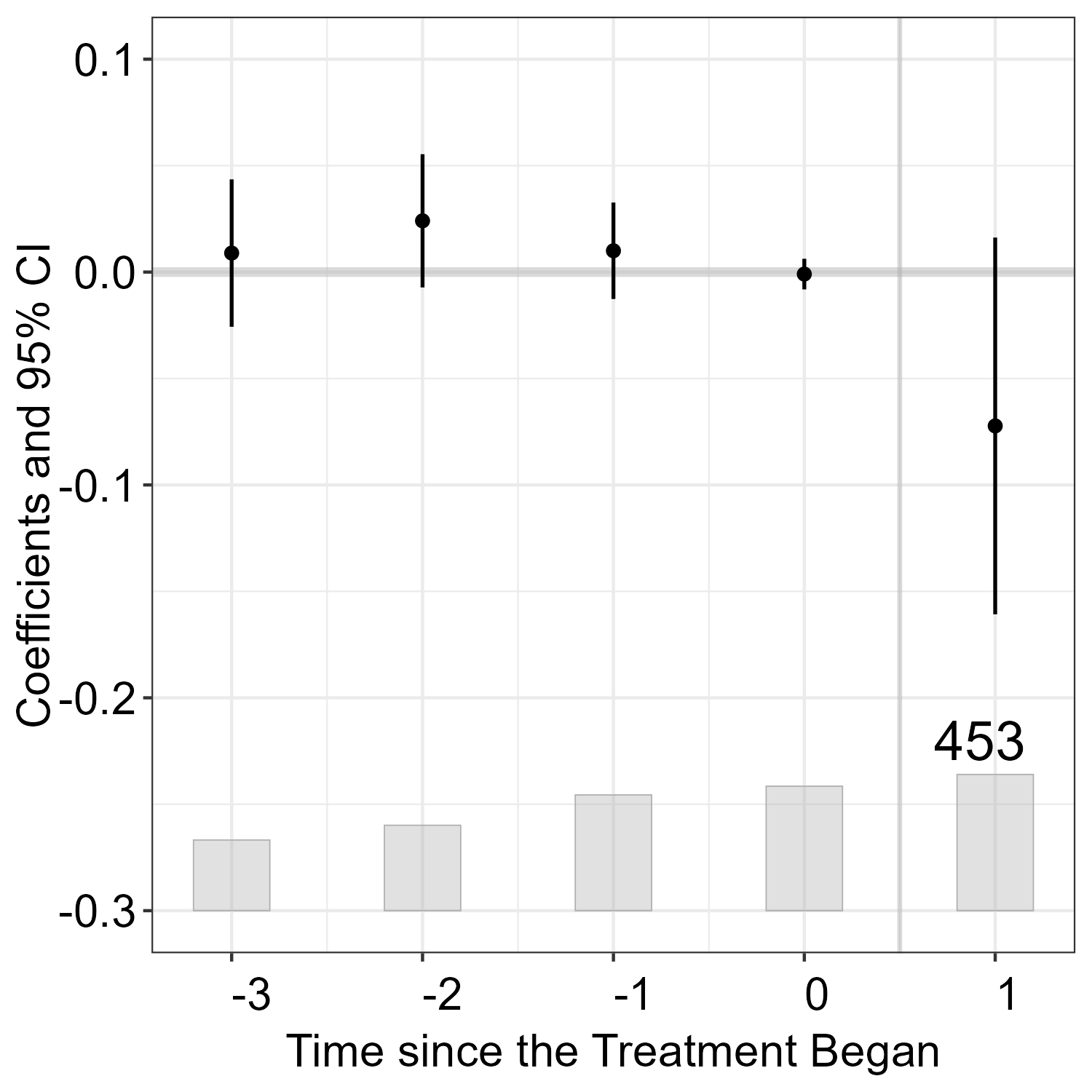} \\ \\
      \citet{Ravanilla2022} \newline ATT:  0.36 (0.16); &   
   \citet{sanford2023democratization} \newline ATT:  -2.16 (0.16) &
      \citet{Schafer2021} \newline  ATT: -0.023 (0.003)  & 
       \citet{Schubiger2021} \newline ATT: 0.047 (0.009)&
       \citet{Schuit2017} \newline  ATT:  -0.16 (0.04) \\ 
   \hspace{-2em}  \includegraphics[width = 0.22\textwidth]{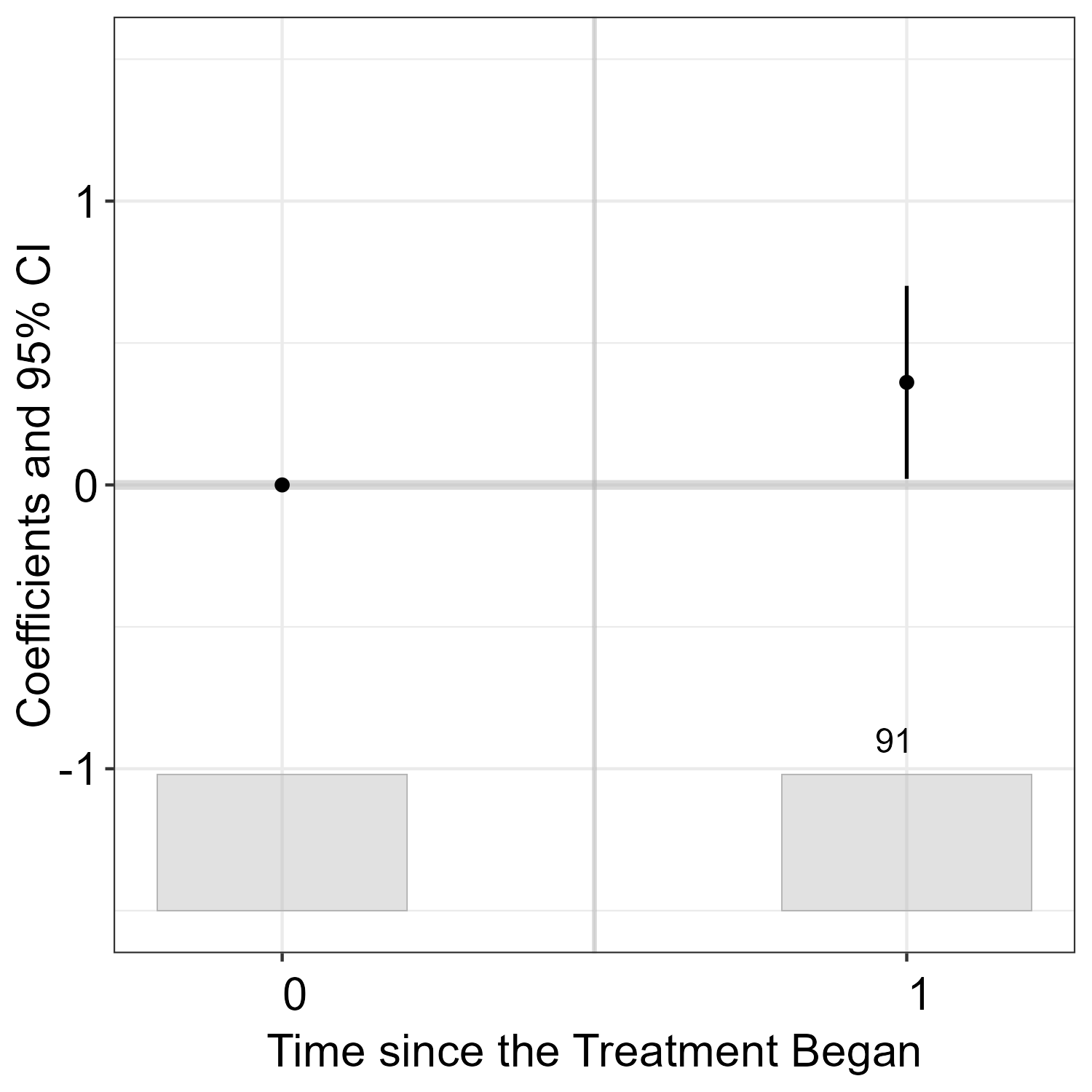} & 
   \hspace{-2em}  \includegraphics[width = 0.22\textwidth]{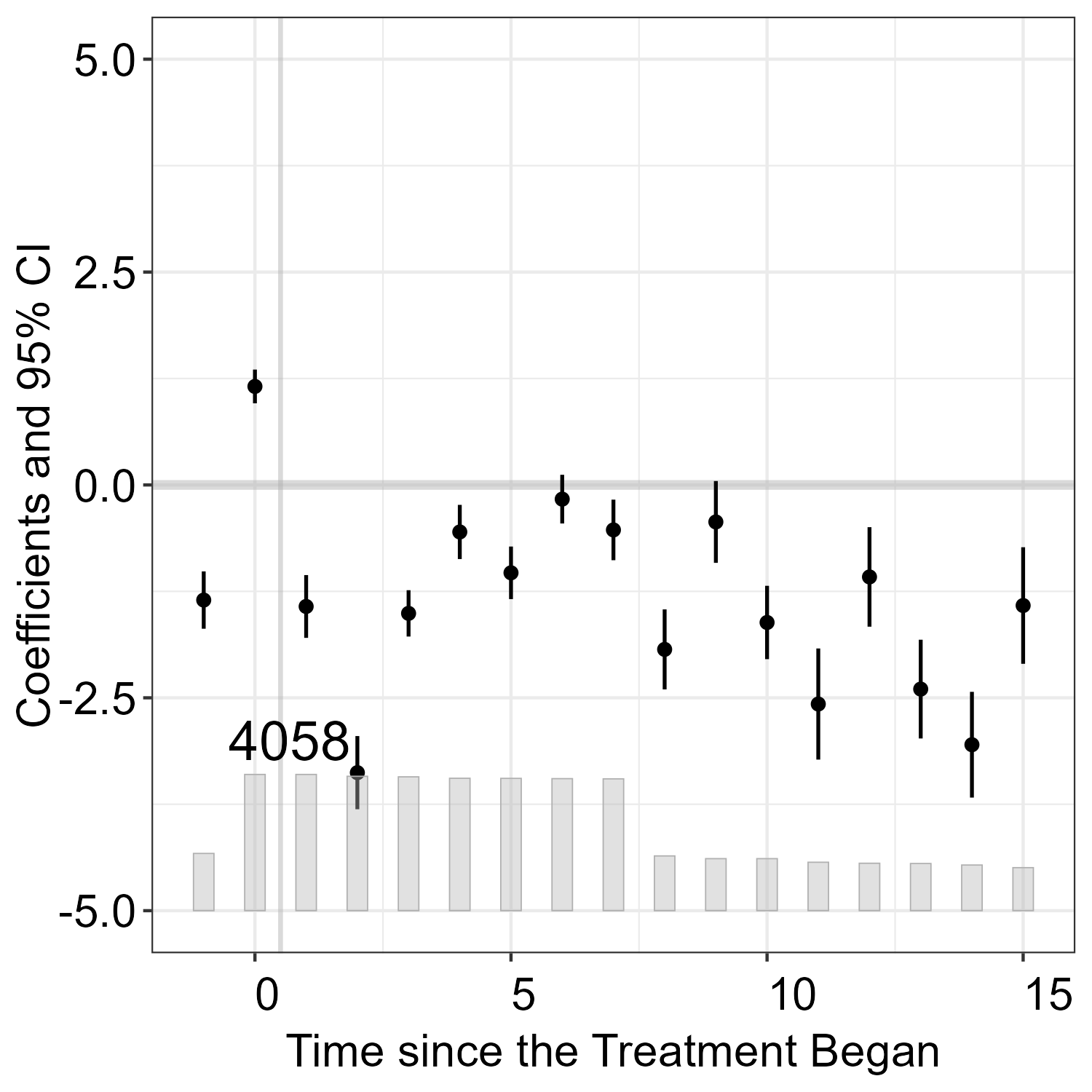} &
   \hspace{-2em}  \includegraphics[width = 0.22\textwidth]{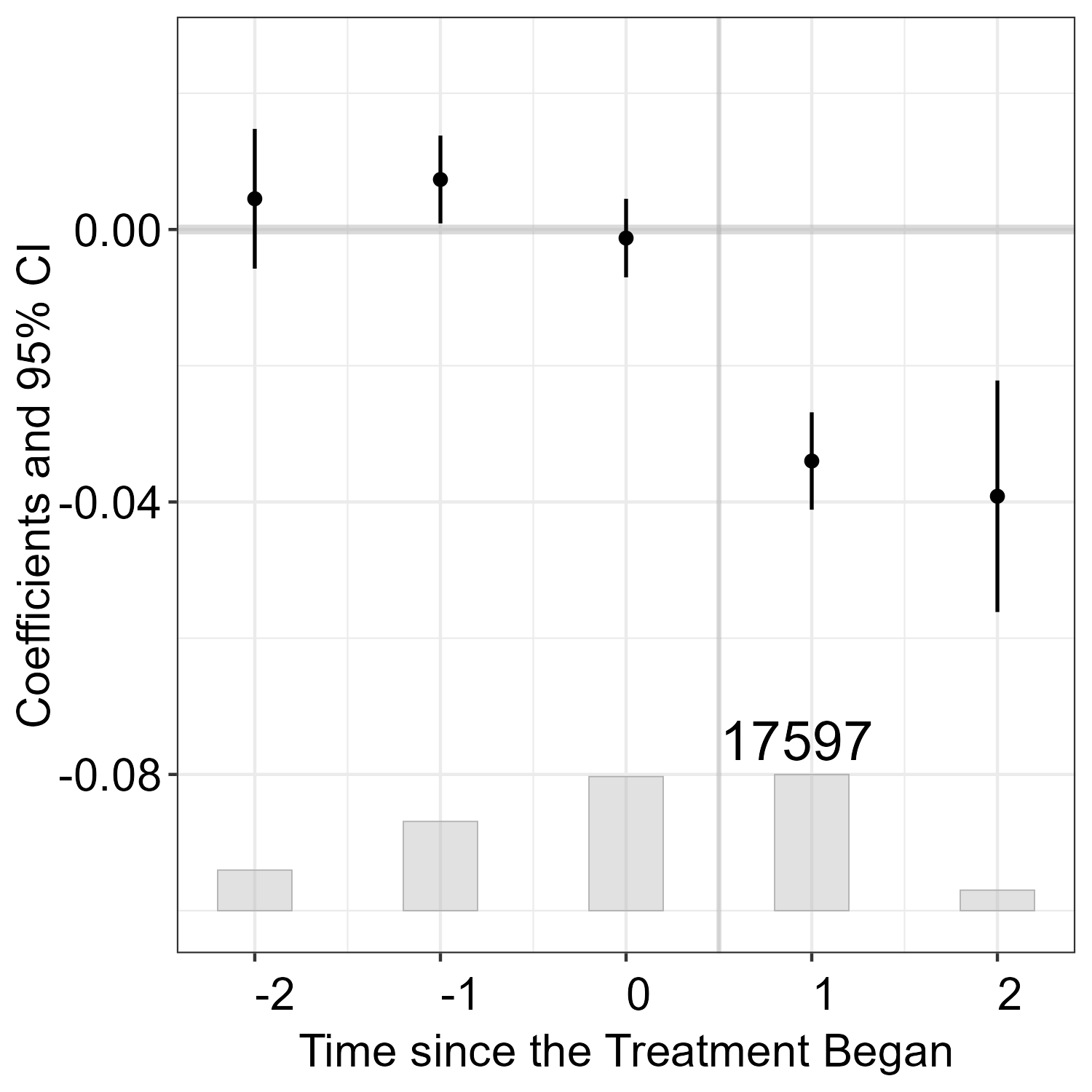} & 
      \hspace{-2em}\includegraphics[width = 0.22\textwidth]{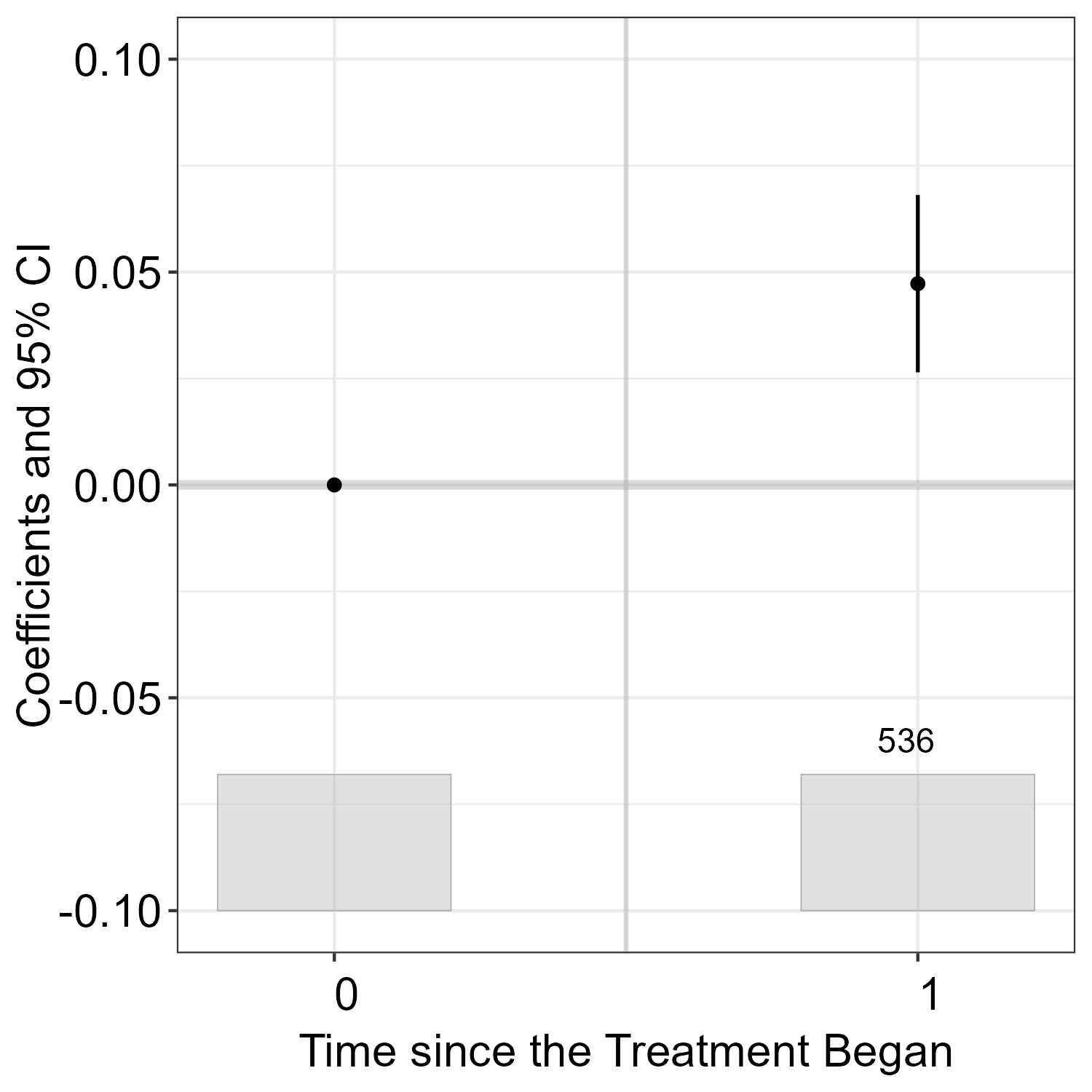} & 
       \hspace{-2em}\includegraphics[width = 0.22\textwidth]{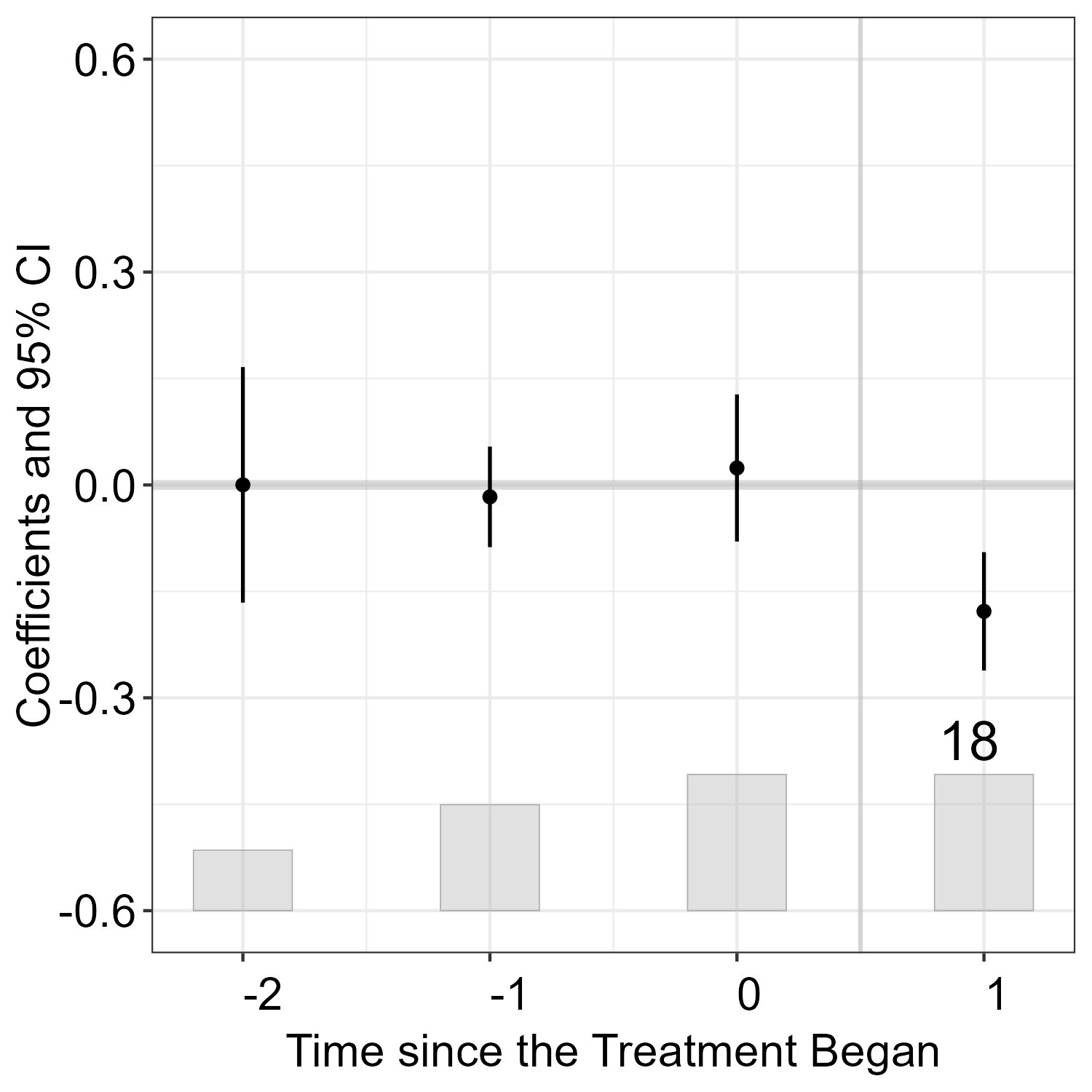} \\ \\  
   \citet{skorge2023mobilizing}  \newline ATT: -0.094 (0.007) &  
   \citet{Trounstine2020} \newline ATT: -0.06 (0.01)&   
   \citet{Weschle2021} \newline ATT: 1.11 (0.77)&
   \citet{Zhang2021jop} \newline ATT: 0.11 (0.06) & \\    
   \hspace{-2em} \includegraphics[width = 0.22\textwidth]{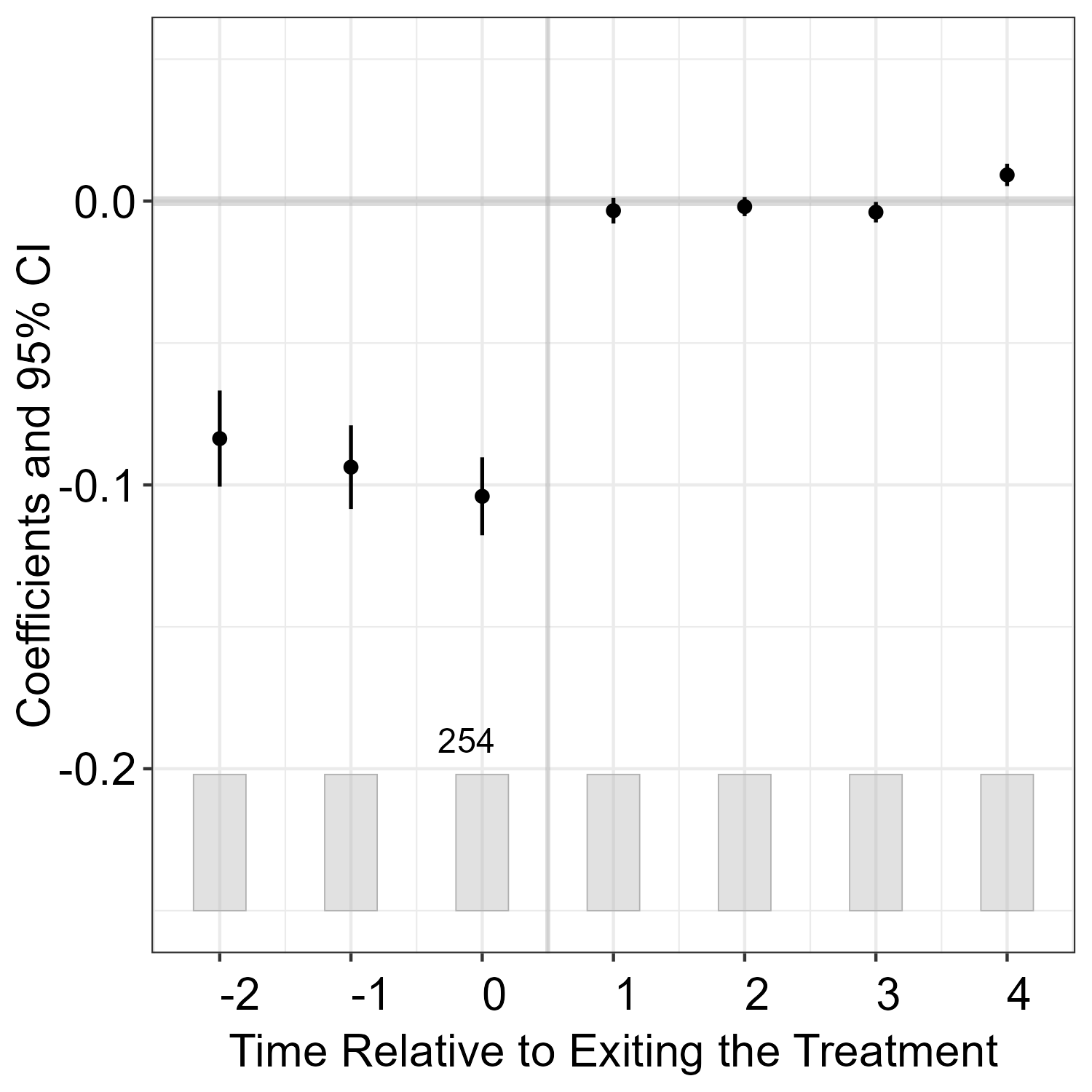} &
   \hspace{-2em} \includegraphics[width = 0.22\textwidth]{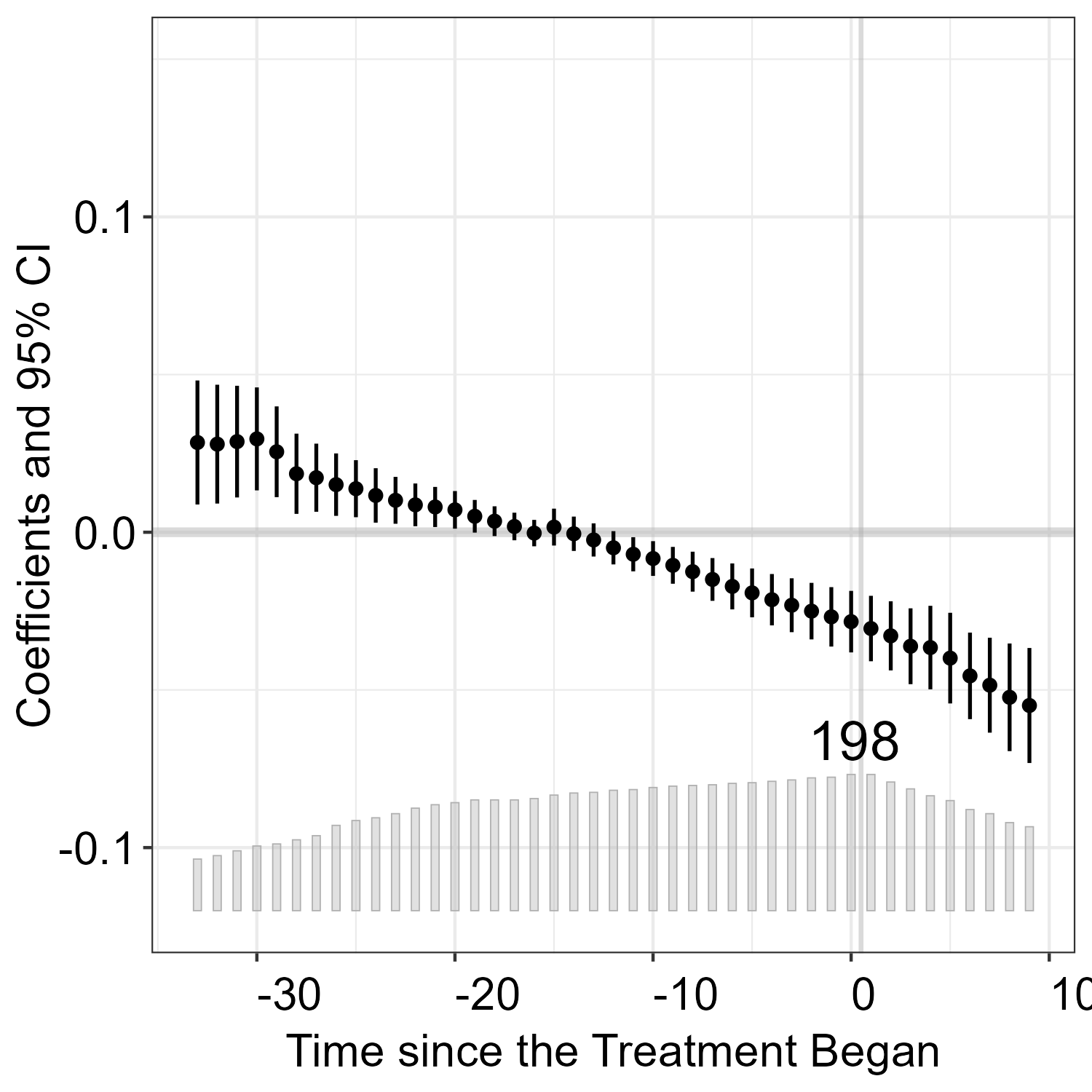} &
      \hspace{-2em} \includegraphics[width = 0.22\textwidth]{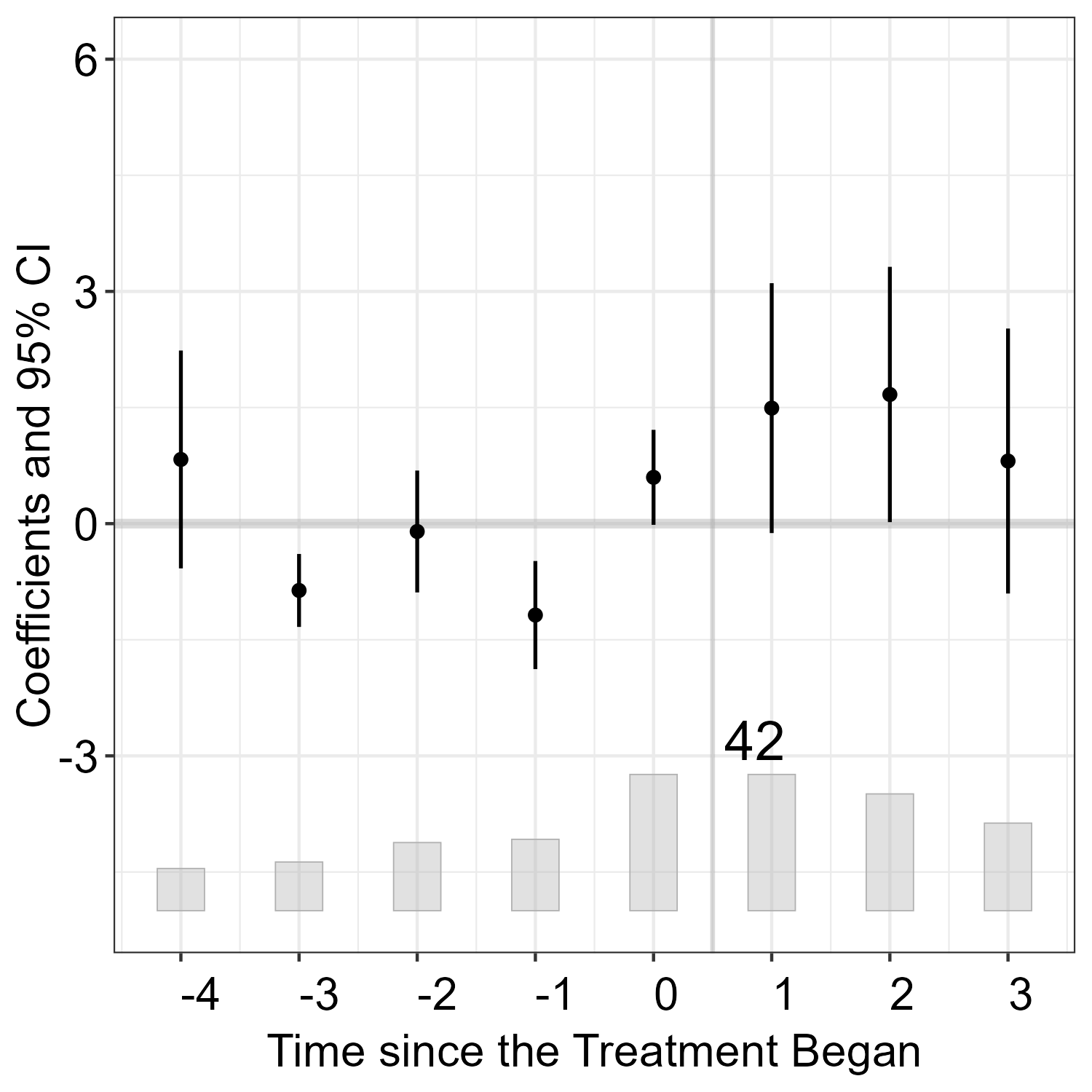} &
      \hspace{-2em} \includegraphics[width = 0.22\textwidth]{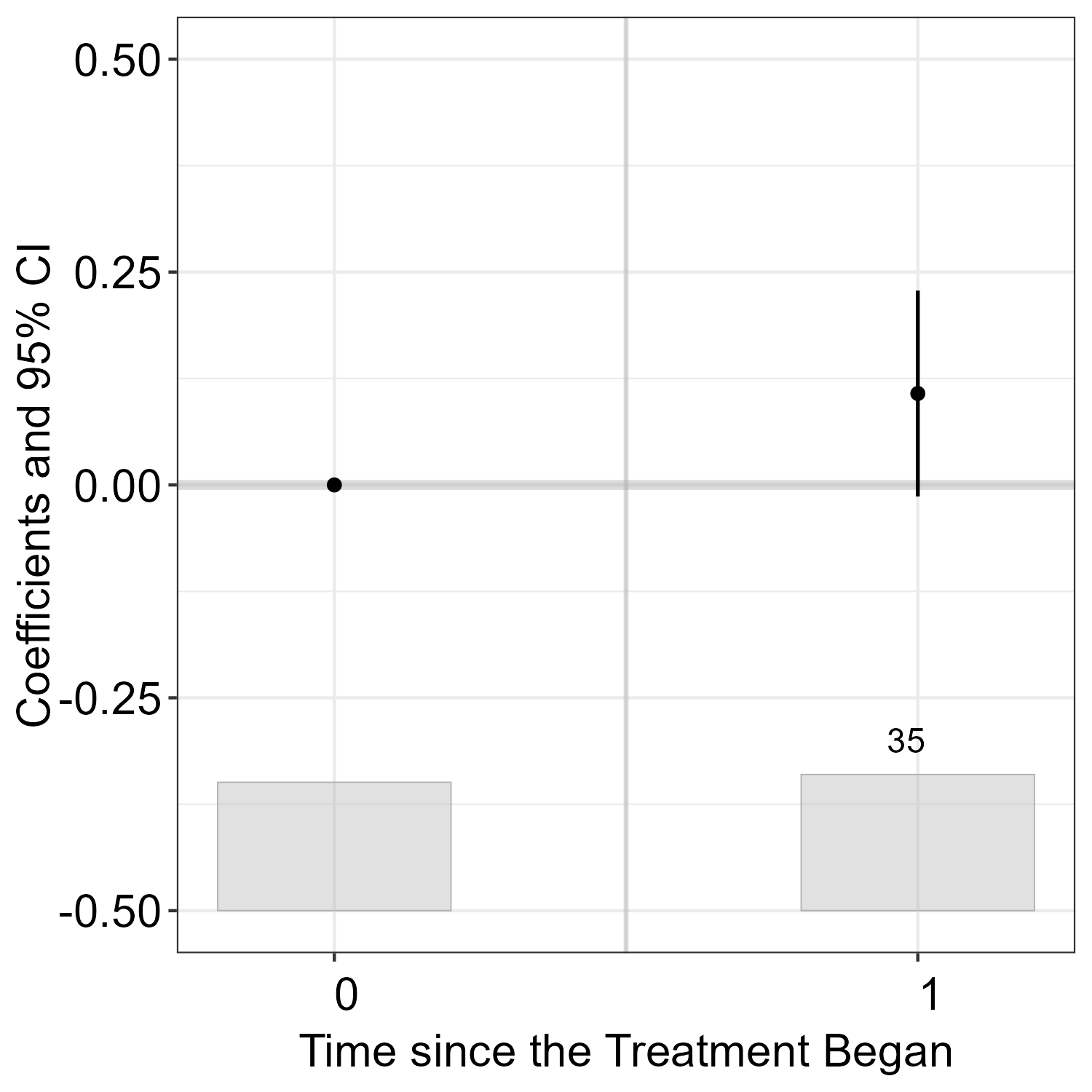}  &
     \\ \\
    \multicolumn{5}{p{20cm}}{\footnotesize\textbf{Note:} We report the estimated ATT and corresponding bootstrap SEs (in parentheses) using \texttt{FEct}. For \citet{skorge2023mobilizing}, we use an ``exit'' plot becuase all treated units receives the treatment in the first period.} 
\end{tabular}}}
\end{minipage}\vspace{1em}
\end{figure}


\paragraph*{Relaxing PT renders most studies unable to reject the null.} Although the recent methodological literature has heavily focused on HTE, PT violations---long recognized as a potential pitfall---remain a primary concern in practice. In Figure~\ref{fg:dynamic}, we present the event-study plot based on estimates from the imputation estimator for each study in our sample. We also report the ATT estimates and their bootstrapped SEs. Due to space limitations, we present the event-study plots from other estimators, as well as results from the placebo tests, robust CSs, and sensitivity analyses, in the SM. 

What is concerning is that, in at least six studies (12\%), the PT assumption seems highly implausible. In these studies, the dynamic estimates in the pre-treatment periods deviate substantially from zero  compared with those in the post-treatment period, and the $F$ test rejects the null.\footnote{There are other cases where the $F$ test rejects, but we do not consider them highly problematic because, with a large sample size, a confounder inconsequential to the ATT estimate can still produce a small $p$-value in the $F$ test. This is why the sensitivity analysis approach is particularly useful.} For the remaining studies, the CIs of pre-treatment estimates mostly cover zero, and the $F$ test and placebo test do not reject. However, this could be simply due to a lack of statistical power. Therefore, to assess the robustness of the findings, we need additional tools that simultaneously account for both the estimated pretrend and statistical power. The sensitivity analysis proposed by \citet{rambachan2023more} addresses this issue. Therefore, we conduct such an analysis with a modified relative magnitude Equation~(\ref{eq:rm}) for 42 studies that have at least three pre-treatment periods.\footnote{For studies with only three pre-treatment periods, we set the number of placebo periods in their placebo tests to 2.}

\FloatBarrier

\begin{figure}[!ht]
  \caption{Allowing PT violations with Robust Confidence Sets}\label{fg:power}
  \centering
  \vspace{-0.5em}
  \begin{minipage}{1\linewidth}{
  \begin{center}  
  \subfigure[Robust infernece under PT violations]{\includegraphics[width = 0.45\textwidth]{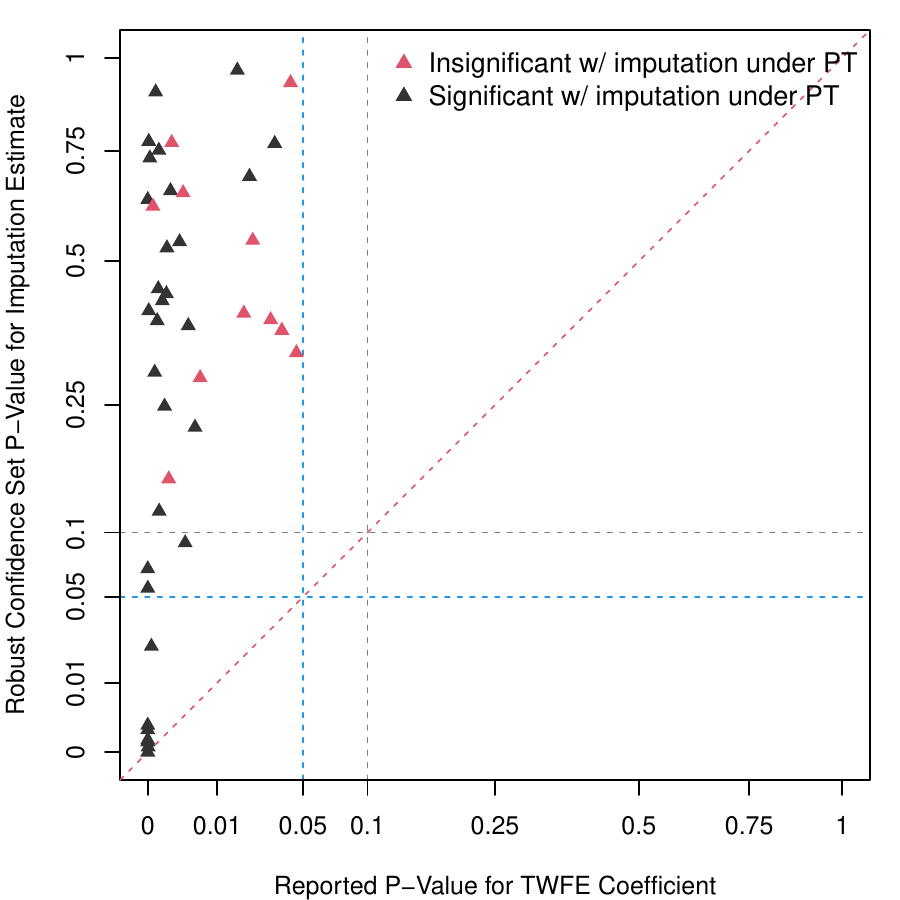}}
  \vspace{-1em}
  \hspace{1em}
  \subfigure[Distribution of breakdown threhsold $\tilde{M}$]{\includegraphics[width = 0.45\textwidth]{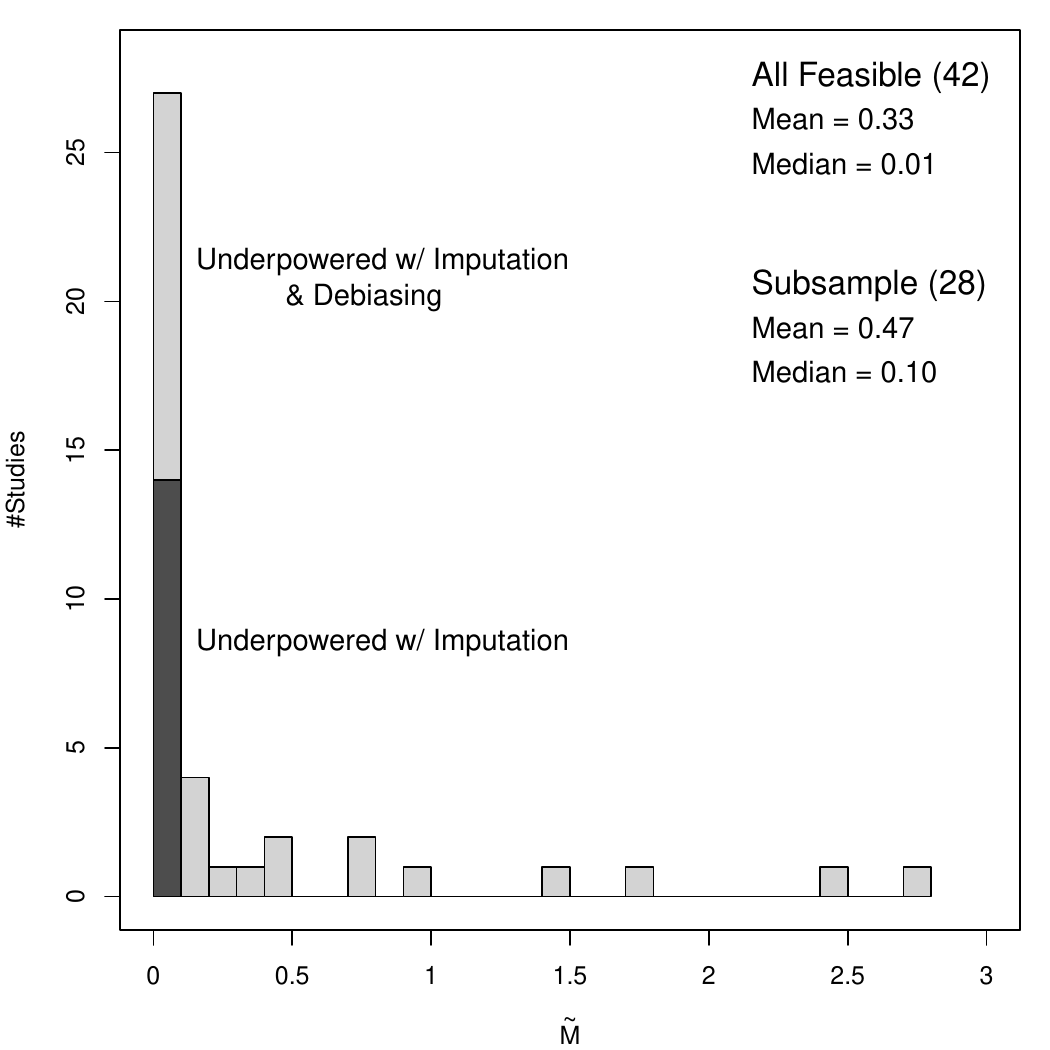}}
  \end{center}
  {\footnotesize\textbf{Note:} The above figures present findings from the sensitivity analysis. The sample consists of 42 studies with more than three pre-treatment periods, allowing for such an analysis. Subfigure (a) displays the $p$-values of partially identified ATT estimates using the imputation method under restricted relative magnitude PT violations with $\bar{M} = 0.5$, compared to reported $p$-values for TWFE coefficients assuming PT. A square root scale is used to facilitate visualization. Black (red) triangles represent studies that are statistically significant (insignificant) at the 5\% level when using the imputation method under PT. Subfigure (b) shows a histogram of $\tilde{M}$, the breakdown values of $\bar{M}$; the dark gray bar represents studies whose ATT estimates are statistically insignificant at the 5\% level when using the imputation method.}}
  \end{minipage}\vspace{-0.5em}
\end{figure}

Figure~\ref{fg:power}(a) shows that when we construct a robust confidence set with \(\bar{M} = 0.5\), the null hypothesis of no effect is rejected at the 5\% level in only eight (19\%) of the 42 studies. Many studies that were robust to HTE now appear underpowered. Figure~\ref{fg:power}(b) displays the distribution of breakdown values \(\tilde{M}\). The spike at \(0\) consists of two types of studies: those that are statistically insignificant with the imputation estimator when three placebo periods are considered (dark gray) and those that become statistically insignificant due to debiasing using the placebo estimate \(\hat\delta_{0}\). In other words, many results are not robust when we take into account \(\hat\delta_{0}\) even without a relative magnitude shift. Among the 42 studies, the median is close to 0 and the mean is 0.33. Focusing only on the studies that remain statistically significant with the imputation estimator, the median and mean are still as low as 0.10 and 0.47, respectively.

These results suggest that although most statistically significant findings are robust to the imputation estimator, in the vast majority of these studies, accounting for potential PT violations—even very mild ones based on estimated pretrends during the placebo periods—prevents us from rejecting the null. In other words, in most studies, the ATT estimates are not substantial enough to be differentiated from realistic PT violations or from estimation and sampling errors.



\paragraph*{Other issues.} Our reanalysis highlights several additional issues. First, the presence of missing values is widespread. Although most methodological work presumes balanced panels without missing data, in reality, many empirical studies encounter varying degrees of data missingness. Substantial differences in results for estimators that are numerically identical in balanced panels suggest that such violations may have important implications. During replication, we generate plots that display the patterns of treatment status for each study \citep{mou2023panelview}. Based on these plots, we also see that in some studies the pattern of missingness is either seemingly nonrandom or extremely prevalent, which weakens our confidence in the respective empirical findings. 

Second, we perform tests for carryover effects for all studies with treatment reversals. If this test is rejected, it suggests that the treatment effects persist beyond the treatment periods. Among 27 studies, five reject the null hypothesis of no carryover effects at 5\%. Part of the concern is that the imputation method and DID extensions will use control observations from previously treated units to fit the potential outcome model or as comparisons for treated observations, and if there are carryover effects, then the comparisons will become tainted. LWX \citeyearpar{liu2024practical} note that the presence of carryover effects for a limited number of periods is less concerning, as researchers can recode treatment to persist for some time after a unit transitions out of treatment. Despite its prevalence, we observe that carryover effects do not substantially alter the findings in most studies. Specifically, in six studies that reject the null of no carryover effects, when we exclude two periods after the treatment switches back to zero, the ATT estimates remain similar in magnitude, and statistically significant results remain significant (Figure A8 in the SM). Nevertheless, we recommend that researchers make it a practice to check for potential carryover effects, considering the low cost of conducting such tests and adjustments.

Finally, many findings are sensitive to model specifications. Some studies that we exclude from our sample employ one-way fixed effects or fixed effects at a level different from that at which treatment is assigned. Many of these findings do not hold when we reanalyze them using a TWFE model. We should clarify that this does not imply that the original results are wrong; rather, these models implicitly operate under different identifying assumptions, and there is substantial variation in how much consideration authors give to this point. Some studies do not provide a rationale for their choice to use one-way fixed effects, while others explicitly outline the type of unobserved confounders they intend to control for. In one instance, the authors inaccurately label their specification as a DID design. The TWFE and DID estimators are generally not equivalent, and we emphasize that this difference becomes even more pronounced when fixed effects are not assigned at the level of treatment. In such cases, a TWFE specification does not correspond to even a broadly defined DID research design.

\FloatBarrier

\paragraph*{Summary.} In Table~\ref{tb:summary} below, we summarize the main findings of our reanalysis. The numbers represent the proportion of studies in a given journal or across all journals that fall under the respective category (hence, a smaller number is better). The first two rows relate to the ratio of the imputation estimate to the TWFE estimate, which proxies the consequences of the weighting problem due to HTE. Across all journals, this ratio is less than 0.8 in 18\% of studies and less than 0.5 in 6\% of studies, indicating that using an HTE-robust estimator does sometimes have a substantively significant impact and is important to use at least as a robustness check.
\begin{table}[!ht]
\caption{Summary of Findings}\label{tb:summary}
    \centering\small
    \begin{tabular}{l|p{4em}<{\centering} p{4em}<{\centering} p{4em}<{\centering} p{4em}<{\centering} p{2em}<{\centering}} \hline\hline
    & \textit{APSR} & \textit{AJPS} & \textit{JOP} & All & $n$ \\ \hline
    Imputation to TWFE Ratio $< 0.8$ & 0.00 & 0.13 & 0.30 & 0.18 & 49 \\
    Imputation to TWFE Ratio $< 0.5$ & 0.00 & 0.00 & 0.13 & 0.06 & 49  \\ \\ 
    $F$ test reject null & 0.30 & 0.21 & 0.30 & 0.27 & 44 \\
    Placebo test reject null & 0.40 & 0.29 & 0.15 & 0.25 & 44 \\ \\ 
    TWFE \emph{not} reject null & 0.00 & 0.00 & 0.04 & 0.02 & 49 \\
    Imputation \emph{not} reject null & 0.00 & 0.13 & 0.44 & 0.25 & 49 \\ \\
    Imputation \emph{not} reject null w/ $\bar{M} = 0$ & 0.40 & 0.23 & 0.74 & 0.50 & 42 \\ 
    Imputation \emph{not} reject null w/ $\bar{M} = 0.5$ & 0.70 & 0.77 & 0.90 & 0.81 & 42 \\ \hline
    \multicolumn{6}{p{15.5cm}}{\footnotesize\textbf{\textit{Note:}} Entries (except in the last column) are proportions of studies satisfying each set of conditions. A null is deemed rejected if $p < 0.05$. Five studies with a single pre-treatment period are not included in the summary statistics for the $F$ test and placebo test. Seven studies are not included in the last two rows due to too few pre-treatment periods.}
    \end{tabular}
\end{table}
The third and fourth rows indicate the proportion of studies with evidence of PT violations based on the $F$ test and placebo tests, respectively. Across all journals, around a quarter of studies reject the null hypothesis at the 5\% level using the $F$ test, with slightly fewer rejecting when using the placebo test due to the loss in power from excluding data from the placebo periods. Note, though, that failure to reject can result from insufficient power and is not sufficient to support that the PT holds. 

The fifth and sixth rows display the proportion of studies in which the null hypothesis of no effect is not rejected using the TWFE and imputation estimators, respectively. When TWFE is employed, this occurs in only one study (2\%); however, when the imputation estimator is used, this number increases to 24\%, suggesting that many studies in our sample are potentially underpowered with an HTE-robust estimator. The last two rows show that in 50\% and 81\% of the studies, the robust confidence sets for the ATT include zero with $\bar{M} = 0$ (debiasing only) and $\bar{M} = 0.5$ (debiasing plus restricted relative magnitude in biases), respectively. The comparison of results in the last four rows highlights that the main source of fragility in the existing literature is potential PT violations, rather than concerns of HTE.

Our reanalysis is not meant to criticize existing studies, many of which were conducted before recent methodological advances.  In fact, we have observed rapid adoption of these new methods and greater statistical power in more recently published studies. However, we want to emphasize two key points: (1) many studies do not adequately assess the plausibility of PT, the key identifying assumption of their research designs; and (2) given recent methodological developments, causal panel analysis under PT requires significantly more statistical power to account for HTE and potential PT violations than previously believed.

\FloatBarrier


\section{Recommendations}

Based on the findings of the reanalysis, we provide the following recommendations.

\paragraph*{Research design is the key.} An important component of a strong research design is a clear understanding of how treatment is assigned. The PT assumption, which both TWFE models and most modern methods rely on, is silent on the assignment mechanism. As a result, many researchers assume that these methods can be applied when the assignment mechanism is unknown and that the absence of a pretrend is sufficient to make PT credible. Another perspective suggests that (quasi-)randomization is required for PT to be credible \citep[e.g.,][]{kahn2020promise}. Our view aligns more closely with the latter, but is less stringent. We argue that for PT to be credible, researchers must justify the following assumption:
\begin{center}
$\Delta_{s,t} Y_{i,t}(0) \indep D_{i,t},\ \forall s, t$ 
\end{center}
where $\Delta_{s,t} Y_{i,t}(0) = Y_{i,t}(0) - Y_{i,s}(0)$ is the before-and-after difference in untreated potential outcomes across any two periods (assuming no covariates, though this naturally extends to include them). This assumption is slightly stronger than PT, which only assumes mean independence, but is more intuitive. It demonstrates both the strength and limitation of causal panel analysis under PT: while the panel structure helps account for time-invariant unobserved confounding through before-and-after differences, the introduction of the treatment must act as a shock in so much as it is orthogonal to the evolution of untreated potential outcomes; hence, any dynamic relationships between past outcomes or covariates and treatment are ruled out. In other words, under PT, a strong causal panel design still requires some (quasi-)random element in treatment assignment. Importantly, the research design issues cannot be resolved simply by applying the novel estimators surveyed in this article.

\paragraph*{Inspecting raw data helps spot obvious issues.} The research design phase should also include inspection of the raw data. We encourage researchers to plot the data at hand to better understand patterns of the treatment status, missingness, and outliers \citep{mou2023panelview}. Treatment status should vary both by unit and time. If the majority of variation occurs over time (across units) with little or no variation between units at any given time period (or across time within a given unit), the TWFE estimand will be likely dominated by impermissible comparisons and thus susceptible to larger biases. Moreover, HTE-robust estimators will estimate the treatment effect using very little data and thus be underpowered. Equally important is the need for researchers to understand the degree and possible origins of data missingness prior to initiating statistical analysis. If missingness does not seem to be random, or if it is too prevalent, leaving insufficient meaningful variation in the data, researchers should consider halting the analysis at this stage. Just as in the cross-sectional case, plotting the raw data can also help researchers to spot outliers and highly skewed distributions, which may require additional preprocessing. At this stage, researchers can also trim the data to make the units in comparison more similar in terms of pre-treatment or time-invariant covariates \citep{SantAnna2018-sj}.

\paragraph*{When estimates diverge, understand why.} At the estimation stage, we recommend using at least one HTE-robust estimator alongside a benchmark TWFE model. While TWFE is often more efficient, its constant treatment effects assumption is too restrictive in many contexts. As shown in this article, TWFE does not produce systematic upward or downward bias compared to the imputation estimator, but it can severely bias causal estimates in individual cases. We recommend the imputation estimator in most settings primarily for its flexibility, though other estimators also have their advantages. The more critical question, however, is why results differ between estimators when they do. If TWFE and an HTE-robust estimator diverge, it could be due to HTE or PT violations. If HTE-robust estimators themselves diverge, it is often because the data are too sparse, leading to high variability, or the PT assumption fails differentially, causing estimators weighting control units differently to produce varying estimates. Plotting raw data and using diagnostics (such as the Goodman-Bacon decomposition) typically clarifies these issues. We also recommend keeping the benchmark TWFE model for its transparency. 

For uncertainty estimates, researchers should use cluster-robust SEs when the number of clusters is large (e.g., exceeds $50$) and opt for cluster-bootstrap or cluster-jackknife procedures when the number of clusters is relatively small. The clustering level should match the higher of either the time-series units or the level of treatment assignment. This follows the rule of thumb to cluster at the level of potential outcome input index, taking into account both treatment assignment and potential temporal spillover \citep{fu2025inference}.  For novel HTE-robust estimators, cluster-bootstrap or jackknife is generally safer than relying on various analytical SEs. In the SM, we show that cluster-bootstrapped SEs are typically larger than analytically derived SEs for five HTE-robust estimators using data from our sample.

\paragraph*{Conduct diagnostics to assess key assumptions and robustness of findings.} With a clear research design, researchers should critically evaluate key identification assumptions and test the robustness of findings when these assumptions are violated. Event-study plots, available for TWFE and most HTE-robust estimators, are valuable tools for assessing whether the no-anticipation and PT assumptions hold. We recommend creating an event-study plot using the chosen estimator(s), followed by both visual inspection and statistical tests to assess how plausible the PT assumption is. Importantly, the absence of statistical significance in pretrend coefficients should not be taken as conclusive evidence for the validity of PT. To avoid the conditional inference problem, we recommend performing a sensitivity analysis with robust confidence sets across different values of $\bar{M}$, regardless of the pretest results. As shown in this article, such tests require more statistical power than rejecting the null that the average treatment effect is zero. Researchers should, therefore, allocate sufficient statistical power for these diagnostic tests during the research design phase.

\bigskip

Panel data provide valuable opportunities for social scientists to tackle complex causal questions; however, these data, especially when analyzed under PT, present distinct challenges. Our findings are not intended to dissuade researchers from employing PT-based research designs in causal panel analysis. Rather, our aim is to guide researchers in conducting their analyses more transparently and credibly. To facilitate this, we have integrated all the procedures described in this paper into the open-source \texttt{R} package \texttt{fect}, and we offer detailed tutorials for these methods.

\FloatBarrier
\onehalfspacing

\section*{Data Availability Statement}
Research documentation and/or data that support the findings of this study are openly available at the \emph{American Political Science Review} Dataverse: \url{https://doi.org/10.7910/DVN/9RJFZF}.

\section*{Conflict of Interest}
 The authors declare no ethical issues or conflicts of interest in this research. 

\section*{Ethical Standards}
The authors affirm that this research did not involve human participants.

\clearpage
\bibliographystyle{apsr}
\bibliography{tscs.bib}

@misc{chiu2025replication,
  author       = {Chiu, Albert and Lan, Xingchen and Liu, Ziyi and Xu, Yiqing},
  year         = {2025},
  title        = {{Replication Data for: Causal Panel Analysis under Parallel Trends: Lessons from A Large Reanalysis Study}},
  howpublished = {\url{https://doi.org/10.7910/DVN/9RJFZF}},
  note         = {Harvard Dataverse}
}

@article{xu2024factorial,
  title={Factorial Difference-in-Differences},
  author={Xu, Yiqing and Zhao, Anqi and Ding, Peng},
  journal={arXiv preprint arXiv:2407.11937},
  year={2024}
}

@article{baker2025difference,
  title={Difference-in-Differences Designs: A Practitioner's Guide},
  author={Baker, Andrew and Callaway, Brantly and Cunningham, Scott and Goodman-Bacon, Andrew and Sant'Anna, Pedro HC},
  journal={arXiv preprint arXiv:2503.13323},
  year={2025}
}

@article{de2023credible,
  title={Credible Answers to Hard Questions: Differences-in-Differences for Natural experiments},
  author={de Chaisemartin, C and D’Haultf{\oe}uille, X},
  journal={Working Paper, SSRN},
  year={2023}
}

@article{roth2022pretest,
  title={Pretest with caution: Event-study estimates after testing for parallel trends},
  author={Roth, Jonathan},
  journal={American Economic Review: Insights},
  volume={4},
  number={3},
  pages={305--322},
  year={2022},
  publisher={American Economic Association 2014 Broadway, Suite 305, Nashville, TN 37203}
}

@article{roth2023s,
  title={What’s Trending in Difference-in-Differences? A Synthesis of the Recent Econometrics Literature},
  author={Roth, Jonathan and Sant’Anna, Pedro HC and Bilinski, Alyssa and Poe, John},
  journal={Journal of Econometrics},
  volume={235},
  number={2},
  pages={2218--2244},
  year={2023},
  publisher={Elsevier}
}

@article{Arkhangelsky2023-zy,
  title={Causal Models for Longitudinal and Panel Data: A Survey},
  author={Arkhangelsky, Dmitry and Imbens, Guido},
  journal={The Econometrics Journal},
  volume={27},
  number={3},
  pages={C1--C61},
  year={2024},
  publisher={Oxford University Press}
}

@article{bai2021matrix,
  title={Matrix Completion, Counterfactuals, and Factor Analysis of Missing Data},
  author={Bai, Jushan and Ng, Serena},
  journal={Journal of the American Statistical Association},
  volume={116},
  number={536},
  pages={1746--1763},
  year={2021},
  publisher={Taylor \& Francis}
}

@article{wing2024stacked,
  title={Stacked difference-in-differences},
  author={Wing, Coady and Freedman, Seth M and Hollingsworth, Alex},
  year={2024},
  journal={Working Paper, NBER}
}

@article{callaway2021-did,
  title={Difference-in-Differences with Multiple Time Periods},
  author={Callaway, Brantly and Sant’Anna, Pedro HC},
  journal={Journal of Econometrics},
  volume={225},
  number={2},
  pages={200--230},
  year={2021},
  publisher={Elsevier}
}

@article{sun2021-event,
  title={Estimating Dynamic Treatment Effects in Event Studies with Heterogeneous Treatment Effects},
  author={Sun, Liyang and Abraham, Sarah},
  journal={Journal of Econometrics},
  volume={225},
  number={2},
  pages={175--199},
  year={2021},
  publisher={Elsevier}
}

@ARTICLE{Green2001-dy,
  title     = "Dirty Pool",
  author    = "Green, Donald P and Kim, Soo Yeon and Yoon, David H",
  journal   = "International Organization",
  publisher = "Cambridge University Press",
  volume    =  55,
  number    =  2,
  pages     = "441--468",
  year      =  2001
}

@ARTICLE{King2001-br,
  title     = "Proper Nouns and Methodological Propriety: Pooling Dyads in
               International Relations Data",
  author    = "King, Gary",
  journal   = "International Organization",
  publisher = "Cambridge University Press",
  volume    =  55,
  number    =  2,
  pages     = "497--507",
  year      =  2001
}

@ARTICLE{Bertrand2004-rc,
  title     = "How Much Should We Trust {Differences-in-Differences} Estimates?",
  author    = "Bertrand, Marianne and Duflo, Esther and Mullainathan, Sendhil",
  journal   = "Quarterly Journal of Economics",
  publisher = "MIT Press",
  volume    =  119,
  number    =  1,
  pages     = "249--275",
  year      =  2004
}

@ARTICLE{Cameron2008-ou,
  title     = "Bootstrap-Based Improvements for Inference with Clustered Errors",
  author    = "Cameron, A Colin and Gelbach, Jonah B and Miller, Douglas L",
  journal   = "Review of Economics and Statistics",
  publisher = "The MIT Press",
  volume    =  90,
  number    =  3,
  pages     = "414--427",
  year      =  2008
}

@article{borusyak2024revisiting,
  title={Revisiting event-study designs: robust and efficient estimation},
  author={Borusyak, Kirill and Jaravel, Xavier and Spiess, Jann},
  journal={Review of Economic Studies},
  volume={91},
  number={6},
  pages={3253--3285},
  year={2024},
  publisher={Oxford University Press UK}
}

@article{Aronow2020-jc,
  title={Design-based Inference for Spatial Experiments under Unknown Interference},
  author={Wang, Ye and Samii, Cyrus and Chang, Haoge and Aronow, PM},
  journal={The Annals of Applied Statistics},
  volume={19},
  number={1},
  pages={744--768},
  year={2025},
  publisher={Institute of Mathematical Statistics}
}

@MISC{Wang2021-xo,
  title        = "Causal Inference under Temporal and Spatial Interference",
  author       = "Wang, Ye",
  howpublished = "Working Paper, New York University",
  year = 2021
}

@misc{fu2025inference,
  author = {Fu, Jiawei and Samii, Cyrus and Wang, Ye},
  year = {2025},
  title = {Inference for Group Interaction Experiments},
  note = {Conference Paper, presented at thh 2024 Annual Conference of the American Political Science Association.}
}

@article{liu2024practical,
  title={A practical guide to counterfactual estimators for causal inference with time-series cross-sectional data},
  author={Liu, Licheng and Wang, Ye and Xu, Yiqing},
  journal={American Journal of Political Science},
  volume={68},
  number={1},
  pages={160--176},
  year={2024},
  publisher={Wiley Online Library}
}

@book{wooldridge2010econometric,
  title={Econometric Analysis of Cross Section and Panel Data},
  author={Wooldridge, Jeffrey M},
  year={2010},
  publisher={MIT press}
}

@ARTICLE{Beck2001-as,
  title     = "Throwing Out the Baby with the Bath Water: A Comment on Green,
               Kim, and Yoon",
  author    = "Beck, Nathaniel and Katz, Jonathan N",
  journal   = "International Organization",
  publisher = "Cambridge University Press",
  volume    =  55,
  number    =  2,
  pages     = "487--495",
  year      =  2001
}

@ARTICLE{Beck1995-am,
  title     = "What to Do (and Not to Do) with {Time-Series} {Cross-Section}
               Data",
  author    = "Beck, Nathaniel and Katz, Jonathan N",
  journal   = "American Political Science Review",
  publisher = "[American Political Science Association, Cambridge University
               Press]",
  volume    =  89,
  number    =  3,
  pages     = "634--647",
  year      =  1995
}

@incollection{Xu2023,
  author      = {Xu, Yiqing},
  title       = {Causal Inference with Time-Series Cross-Sectional Data: A Reflection},
  editor      = {Box-Steffensmeier, Janet M. and Christenson, Dino P. and Sinclair-Chapman, Valeria},
  booktitle   = {The Oxford Handbook of Methodological Pluralism in Political Science},
  publisher   = {Oxford University Press},
  address     = {Oxford},
  year        = {2023},
  chapter     = {30}
}

@article{BLW2022,
  title={How Much Should We Trust Staggered Difference-in-Differences Estimates?},
  author={Baker, Andrew C and Larcker, David F and Wang, Charles CY},
  journal={Journal of Financial Economics},
  volume={144},
  number={2},
  pages={370--395},
  year={2022}
}

@article{Roth2021-hl,
  title={When is Parallel Trends Sensitive to Functional Form?},
  author={Roth, Jonathan and Sant'Anna, Pedro HC},
  journal={Econometrica},
  volume={91},
  number={2},
  pages={737--747},
  year={2023},
  publisher={Wiley Online Library}
}

@ARTICLE{SantAnna2018-sj,
  title         = "Doubly Robust {Difference-in-Differences} Estimators",
  author        = "Sant'Anna, Pedro H C and Zhao, Jun B",
  month         =  nov,
  year          =  2020,
  journal = "Journal of Econometrics", 
  volume = 219, 
  number = 1, 
  pages = "101--122"  
}

@article{eggers2015validity,
  title={On the Validity of the Regression Discontinuity Design for Estimating Electoral Effects: New Evidence from Over 40,000 Close Races},
  author={Eggers, Andrew C and Fowler, Anthony and Hainmueller, Jens and Hall, Andrew B and Snyder Jr, James M},
  journal={American Journal of Political Science},
  volume={59},
  number={1},
  pages={259--274},
  year={2015},
  publisher={Wiley Online Library}
}

@article{hainmueller2019much,
  title={How Much Should We Trust Estimates from Multiplicative Interaction Models? Simple Tools to Improve Empirical Practice},
  author={Hainmueller, Jens and Mummolo, Jonathan and Xu, Yiqing},
  journal={Political Analysis},
  volume={27},
  number={2},
  pages={163--192},
  year={2019},
  publisher={Cambridge University Press}
}

@ARTICLE{Roth2024interpret,
  title         = "Interpreting {Event-Studies} from Recent
                   {Difference-in-Differences} Methods",
  journal = "Working Paper",
  author        = "Roth, Jonathan",
  month         =  jan,
  year          =  2024,
  archivePrefix = "arXiv",
  primaryClass  = "econ.EM",
  eprint        = "2401.12309"
}

@article{lal2021much, title={How Much Should We Trust Instrumental Variable Estimates in Political Science? Practical Advice Based on 67 Replicated Studies}, volume={32}, DOI={10.1017/pan.2024.2}, number={4}, journal={Political Analysis}, author={Lal, Apoorva and Lockhart, Mackenzie and Xu, Yiqing and Zu, Ziwen}, year={2024}, pages={521–540}}

@ARTICLE{rambachan2023more,
  title     = "A More Credible Approach to Parallel Trends",
  author    = "Rambachan, Ashesh and Roth, Jonathan",
  journal   = "The Review of Economic Studies",
  publisher = "Oxford Academic",
  volume    =  90,
  number    =  5,
  pages     = "2555--2591",
  month     =  sep,
  year      =  2023
}

@article{cengiz2019effect,
  title={The Effect of Minimum Wages on {Low-Wage} Jobs},
  author={Cengiz, Doruk and Dube, Arindrajit and Lindner, Attila and Zipperer, Ben},
  journal={Quarterly Journal of Economics},
  volume={134},
  number={3},
  pages={1405--1454},
  year={2019},
  publisher={Oxford Academic}
}

@article{mou2023panelview,
  title={Panel Data Visualization in R (panelView) and Stata (panelview)},
  author={Mou, Hongyu and Liu, Licheng and Xu, Yiqing},
  journal={Journal of Statistical Software},
 volume={107},
 number={7},
 pages={1–20},
  year={2023}
}

@article{Goodman-Bacon2021-xb,
  title={Difference-in-Differences with Variation in Treatment Timing},
  author={Goodman-Bacon, Andrew},
  journal={Journal of Econometrics},
  volume={225},
  number={2},
  pages={254--277},
  year={2021},
  publisher={Elsevier}
}

@ARTICLE{CDH2020,
  title   = "Two-Way Fixed Effects Estimators with Heterogeneous Treatment
             Effects",
  author  = "de Chaisemartin, Cl{\'e}ment and D'Haultf{\oe}uille, Xavier",
  journal = "American Economic Review",
  volume  =  110,
  number  =  9,
  pages   = "2964--2996",
  year    =  2020
}

@article{athey2022design,
  title={Design-based analysis in difference-in-differences settings with staggered adoption},
  author={Athey, Susan and Imbens, Guido W},
  journal={Journal of Econometrics},
  volume={226},
  number={1},
  pages={62--79},
  year={2022},
  publisher={Elsevier}
}

@ARTICLE{Blackwell2018-br,
  title   = "How to Make Causal Inferences with {Time-Series} {Cross-Sectional}
             Data Under Selection on Observables",
  author  = "Blackwell, Matthew and Glynn, Adam",
  journal = "American Political Science Review",
  volume  =  112,
  number  =  2,
  pages   = "1067--1082",
  year    =  2018
}

@article{IKW2021,
  title={Matching Methods for Causal Inference with Time-Series Cross-Sectional Data},
  author={Imai, Kosuke and Kim, In Song and Wang, Erik H},
  journal={American Journal of Political Science},
  volume={67},
  number={3},
  pages={587--605},
  year={2023},
  publisher={Wiley Online Library}
}

@ARTICLE{Imai2019-nw,
  title   = "When Should We Use Unit Fixed Effects Regression Models for Causal
             Inference with Longitudinal Data?",
  author  = "Imai, Kosuke and Kim, In Song",
  journal = "American Journal of Political Science",
  volume  =  63,
  number  =  2,
  pages   = "467--490",
  month   =  apr,
  year    =  2019
}

@MISC{Strezhnev2018-ku,
  title        = "Semiparametric Weighting Estimators for {Multi-Period}
                  {Difference-in-Differences} Designs",
  author       = "Strezhnev, Anton",
  year         =  2018,
  howpublished = "Mimeo, New York University"
}

@article{Bischof2019,
author = {Bischof, Daniel and Wagner, Markus},
doi = {10.1111/ajps.12449},
issn = {15405907},
journal = {American Journal of Political Science},
number = {4},
pages = {888--904},
title = {{Do Voters Polarize When Radical Parties Enter Parliament?}},
volume = {63},
year = {2019}
}

@article{Bisgaard2018,
author = {Bisgaard, Martin and Slothuus, Rune},
doi = {10.1111/ajps.12349},
issn = {15405907},
journal = {American Journal of Political Science},
number = {2},
pages = {456--469},
title = {{Partisan Elites as Culprits? How Party Cues Shape Partisan Perceptual Gaps}},
volume = {62},
year = {2018}
}

@article{Caughey2017,
author = {Caughey, Devin and Warshaw, Christopher and Xu, Yiqing},
doi = {10.1086/692669},
issn = {14682508},
journal = {The Journal of Politics},
number = {4},
pages = {1342--1358},
title = {{Incremental Democracy: The Policy Effects of Partisan Control of State Government}},
volume = {79},
year = {2017}
}

@article{Christensen2021,
  title={The Politics of Property Taxation: Fiscal Infrastructure and Electoral Incentives in Brazil},
  author={Christensen, Darin and Garfias, Francisco},
  journal={The Journal of Politics},
  volume={83},
  number={4},
  pages={1399--1416},
  year={2021},
  publisher={The University of Chicago Press Chicago, IL}
}

@article{Clarke2020,
author = {Clarke, Andrew J.},
doi = {10.1111/ajps.12504},
issn = {15405907},
journal = {American Journal of Political Science},
number = {3},
pages = {452--470},
title = {{Party Sub-Brands and American Party Factions}},
volume = {64},
year = {2020}
}

@article{Clayton2018,
author = {Clayton, Amanda and Zetterberg, P{\"{a}}r},
doi = {10.1086/697251},
issn = {14682508},
journal = {The Journal of Politics},
number = {3},
pages = {916--932},
title = {{Quota Shocks: Electoral Gender Quotas and Government Spending Priorities Worldwide}},
volume = {80},
year = {2018}
}

@article{Distelhorst2018,
author = {Distelhorst, Greg and Locke, Richard M.},
doi = {10.1111/ajps.12372},
issn = {15405907},
journal = {American Journal of Political Science},
number = {3},
pages = {695--711},
title = {{Does Compliance Pay? Social Standards and Firm-Level Trade}},
volume = {62},
year = {2018}
}

@article{Fouirnaies2018ajps,
author = {Fouirnaies, Alexander},
doi = {10.1111/ajps.12316},
journal = {American Journal of Political Science},
number = {1},
pages = {176--191},
title = {{When Are Agenda Setters Valuable?}},
volume = {62},
year = {2018}
}

@article{fh2018,
author = {Fouirnaies, Alexander and Hall, Andrew B.},
doi = {10.1111/ajps.12323},
journal = {American Journal of Political Science},
number = {1},
pages = {132--147},
title = {{How Do Interest Groups Seek Access to Committees?}},
volume = {62},
year = {2018}
}

@article{Fresh2018,
author = {Fresh, Adriane},
doi = {10.1086/695852},
issn = {14682508},
journal = {The Journal of Politics},
number = {2},
pages = {713--718},
title = {{The Effect of the Voting Rights Act on Enfranchisement: Evidence from North Carolina}},
volume = {80},
year = {2018}
}

@article{Garfias2019jop,
author = {Garfias, Francisco},
doi = {10.1086/700105},
issn = {14682508},
journal = {The Journal of Politics},
number = {1},
pages = {95--111},
title = {{Elite Coalitions, Limited Government, and Fiscal Capacity Development: Evidence from Bourbon Mexico}},
volume = {81},
year = {2019}
}

@article{Grumbach2020,
author = {Grumbach, Jacob M. and Sahn, Alexander},
doi = {10.1017/S0003055419000637},
issn = {15375943},
journal = {American Political Science Review},
number = {1},
pages = {206--221},
title = {{Race and Representation in Campaign Finance}},
volume = {114},
year = {2020}
}

@article{Jiang2018,
author = {Jiang, Junyan},
doi = {10.1111/ajps.12394},
issn = {15405907},
journal = {American Journal of Political Science},
number = {4},
pages = {982--999},
title = {{Making Bureaucracy Work: Patronage Networks, Performance Incentives, and Economic Development in China}},
volume = {62},
year = {2018}
}

@article{Kogan2021,
author = {Kogan, Vladimir},
doi = {10.1086/708914},
issn = {14682508},
journal = {The Journal of Politics},
number = {1},
pages = {20--70},
title = {{Do Welfare Benefits Pay Electoral Dividends? Evidence from the National Food Stamp Program Rollout}},
volume = {83},
year = {2021}
}

@article{Payson2020apsr,
author = {Payson, Julia A. },
doi = {10.1017/S0003055420000118},
issn = {15375943},
journal = {American Political Science Review},
number = {3},
pages = {677--690},
title = {{The Partisan Logic of City Mobilization: Evidence from State Lobbying Disclosures}},
volume = {114},
year = {2020a}
}

@article{Payson2020jop,
author = {Payson, Julia A. },
doi = {10.1086/706767},
issn = {14682508},
journal = {The Journal of Politics},
number = {2},
pages = {403--417},
title = {{Cities in the Statehouse: How Local Governments Use Lobbyists to Secure State Funding}},
volume = {82},
year = {2020b}
}

@article{magaloni2020killing,
  title={Killing in the Slums: Social Order, Criminal Governance, and Police Violence in Rio de Janeiro},
  author={Magaloni, Beatriz and Franco-Vivanco, Edgar and Melo, Vanessa},
  journal={American Political Science Review},
  volume={114},
  number={2},
  pages={552--572},
  year={2020},
  publisher={Cambridge University Press}
}

@article{Pierskalla2018,
author = {Pierskalla, Jan H. and Sacks, Audrey},
doi = {10.1086/694547},
issn = {14682508},
journal = {The Journal of Politics},
number = {2},
pages = {510--524},
title = {{Unpaved Road Ahead: The Consequences of Election Cycles for Capital Expenditures}},
volume = {80},
year = {2018}
}

@article{Schubiger2021,
author = {Schubiger, Livia Isabella},
doi = {10.1086/711720},
issn = {14682508},
journal = {The Journal of Politics},
number = {4},
pages = {1383--1398},
title = {{State Violence and Wartime Civilian Agency: Evidence from Peru}},
volume = {83},
year = {2021}
}

@article{Schuit2017,
author = {Schuit, Sophie and Rogowski, Jon C.},
doi = {10.1111/ajps.12284},
issn = {15405907},
journal = {American Journal of Political Science},
number = {3},
pages = {513--526},
title = {{Race, Representation, and the Voting Rights Act}},
volume = {61},
year = {2017}
}

@article{Trounstine2020,
author = {Trounstine, Jessica},
doi = {10.1017/S0003055419000844},
issn = {15375943},
journal = {American Political Science Review},
number = {2},
pages = {443--455},
title = {{The Geography of Inequality: How Land Use Regulation Produces Segregation}},
volume = {114},
year = {2020}
}

@article{Weschle2021,
author = {Weschle, Simon},
doi = {10.1086/710087},
issn = {14682508},
journal = {The Journal of Politics},
number = {2},
pages = {706--721},
title = {{Parliamentary Positions and Politicians' Private Sector Earnings: Evidence from the UK House of Commons}},
volume = {83},
year = {2021}
}

@article{lall2016multiple,
  title={How Multiple Imputation Makes a Difference},
  author={Lall, Ranjit},
  journal={Political Analysis},
  volume={24},
  number={4},
  pages={414--433},
  year={2016},
  publisher={Cambridge University Press}
}

@article{Zhang2021jop,
author = {Zhang, Qi and Zhang, Dong and Liu, Mingxing and Shih, Victor},
doi = {10.1086/711131},
issn = {14682508},
journal = {The Journal of Politics},
number = {3},
pages = {1010--1023},
title = {{Elite Cleavage and the Rise of Capitalism under Authoritarianism: A Tale of Two Provinces in China}},
volume = {83},
year = {2021}
}

@article{cox2021budgetary,
  title={The Budgetary Origins of Fiscal-Military Prowess},
  author={Cox, Gary W and Dincecco, Mark},
  journal={The Journal of Politics},
  volume={83},
  number={3},
  pages={851--866},
  year={2021},
  publisher={The University of Chicago Press Chicago, IL}
}

@article{hainmueller2019does,
  title={Does Direct Democracy Hurt Immigrant Minorities? Evidence from Naturalization Decisions in Switzerland},
  author={Hainmueller, Jens and Hangartner, Dominik},
  journal={American Journal of Political Science},
  volume={63},
  number={3},
  pages={530--547},
  year={2019},
  publisher={Wiley Online Library}
}

@article{Grumbach2022,
  title={Rock the Registration: Same Day Registration Increases Turnout of Young Voters},
  author={Grumbach, Jacob M and Hill, Charlotte},
  journal={The Journal of Politics},
  volume={84},
  number={1},
  pages={405--417},
  year={2022},
  publisher={The University of Chicago Press Chicago, IL}
}

@article{Ravanilla2022,
  title={Deadly Populism: How Local Political Outsiders Drive Duterte’s War on Drugs in the Philippines},
  author={Ravanilla, Nico and Sexton, Renard and Haim, Dotan},
  journal={The Journal of Politics},
  volume={84},
  number={2},
  pages={1035--1056},
  year={2022},
  publisher={The University of Chicago Press Chicago, IL}
}

@article{Bokobza2022,
   author = {Laure Bokobza and Suthan Krishnarajan and Jacob Nyrup and Casper Sakstrup and Lasse Aaskoven},
   doi = {10.1086/716952},
   issn = {14682508},
   number = {3},
   journal = {The Journal of Politics},
   month = {7},
   pages = {1437-1452},
   publisher = {University of Chicago Press},
   title = {The Morning After: Cabinet Instability and the Purging of Ministers after Failed Coup Attempts in Autocracies},
   volume = {84},
   year = {2022},
}

@article{Blair2022,
   author = {Graeme Blair and Darin Christensen and Valerie Wirtschafter},
   doi = {10.1086/715255},
   issn = {14682508},
   number = {1},
   journal = {The Journal of Politics},
   month = {1},
   pages = {116-133},
   publisher = {University of Chicago Press},
   title = {How Does Armed Conflict Shape Investment? Evidence from the Mining Sector},
   volume = {84},
   year = {2022},
}

@article{eckhouse2022metrics,
  title={Metrics Management and Bureaucratic Accountability: Evidence from Policing},
  author={Eckhouse, Laurel},
  journal={American Journal of Political Science},
  volume={66},
  number={2},
  pages={385--401},
  year={2022},
  publisher={Wiley Online Library}
}

@article{Schafer2021,
  title={Making Unequal Democracy Work? The Effects of Income on Voter Turnout in Northern Italy},
  author={Schafer, Jerome and Cantoni, Enrico and Bellettini, Giorgio and Berti Ceroni, Carlotta},
  journal={American Journal of Political Science},
  volume={66},
  number={3},
  pages={745--761},
  year={2022},
  publisher={Wiley Online Library}
}

@article{kilborn2022public,
  title={Public Money Talks Too: How Public Campaign Financing Degrades Representation},
  author={Kilborn, Mitchell and Vishwanath, Arjun},
  journal={American Journal of Political Science},
  volume={66},
  number={3},
  pages={730--744},
  year={2022},
  publisher={Wiley Online Library}
}

@article{Fouirnaies2022,
  title={How Do Electoral Incentives Affect Legislator Behavior? Evidence from U.S. State Legislatures},
  author={Fouirnaies, Alexander and Hall, Andrew B},
  journal={American Political Science Review},
  volume={116},
  number={2},
  pages={662--676},
  year={2022},
  publisher={Cambridge University Press}
}

@article{Hirano2022,
   author = {Shigeo Hirano and Jaclyn Kaslovsky and Michael P. Olson and James M. Snyder},
   doi = {10.1086/719008},
   issn = {14682508},
   number = {3},
   journal = {The Journal of Politics},
   month = {7},
   pages = {1482-1496},
   publisher = {University of Chicago Press},
   title = {The Growth of Campaign Advertising in the United States, 1880–1930},
   volume = {84},
   year = {2022},
}

@article{Paglayan2022,
  title={Education or Indoctrination? The Violent Origins of Public School Systems in an Era of State-Building},
  author={Paglayan, Agustina S},
  journal={American Political Science Review},
  volume={116},
  number={4},
  pages={1242--1257},
  year={2022},
  publisher={Cambridge University Press}
}

@article{Beazer2022,
   author = {Quintin H. Beazer and Ora John Reuter},
   doi = {10.1086/714775},
   issn = {14682508},
   number = {1},
   journal = {The Journal of Politics},
   month = {1},
   pages = {437-454},
   publisher = {University of Chicago Press},
   title = {Do Authoritarian Elections Help the Poor? Evidence from Russian Cities},
   volume = {84},
   year = {2022},
}

@article{Hall2022,
   author = {Andrew B. Hall and Jesse Yoder},
   doi = {10.1086/714932},
   issn = {14682508},
   number = {1},
   journal = {The Journal of Politics},
   month = {1},
   pages = {351-366},
   publisher = {University of Chicago Press},
   title = {Does Homeownership Influence Political Behavior? Evidence from Administrative Data},
   volume = {84},
   year = {2022},
}

@article{Liao2023,
   author = {Steven Liao},
   doi = {10.1086/723984},
   issn = {14682508},
   number = {4},
   journal = {The Journal of Politics},
   month = {10},
   pages = {1416-1429},
   publisher = {University of Chicago Press},
   title = {The Effect of Firm Lobbying on High-Skilled Visa Adjudication},
   volume = {85},
   year = {2023},
}

@article{Grumbach2023,
  title={Laboratories of Democratic Backsliding},
  author={Grumbach, Jacob M},
  journal={American Political Science Review},
  volume={117},
  number={3},
  pages={967--984},
  year={2023}
}

@article{Marsh2023,
  title={Trauma and Turnout: The Political Consequences of Traumatic Events},
  author={Marsh, Wayde ZC},
  journal={American Political Science Review},
  volume={117},
  number={3},
  pages={1036--1052},
  year={2023}
}

@article{Kroeger2023,
  title={Motivated Corporate Political Action: Evidence from an SEC Experiment},
  author={Kroeger, Mary and Silfa, Maria},
  journal={The Journal of Politics},
  volume={85},
  number={3},
  pages={1139--1144},
  year={2023},
  publisher={The University of Chicago Press Chicago, IL}
}

@article{Dipoppa2023,
   author = {Gemma Dipoppa and Guy Grossman and Stephanie Zonszein},
   doi = {10.1086/722346},
   issn = {14682508},
   number = {2},
   journal = {The Journal of Politics},
   month = {4},
   pages = {389-404},
   publisher = {University of Chicago Press},
   title = {Locked Down, Lashing Out: COVID-19 Effects on Asian Hate Crimes in Italy},
   volume = {85},
   year = {2023},
}

@article{Kuipers2023,
  title={The Representational Consequences of Municipal Civil Service Reform},
  author={Kuipers, Nicholas and Sahn, Alexander},
  journal={American Political Science Review},
  volume={117},
  number={1},
  pages={200--216},
  year={2023},
  publisher={Cambridge University Press}
}

@article{sanford2023democratization,
  title={Democratization, Elections, and Public Goods: The Evidence from Deforestation},
  author={Sanford, Luke},
  journal={American Journal of Political Science},
  volume={67},
  number={3},
  pages={748--763},
  year={2023},
  publisher={Wiley Online Library}
}

@article{skorge2023mobilizing,
  title={Mobilizing the Underrepresented: Electoral Systems and Gender Inequality in Political Participation},
  author={Skorge, {\O}yvind S{\o}raas},
  journal={American Journal of Political Science},
  volume={67},
  number={3},
  pages={538--552},
  year={2023},
  publisher={Wiley Online Library}
}

@article{latura2023corporate,
  title={Corporate Board Quotas and Gender Equality Policies in the Workplace},
  author={Latura, Audrey and Weeks, Ana Catalano},
  journal={American Journal of Political Science},
  volume={67},
  number={3},
  pages={606--622},
  year={2023},
  publisher={Wiley Online Library}
}

@article{Esberg2023,
   author = {Jane Esberg and Alexandra A. Siegel},
   doi = {10.1017/S0003055422001290},
   issn = {15375943},
   number = {4},
   journal = {American Political Science Review},
   month = {11},
   pages = {1361-1378},
   publisher = {Cambridge University Press},
   title = {How Exile Shapes Online Opposition: Evidence from Venezuela},
   volume = {117},
   year = {2023},
}

@article{Dahlstrom2023,
   author = {Carl Dahlström and Mikael Holmgren},
   doi = {10.1086/717756},
   issn = {14682508},
   number = {2},
   journal = {The Journal of Politics},
   month = {4},
   pages = {640-653},
   publisher = {University of Chicago Press},
   title = {Loyal Leaders, Affluent Agencies: The Budgetary Implications of Political Appointments in the Executive Branch},
   volume = {85},
   year = {2023},
}

@article{kahn2020promise,
  title={The Promise and Pitfalls of Differences-in-Differences: Reflections on \textit{16 and Pregnant} and Other Applications},
  author={Kahn-Lang, Ariella and Lang, Kevin},
  journal={Journal of Business \& Economic Statistics},
  volume={38},
  number={3},
  pages={613--620},
  year={2020},
  publisher={Taylor \& Francis}
}

@article{Hankinson2023,
   author = {Michael Hankinson and Asya Magazinnik},
   doi = {10.1086/723818},
   issn = {14682508},
   number = {3},
   journal = {The Journal of Politics},
   month = {7},
   pages = {1033-1047},
   publisher = {University of Chicago Press},
   title = {The Supply-Equity Trade-Off: The Effect of Spatial Representation on the Local Housing Supply},
   volume = {85},
   year = {2023},
}

@article{de2024difference,
  title={Difference-in-Differences Estimators of Intertemporal Treatment Effects},
  author={de Chaisemartin, Cl{\'e}ment and D'Haultf{\oe}uille, Xavier},
  journal={Review of Economics and Statistics},
  pages={1--45},
  year={2024},
  publisher={MIT Press One Rogers Street, Cambridge, MA 02142-1209, USA journals-info~…}
}
\clearpage

\end{document}